\documentclass[a4paper,10pt]{article}
\pdfoutput=1
\usepackage{jheppub}
\usepackage{dsfont}
\usepackage{graphicx}
\usepackage{amsmath}
\usepackage{empheq}
\usepackage{subcaption}
\usepackage{todonotes}
\usepackage{mathtools}

\newlength\dlf  

\usepackage{tikz}
\usetikzlibrary{arrows,shapes}
\usetikzlibrary{trees}
\usetikzlibrary{patterns}
\usetikzlibrary{matrix,arrows} 				
\usetikzlibrary{positioning}				
\usetikzlibrary{calc,through}				
\usetikzlibrary{decorations.pathreplacing}  
\usepackage{pgffor}							

\usetikzlibrary{decorations.pathmorphing}	
\usetikzlibrary{decorations.markings}

\newcommand{\ang}[1]{\langle #1\rangle}
\newcommand{\sq}[1]{[ #1]}

\usepackage{xcolor}
\renewcommand{\imath}{\mathrm{i}}

\let\originalleft\left
\let\originalright\right
\renewcommand{\left}{\mathopen{}\mathclose\bgroup\originalleft}
\renewcommand{\right}{\aftergroup\egroup\originalright}

\begin{document}

\title{Smooth Splitting and Zeros from On-Shell Recursion}
\author[a]{Callum R. T. Jones,}
\emailAdd{c.r.t.jones@uva.nl}
\author[b]{Shruti Paranjape,}
\emailAdd{shruti\_paranjape@brown.edu}
\affiliation[a]{
Institute of Physics, University of Amsterdam, Amsterdam, 1098 XH, The Netherlands
}
\affiliation[b]{
Department of Physics, Brown University, 182 Hope Street, Providence, RI 02912, USA.
}

\abstract{
We describe a new approach to understanding the origins of recently discovered ``hidden zeros" and ``smooth splitting" of tree-level amplitudes in $\text{Tr}\phi^3$, Non-Linear Sigma Model (NLSM), Yang-Mill-Scalar (YMS) and the special Galileon. Introducing a new type of linear shift in kinematic space we demonstrate that the mysterious splitting formulae follow from a simple contour integration argument in the style of on-shell recursion. The argument makes use of only standard notions of tree-level factorization on propagators, but assumes improved UV behavior in the form of the absence of a residue at infinity. In the case of $\text{Tr}\phi^3$  and NLSM this is proven by identifying our shift as a special case of a more general construction called a $g$-vector shift; in the case of YMS it remains an unproven conjecture. This recursive perspective leads to numerous new results: we derive generalizations of the splitting formulae on more relaxed near-zero kinematics, including interesting new kinematic limits in which the amplitude splits into a triple-product; we also demonstrate that the uncolored special Galileon model has improved UV scaling and hence also splits. We also investigate the possible realization of hidden zeros in four dimensions. The conditions under which the dimensionality constraints are compatible with zero kinematics is investigated in detail for $\text{Tr}\phi^3$ and YMS; for the latter we find they can be realized only with certain restrictions on external helicity states. The realizable 4d zeros are proven by a similar recursive argument based on BCFW and is found to generalize to a new class of intrinsically 4d ``helicity zeros" present in all sectors of YM and also gravity. 
}

\maketitle
\flushbottom


\section{Introduction}
\label{sec:intro}
Studying the analytic properties of scattering amplitudes is crucial in order to understand, bootstrap and calculate observables in quantum field theory. At tree-level, amplitudes are rational functions of the external data, fully characterized by their zeros and poles. Zeros are often related to the existence of a Ward identity, for example of a spontaneously broken symmetry \cite{Adler:1964um}. Poles on the other hand, are associated with locality and unitarity, which determine that the associated residue factorizes. Yet recently a new class of ``hidden zeros'' \cite{Arkani-Hamed:2023swr} were reported in a large class of models such as Tr$\phi^3$, Yang-Mills Scalar (YMS), Non-Linear Sigma Model (NLSM) and the special Galileon \cite{Bartsch:2024amu,Li:2024qfp}, and the amplitudes were seen to split or factorize near these zeros \cite{Arkani-Hamed:2024fyd,Cao:2024gln, Cao:2024qpp}. Despite the superficial similarity with ordinary factorization, these splitting relations are no longer an obvious consequence of the unitarity of the theory. The main result of this paper is a new perspective on these mysterious properties; we demonstrate that assuming only standard analyticity and factorization properties of tree-amplitudes, the splitting relations are equivalent to a certain kind of improved UV behavior. 

In so-called on-shell constructible theories, unitarity dictates that all the information required to construct an amplitude is localized to its poles and residues. This allows for the efficient calculation of amplitudes via on-shell recursion relations that build higher-point amplitudes from lower-point ones \cite{Britto:2004ap, Britto:2005fq, Cheung:2014dqa, Cheung:2015ota, Berends:1987me, Cachazo:2004kj, He:2018svj,Yang:2019esm}. One important example is the BCFW recursion relation \cite{Britto:2005fq,Britto:2004ap} that calculates Yang-Mills (YM) and gravity amplitudes. This recursion is derived from applying Cauchy's residue theorem to an amplitude evaluated on complex-shifted kinematics. Thus it not only relies on knowing the residues on the poles, but also on the fact that the amplitudes do not have a pole at infinity \cite{Britto:2004ap,Britto:2005fq,Benincasa:2007qj}. This ``good UV behavior'' is closely related to the fact that these theories are often, but not always, power-counting renormalizable. Having a recursive construction of amplitudes allows us to prove many properties such as existence \cite{Cheung:2016drk}, supersymmetrizability \cite{Elvang:2018dco} and more recently the existence of hidden zeros in Tr$\phi^3$ \cite{Feng:2025ofq}. Other relevant discussions of the importance of improved UV behavior and the role of residues at infinity for tree-amplitudes and loop integrands include \cite{Herrmann:2016qea,Herrmann:2018dja,Trnka:2020dxl,Paranjape:2023qsq,Carrasco:2019qwr,Jin:2015pua,Belayneh:2024lzq,Cachazo:2024mdn}.

In this paper, we ask the question: does there exist a recursive proof of the existence of smooth splitting in theories with hidden zeroes? This includes not just Tr$\phi^3$, but also YMS and effective field theories like NLSM and special Galileon. As with other residue theorems, our study of the origin of such smooth splitting relies on both unitarity and good behavior of the amplitude in the UV. We find that this leads to a variety of generalized splitting theorems in these theories, of which near-zero splitting is a special case.
	
Understanding the pole at infinity has been a topic of study in a variety of theories \cite{Bern:2012gh, Bern:2014sna, Bern:2017lpv, Edison:2019ovj, Brown:2022wqr, Bourjaily:2018omh, Drummond:2008vq, Cheung:2014dqa}. Here we take two different approaches to the behavior at infinity. In models like YMS and special Galileon, the lack of a direct surface description makes a proof of enhanced fall-off at infinity difficult. Instead, we use the fact that the amplitude splits to conjecture good UV behavior. In theories like Tr$\phi^3$ and NLSM, we utilize the recently introduced surface description \cite{Arkani-Hamed:2023lbd, Arkani-Hamed:2024vna, Arkani-Hamed:2023mvg} of amplitudes to prove the enhanced fall-off at infinity. 

Surfaceology uses chords on a surface to encode the combinatorics of propagators in an amplitude. It also provides a unified view of Tr$\phi^3$, YM, YMS, NLSM and bosonic strings \cite{Arkani-Hamed:2023jry, Arkani-Hamed:2024nhp, Arkani-Hamed:2024yvu}. This extends some properties of the closely related positive geometry description of Tr$\phi^3$ (via the ABHY realization of the associahedron \cite{Arkani-Hamed:2019vag}) to YM, YMS, NLSM and bosonic strings. In particular, the property of hidden zeros and near-zero splitting, which were first discovered in Tr$\phi^3$ as flattenings of the ABHY polytope and the splitting of its corresponding canonical form. While the hidden zeros have a clear geometric meaning in the form of sending Minkowski summands of simple polytopes to zero, the near-zero splitting is not apparent from the geometric construction, providing another motivation for understanding the origin of smooth splitting.

Using the residue theorem approach, we not only prove the existence of near-zero splitting but a larger class of generalized splitting formulae. Just as residues on poles are fixed products of lower-point amplitudes, these splitting formulae show that tuning certain non-pole kinematic invariants to zero can also be used to isolate a fixed subset of Feynman diagrams. It is also interesting that some of these splitting theorems involve the amplitude evaluated on kinematics that are far outside the positive orthant on which the ABHY associahedron lives, making these very different from the existing near-zero splitting theorems. 

Note that all of the discussion about hidden zeros so far has been in an arbitrary number of dimensions, high enough so that no dimension-dependent identities have to be taken into account. In this paper, we discuss the restriction of hidden zeros to four dimensions, the presumed number of dimensions of our universe. We find that hidden zero conditions are more subtle in 4d. In particular, the existence of a vast number of dimension-dependent identities requires that many additional invariants (not in the original zero locus) be set to zero. This makes avoiding a pole non-trivial and indeed can only be achieved in special cases. In addition, for theories with spin like YM and gravity, only certain helicity configurations can realize the zero conditions. 

This paper is organized as follows. In Section \ref{subsec:mesh}, we provide an introduction to the kinematic mesh, hidden zeros and near-zero splitting. Section \ref{sec:shifts} introduces the kinematic shift that we use throughout this work, while Section \ref{sec:gvector} connects it to known shifts and the Feynman fan. In Section \ref{sec:proofofzeros} we prove the existence of zeros and splitting via a residue theorem. In addition, Section \ref{sec:highersplits} contains novel generalized splitting theorems, including triple-splitting for NLSM and Tr$\phi^3$. In Section \ref{sec:galileon}, we discuss the generalization of our methods to theories without color, in particular the special Galileon. Finally, in Section \ref{sec:4d}, we discuss how hidden zeros manifest in four dimensions. We prove their existence in YM and gravity via BCFW recursion in Section \ref{sec:BCFW}. We end with the Discussion.

\section{Shifting the Kinematic Mesh}
\label{sec:contour}

\subsection{Primer on the kinematic mesh}
\label{subsec:mesh}

To describe the hidden properties of ordered scalar amplitudes it has proven to be very useful to organize the Mandelstam invariants graphically into a so-called \textit{kinematic mesh}. Details of the motivation behind this construction have been given at length elsewhere \cite{Arkani-Hamed:2019vag}, so in this subsection we will only review some essential properties of ordered amplitudes and their realization in the mesh that will be useful for the rest of the paper. 

\begin{itemize}
    \item \textbf{Kinematic variables:} In $d>n-2$ dimensions, scalar amplitudes are functions of $\frac{n(n-3)}{2}$ independent Mandelstam invariants. For ordered amplitudes a natural choice of variables is given by the \textit{planar} variables
    \begin{equation}
        X_{ij} \equiv \left(p_i+p_{i+1}+...+p_{j-2}+p_{j-1}\right)^2,
    \end{equation}
    where it is always understood that the subscripts are defined modulo $n$. The (dependent) non-planar variables can be expressed in terms of the 2-particle invariants
    \begin{equation}
        c_{ij} \equiv -2p_i\cdot p_j.
    \end{equation}
    The graphical arrangement of these variables in the mesh is shown in Figure \ref{fig:mesh1}; the $X_{ij}$ variables are associated with the node at the intersection of the upward diagonal rays labeled $i$ and $j$ and the $c_{ij}$ associated with the plaquette immediately above the corresponding node. The $n$-point amplitude is then a function of $X_{ij}$'s associated to all nodes belonging to the principal domain i.e. the triangular region in Figure \ref{fig:mesh1}. 
    
    \item \textbf{Rectangle rule:} The planar and non-planar invariants are related in a simple way
    \begin{equation}
        c_{ij} = X_{i,j} + X_{i+1,j+1}- X_{i,j+1} - X_{i+1,j}.
    \end{equation}
    This relation generalizes in the mesh according to a simple graphical rule that we will make repeated use of in later sections. In the mesh, we can draw any rectangular region anchored by the nodes $X_T$, $X_B$, $X_L$ and $X_R$ at the \textit{top}, \textit{bottom}, \textit{left} and \textit{right} respectively, as shown in Figure \ref{fig:RectRule}. For any such region we have
    \begin{equation}
        X_T+X_B - X_L - X_R = \sum_{(ij)\in\text{interior}} c_{ij}.
    \end{equation}
    
    \item \textbf{Factorization:} Tree-level amplitudes have simple poles where Feynman propagators vanish, the residue of these poles are related by unitarity to the product of lower-point amplitudes\footnote{Throughout this paper the labels $\sigma$ in $A\left[\{\sigma\}\right]$ refer to the set of $X$-variables on which the amplitude depends, not directly to the external momenta. In (\ref{factor}) the amplitude $A\left[i,i+1,...,j-1,j\right]$ is a function of the cyclic Mandelstam invariants formed from momenta in the ordered set $\{p_i,p_{i+1},...,p_{j-1}\}$. Invariants of the form $\left(p_k+p_{k+1}+...+p_{j-1}\right)^2$ are \textit{formally} identical to the invariant $X_{k,j}$; exploiting this trivial fact we find it convenient to label the sub-amplitudes in this way without explicit reference to an ``internal" momentum.  }
    \begin{equation}
        \label{factor}
        \lim_{X_{ij}\rightarrow 0} X_{ij}A\left[1,2,...,n\right] = A\left[i,i+1,...,j-1,j\right]A\left[j,j+1,...,i-1,i\right].
    \end{equation}
    In the mesh, as depicted in Figure \ref{fig:Fact}, the sub-amplitudes that appear in the factorization on a pole $X_{ij}$ correspond to the smaller triangular regions inside of the rays extending from the $(i,j)$ node to either boundary. 

    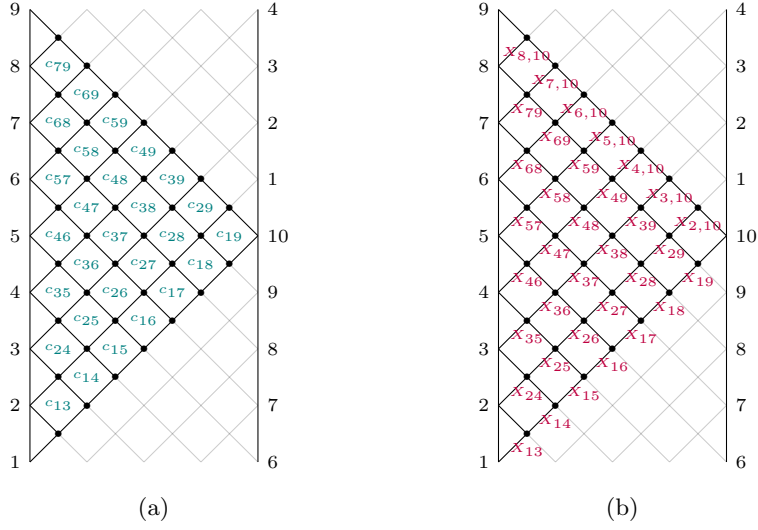
\begin{figure}
    \centering
            \begin{subfigure}[t]{0.4\textwidth}
                \centering
                \begin{tikzpicture}[scale=0.75]
                \draw (0,0)--(4,4);
			\draw (0,1)--(3.5,4.5);
			\draw(0,2)--(3,5);
			\draw(0,3)--(2.5,5.5);
			\draw(0,4)--(2,6);
			\draw (0,5)--(1.5,6.5);
			\draw(0,6)--(1,7);
			\draw(0,7)--(0.5,7.5);
			\draw(0,0)--(0,8);
			\draw (4,0)--(4,8);
			\draw(0,8)--(4,4);
			\draw(0,7)--(3.5,3.5);
			\draw(0,6)--(3,3);
			\draw(0,5)--(2.5,2.5);
			\draw(0,4)--(2,2);
			\draw(0,3)--(1.5,1.5);
			\draw(0,2)--(1,1);
			\draw(0,1)--(0.5,0.5); 
                \draw[opacity=0.2](0.5,0.5)--(1,0); 
                \draw[opacity=0.2](1,1)--(2,0); 
                \draw[opacity=0.2](1.5,1.5)--(3,0); 
                \draw[opacity=0.2](2,2)--(4,0);
                \draw[opacity=0.2](2.5,2.5)--(4,1);
                \draw[opacity=0.2](3,3)--(4,2);
                \draw[opacity=0.2](3.5,3.5)--(4,3);
                \draw[opacity=0.2](3.5,4.5)--(4,5); 
                \draw[opacity=0.2](3,5)--(4,6);
                \draw[opacity=0.2](2.5,5.5)--(4,7);
                \draw[opacity=0.2](2,6)--(4,8);
                \draw[opacity=0.2](1.5,6.5)--(3,8);
                \draw[opacity=0.2](1,7)--(2,8);
                \draw[opacity=0.2](0.5,7.5)--(1,8);
                \draw[opacity=0.2](1,0)--(4,3);
                \draw[opacity=0.2](2,0)--(4,2);
                \draw[opacity=0.2](3,0)--(4,1);
                \draw[opacity=0.2](4,5)--(1,8);
                \draw[opacity=0.2](4,6)--(2,8);
                \draw[opacity=0.2](4,7)--(3,8);
			\node at (0.5,0.5) {\tiny$\bullet$};
                \node at (1,1) {\tiny$\bullet$};
			\node at (1.5,1.5) {\tiny$\bullet$};
                \node at (2,2) {\tiny$\bullet$};
			\node at (2.5,2.5) {\tiny$\bullet$};
                \node at (3,3) {\tiny$\bullet$};
			\node at (3.5,3.5) {\tiny$\bullet$};
                \node at (0.5,1.5) {\tiny$\bullet$};
                \node at (1,2) {\tiny$\bullet$};
			\node at (1.5,2.5) {\tiny$\bullet$};
                \node at (2,3) {\tiny$\bullet$};
			\node at (2.5,3.5) {\tiny$\bullet$};
                \node at (3,4) {\tiny$\bullet$};
			\node at (3.5,4.5) {\tiny$\bullet$};
                \node at (0.5,2.5) {\tiny$\bullet$};
                \node at (1,3) {\tiny$\bullet$};
			\node at (1.5,3.5) {\tiny$\bullet$};
                \node at (2,4) {\tiny$\bullet$};
			\node at (2.5,4.5) {\tiny$\bullet$};
                \node at (3,5) {\tiny$\bullet$};
                \node at (0.5,3.5) {\tiny$\bullet$};
                \node at (1,4) {\tiny$\bullet$};
			\node at (1.5,4.5) {\tiny$\bullet$};
                \node at (2,5) {\tiny$\bullet$};
			\node at (2.5,5.5) {\tiny$\bullet$};
                \node at (0.5,4.5) {\tiny$\bullet$};
                \node at (1,5) {\tiny$\bullet$};
			\node at (1.5,5.5) {\tiny$\bullet$};
                \node at (2,6) {\tiny$\bullet$};
                \node at (0.5,5.5) {\tiny$\bullet$};
                \node at (1,6) {\tiny$\bullet$};
			\node at (1.5,6.5) {\tiny$\bullet$};
                \node at (0.5,6.5) {\tiny$\bullet$};
                \node at (1,7) {\tiny$\bullet$};
                \node at (0.5,7.5) {\tiny$\bullet$};
                \node at (0.5,1) {\color{teal}\tiny $c_{13}$};
			\node at (1,1.5) {\color{teal}\tiny $c_{14}$};
			\node at (1.5,2) {\color{teal}\tiny $c_{15}$};
			\node at (2,2.5) {\color{teal}\tiny $c_{16}$};
			\node at (2.5,3) {\color{teal}\tiny $c_{17}$};
			\node at (3,3.5) {\color{teal}\tiny $c_{18}$};
			\node at (3.5,4) {\color{teal}\tiny $c_{19}$};
			\node at (0.5,2) {\color{teal}\tiny $c_{24}$};
			\node at (1,2.5) {\color{teal}\tiny $c_{25}$};
			\node at (1.5,3) {\color{teal}\tiny $c_{26}$};
			\node at (2,3.5) {\color{teal}\tiny $c_{27}$};
			\node at (2.5,4) {\color{teal}\tiny $c_{28}$};
			\node at (3,4.5) {\color{teal}\tiny $c_{29}$};
			\node at (0.5,3) {\color{teal}\tiny $c_{35}$};
			\node at (1,3.5) {\color{teal}\tiny $c_{36}$};
			\node at (1.5,4) {\color{teal}\tiny $c_{37}$};
			\node at (2,4.5) {\color{teal}\tiny $c_{38}$};
			\node at (2.5,5) {\color{teal}\tiny $c_{39}$};
			\node at (0.5,4) {\color{teal}\tiny $c_{46}$};
			\node at (1,4.5) {\color{teal}\tiny $c_{47}$};
			\node at (1.5,5) {\color{teal}\tiny $c_{48}$};
			\node at (2,5.5) {\color{teal}\tiny $c_{49}$};
			\node at (0.5,5) {\color{teal}\tiny $c_{57}$};
			\node at (1,5.5) {\color{teal}\tiny $c_{58}$};
			\node at (1.5,6) {\color{teal}\tiny $c_{59}$};
			\node at (0.5,6) {\color{teal}\tiny $c_{68}$};
			\node at (1,6.5) {\color{teal}\tiny $c_{69}$};
			\node at (0.5,7) {\color{teal}\tiny $c_{79}$};
                \node[left] at (0,0) {\scriptsize $1$};
                \node[left] at (0,1) {\scriptsize $2$};
                \node[left] at (0,2) {\scriptsize $3$};
                \node[left] at (0,3) {\scriptsize $4$};
                \node[left] at (0,4) {\scriptsize $5$};
                \node[left] at (0,5) {\scriptsize $6$};
                \node[left] at (0,6) {\scriptsize $7$};
                \node[left] at (0,7) {\scriptsize $8$};
                \node[left] at (0,8) {\scriptsize $9$};
                \node[right] at (4,0) {\scriptsize $6$};
                \node[right] at (4,1) {\scriptsize $7$};
                \node[right] at (4,2) {\scriptsize $8$};
                \node[right] at (4,3) {\scriptsize $9$};
                \node[right] at (4,4) {\scriptsize $10$};
                \node[right] at (4,5) {\scriptsize $1$};
                \node[right] at (4,6) {\scriptsize $2$};
                \node[right] at (4,7) {\scriptsize $3$};
                \node[right] at (4,8) {\scriptsize $4$};
		      \end{tikzpicture}
                \caption{}
                \label{fig:MeshWithCs}
            \end{subfigure}
            \begin{subfigure}[t]{0.4\textwidth}
                \centering
    	   \begin{tikzpicture}[scale=0.75]
                \draw (0,0)--(4,4);
			\draw (0,1)--(3.5,4.5);
			\draw(0,2)--(3,5);
			\draw(0,3)--(2.5,5.5);
			\draw(0,4)--(2,6);
			\draw (0,5)--(1.5,6.5);
			\draw(0,6)--(1,7);
			\draw(0,7)--(0.5,7.5);
			\draw(0,0)--(0,8);
			\draw (4,0)--(4,8);
			\draw(0,8)--(4,4);
			\draw(0,7)--(3.5,3.5);
			\draw(0,6)--(3,3);
			\draw(0,5)--(2.5,2.5);
			\draw(0,4)--(2,2);
			\draw(0,3)--(1.5,1.5);
			\draw(0,2)--(1,1);
			\draw(0,1)--(0.5,0.5); 
                \draw[opacity=0.2](0.5,0.5)--(1,0); 
                \draw[opacity=0.2](1,1)--(2,0); 
                \draw[opacity=0.2](1.5,1.5)--(3,0); 
                \draw[opacity=0.2](2,2)--(4,0);
                \draw[opacity=0.2](2.5,2.5)--(4,1);
                \draw[opacity=0.2](3,3)--(4,2);
                \draw[opacity=0.2](3.5,3.5)--(4,3);
                \draw[opacity=0.2](3.5,4.5)--(4,5); 
                \draw[opacity=0.2](3,5)--(4,6);
                \draw[opacity=0.2](2.5,5.5)--(4,7);
                \draw[opacity=0.2](2,6)--(4,8);
                \draw[opacity=0.2](1.5,6.5)--(3,8);
                \draw[opacity=0.2](1,7)--(2,8);
                \draw[opacity=0.2](0.5,7.5)--(1,8);
                \draw[opacity=0.2](1,0)--(4,3);
                \draw[opacity=0.2](2,0)--(4,2);
                \draw[opacity=0.2](3,0)--(4,1);
                \draw[opacity=0.2](4,5)--(1,8);
                \draw[opacity=0.2](4,6)--(2,8);
                \draw[opacity=0.2](4,7)--(3,8);
			\node at (0.5,0.5) {\tiny$\bullet$};
                \node at (1,1) {\tiny$\bullet$};
			\node at (1.5,1.5) {\tiny$\bullet$};
                \node at (2,2) {\tiny$\bullet$};
			\node at (2.5,2.5) {\tiny$\bullet$};
                \node at (3,3) {\tiny$\bullet$};
			\node at (3.5,3.5) {\tiny$\bullet$};
                \node at (0.5,1.5) {\tiny$\bullet$};
                \node at (1,2) {\tiny$\bullet$};
			\node at (1.5,2.5) {\tiny$\bullet$};
                \node at (2,3) {\tiny$\bullet$};
			\node at (2.5,3.5) {\tiny$\bullet$};
                \node at (3,4) {\tiny$\bullet$};
			\node at (3.5,4.5) {\tiny$\bullet$};
                \node at (0.5,2.5) {\tiny$\bullet$};
                \node at (1,3) {\tiny$\bullet$};
			\node at (1.5,3.5) {\tiny$\bullet$};
                \node at (2,4) {\tiny$\bullet$};
			\node at (2.5,4.5) {\tiny$\bullet$};
                \node at (3,5) {\tiny$\bullet$};
                \node at (0.5,3.5) {\tiny$\bullet$};
                \node at (1,4) {\tiny$\bullet$};
			\node at (1.5,4.5) {\tiny$\bullet$};
                \node at (2,5) {\tiny$\bullet$};
			\node at (2.5,5.5) {\tiny$\bullet$};
                \node at (0.5,4.5) {\tiny$\bullet$};
                \node at (1,5) {\tiny$\bullet$};
			\node at (1.5,5.5) {\tiny$\bullet$};
                \node at (2,6) {\tiny$\bullet$};
                \node at (0.5,5.5) {\tiny$\bullet$};
                \node at (1,6) {\tiny$\bullet$};
			\node at (1.5,6.5) {\tiny$\bullet$};
                \node at (0.5,6.5) {\tiny$\bullet$};
                \node at (1,7) {\tiny$\bullet$};
                \node at (0.5,7.5) {\tiny$\bullet$};
                \node[below] at (0.5,0.5) {\color{purple}\tiny $X_{13}$};
			\node[below] at (1,1) {\color{purple}\tiny $X_{14}$};
			\node[below] at (1.5,1.5) {\color{purple}\tiny $X_{15}$};
			\node[below] at (2,2) {\color{purple}\tiny $X_{16}$};
			\node[below] at (2.5,2.5) {\color{purple}\tiny $X_{17}$};
			\node[below] at (3,3) {\color{purple}\tiny $X_{18}$};
			\node[below] at (3.5,3.5) {\color{purple}\tiny $X_{19}$};
			\node[below] at (0.5,1.5) {\color{purple}\tiny $X_{24}$};
			\node[below] at (1,2) {\color{purple}\tiny $X_{25}$};
			\node[below] at (1.5,2.5) {\color{purple}\tiny $X_{26}$};
			\node[below] at (2,3) {\color{purple}\tiny $X_{27}$};
			\node[below] at (2.5,3.5) {\color{purple}\tiny $X_{28}$};
			\node[below] at (3,4) {\color{purple}\tiny $X_{29}$};
			\node[below] at (3.5,4.5) {\color{purple}\tiny $X_{2,10}$};
			\node[below] at (0.5,2.5) {\color{purple}\tiny $X_{35}$};
			\node[below] at (1,3) {\color{purple}\tiny $X_{36}$};
			\node[below] at (1.5,3.5) {\color{purple}\tiny $X_{37}$};
			\node[below] at (2,4) {\color{purple}\tiny $X_{38}$};
			\node[below] at (2.5,4.5) {\color{purple}\tiny $X_{39}$};
			\node[below] at (3,5) {\color{purple}\tiny $X_{3,10}$};
			\node[below] at (0.5,3.5) {\color{purple}\tiny $X_{46}$};
			\node[below] at (1,4) {\color{purple}\tiny $X_{47}$};
			\node[below] at (1.5,4.5) {\color{purple}\tiny $X_{48}$};
			\node[below] at (2,5) {\color{purple}\tiny $X_{49}$};
			\node[below] at (2.5,5.5) {\color{purple}\tiny $X_{4,10}$};
			\node[below] at (0.5,4.5) {\color{purple}\tiny $X_{57}$};
			\node[below] at (1,5) {\color{purple}\tiny $X_{58}$};
			\node[below] at (1.5,5.5) {\color{purple}\tiny $X_{59}$};
			\node[below] at (2,6) {\color{purple}\tiny $X_{5,10}$};
			\node[below] at (0.5,5.5) {\color{purple}\tiny $X_{68}$};
			\node[below] at (1,6) {\color{purple}\tiny $X_{69}$};
			\node[below] at (1.5,6.5) {\color{purple}\tiny $X_{6,10}$};
			\node[below] at (0.5,6.5) {\color{purple}\tiny $X_{79}$};
			\node[below] at (1,7) {\color{purple}\tiny $X_{7,10}$};
			\node[below] at (0.5,7.5) {\color{purple}\tiny $X_{8,10}$};
                \node[left] at (0,0) {\scriptsize $1$};
                \node[left] at (0,1) {\scriptsize $2$};
                \node[left] at (0,2) {\scriptsize $3$};
                \node[left] at (0,3) {\scriptsize $4$};
                \node[left] at (0,4) {\scriptsize $5$};
                \node[left] at (0,5) {\scriptsize $6$};
                \node[left] at (0,6) {\scriptsize $7$};
                \node[left] at (0,7) {\scriptsize $8$};
                \node[left] at (0,8) {\scriptsize $9$};
                \node[right] at (4,0) {\scriptsize $6$};
                \node[right] at (4,1) {\scriptsize $7$};
                \node[right] at (4,2) {\scriptsize $8$};
                \node[right] at (4,3) {\scriptsize $9$};
                \node[right] at (4,4) {\scriptsize $10$};
                \node[right] at (4,5) {\scriptsize $1$};
                \node[right] at (4,6) {\scriptsize $2$};
                \node[right] at (4,7) {\scriptsize $3$};
                \node[right] at (4,8) {\scriptsize $4$};
		      \end{tikzpicture}
                \caption{}
           \label{fig:MeshWithXs}
            \end{subfigure}
    \caption{Arrangement of Mandelstam variables in a 10-point kinematic mesh. (a) on the interior of each plaquette is associated a non-cyclic 2-particle invariant $c_{ij}$ and (b) to the node at the bottom of the $c_{ij}$ plaquette is associated a cyclic Mandelstam invariant $X_{ij}$. In our convention the node labeled $X_{ij}$ corresponds to the intersection of the rays extending diagonally upward from the exterior labels $i$ and $j$. The mesh is understood to infinitely cyclically repeat above and below; the triangular region highlighted above forms a (non-unique) ``principal domain" of the mesh, and the corresponding $X_{ij}$ a complete set of kinematic invariants.}
    \label{fig:mesh1}
\end{figure}

\begin{figure}
    \centering
            \begin{subfigure}[t]{0.4\textwidth}
                \centering
                    		\begin{tikzpicture}[scale=0.75]
                \draw (0,0)--(4,4);
			\draw (0,1)--(3.5,4.5);
			\draw(0,2)--(3,5);
			\draw(0,3)--(2.5,5.5);
			\draw(0,4)--(2,6);
			\draw (0,5)--(1.5,6.5);
			\draw(0,6)--(1,7);
			\draw(0,7)--(0.5,7.5);
			\draw(0,0)--(0,8);
			\draw (4,0)--(4,8);
			\draw(0,8)--(4,4);
			\draw(0,7)--(3.5,3.5);
			\draw(0,6)--(3,3);
			\draw(0,5)--(2.5,2.5);
			\draw(0,4)--(2,2);
			\draw(0,3)--(1.5,1.5);
			\draw(0,2)--(1,1);
			\draw(0,1)--(0.5,0.5); 
                \draw[opacity=0.2](0.5,0.5)--(1,0); 
                \draw[opacity=0.2](1,1)--(2,0); 
                \draw[opacity=0.2](1.5,1.5)--(3,0); 
                \draw[opacity=0.2](2,2)--(4,0);
                \draw[opacity=0.2](2.5,2.5)--(4,1);
                \draw[opacity=0.2](3,3)--(4,2);
                \draw[opacity=0.2](3.5,3.5)--(4,3);
                \draw[opacity=0.2](3.5,4.5)--(4,5); 
                \draw[opacity=0.2](3,5)--(4,6);
                \draw[opacity=0.2](2.5,5.5)--(4,7);
                \draw[opacity=0.2](2,6)--(4,8);
                \draw[opacity=0.2](1.5,6.5)--(3,8);
                \draw[opacity=0.2](1,7)--(2,8);
                \draw[opacity=0.2](0.5,7.5)--(1,8);
                \draw[opacity=0.2](1,0)--(4,3);
                \draw[opacity=0.2](2,0)--(4,2);
                \draw[opacity=0.2](3,0)--(4,1);
                \draw[opacity=0.2](4,5)--(1,8);
                \draw[opacity=0.2](4,6)--(2,8);
                \draw[opacity=0.2](4,7)--(3,8);
			\node at (0.5,0.5) {\tiny$\bullet$};
                \node at (1,1) {\tiny$\bullet$};
			\node at (1.5,1.5) {\tiny$\bullet$};
                \node at (2,2) {\tiny$\bullet$};
			\node at (2.5,2.5) {\tiny$\bullet$};
                \node at (3,3) {\tiny$\bullet$};
			\node at (3.5,3.5) {\tiny$\bullet$};
                \node at (0.5,1.5) {\tiny$\bullet$};
                \node at (1,2) {\tiny$\bullet$};
			\node at (1.5,2.5) {\tiny$\bullet$};
                \node at (2,3) {\tiny$\bullet$};
			\node at (2.5,3.5) {\tiny$\bullet$};
                \node at (3,4) {\tiny$\bullet$};
			\node at (3.5,4.5) {\tiny$\bullet$};
                \node at (0.5,2.5) {\tiny$\bullet$};
                \node at (1,3) {\tiny$\bullet$};
			\node at (1.5,3.5) {\tiny$\bullet$};
                \node at (2,4) {\tiny$\bullet$};
			\node at (2.5,4.5) {\tiny$\bullet$};
                \node at (3,5) {\tiny$\bullet$};
                \node at (0.5,3.5) {\tiny$\bullet$};
                \node at (1,4) {\tiny$\bullet$};
			\node at (1.5,4.5) {\tiny$\bullet$};
                \node at (2,5) {\tiny$\bullet$};
			\node at (2.5,5.5) {\tiny$\bullet$};
                \node at (0.5,4.5) {\tiny$\bullet$};
                \node at (1,5) {\tiny$\bullet$};
			\node at (1.5,5.5) {\tiny$\bullet$};
                \node at (2,6) {\tiny$\bullet$};
                \node at (0.5,5.5) {\tiny$\bullet$};
                \node at (1,6) {\tiny$\bullet$};
			\node at (1.5,6.5) {\tiny$\bullet$};
                \node at (0.5,6.5) {\tiny$\bullet$};
                \node at (1,7) {\tiny$\bullet$};
                \node at (0.5,7.5) {\tiny$\bullet$};
                \node[left] at (0,0) {\scriptsize $1$};
                \node[left] at (0,1) {\scriptsize $2$};
                \node[left] at (0,2) {\scriptsize $3$};
                \node[left] at (0,3) {\scriptsize $4$};
                \node[left] at (0,4) {\scriptsize $5$};
                \node[left] at (0,5) {\scriptsize $6$};
                \node[left] at (0,6) {\scriptsize $7$};
                \node[left] at (0,7) {\scriptsize $8$};
                \node[left] at (0,8) {\scriptsize $9$};
                \node[right] at (4,0) {\scriptsize $6$};
                \node[right] at (4,1) {\scriptsize $7$};
                \node[right] at (4,2) {\scriptsize $8$};
                \node[right] at (4,3) {\scriptsize $9$};
                \node[right] at (4,4) {\scriptsize $10$};
                \node[right] at (4,5) {\scriptsize $1$};
                \node[right] at (4,6) {\scriptsize $2$};
                \node[right] at (4,7) {\scriptsize $3$};
                \node[right] at (4,8) {\scriptsize $4$};
                \fill[color=teal,opacity=0.2] (0.5,3.5)--(1.5,2.5)--(3,4)--(2,5);
                \node[below] at (0.5,3.5) {\color{purple}\tiny $X_{\text{L}}$};
                \node[below] at (1.5,2.5) {\color{purple}\tiny $X_{\text{B}}$};
                \node[below] at (3,4) {\color{purple}\tiny $X_{\text{R}}$};
                \node[above] at (2,5) {\color{purple}\tiny $X_{\text{T}}$};
		      \end{tikzpicture}
                \caption{}
                \label{fig:RectRule}
            \end{subfigure}
            \begin{subfigure}[t]{0.4\textwidth}
                \centering
                    		\begin{tikzpicture}[scale=0.75]
                \draw (0,0)--(4,4);
			\draw (0,1)--(3.5,4.5);
			\draw(0,2)--(3,5);
			\draw(0,3)--(2.5,5.5);
			\draw(0,4)--(2,6);
			\draw (0,5)--(1.5,6.5);
			\draw(0,6)--(1,7);
			\draw(0,7)--(0.5,7.5);
			\draw(0,0)--(0,8);
			\draw (4,0)--(4,8);
			\draw(0,8)--(4,4);
			\draw(0,7)--(3.5,3.5);
			\draw(0,6)--(3,3);
			\draw(0,5)--(2.5,2.5);
			\draw(0,4)--(2,2);
			\draw(0,3)--(1.5,1.5);
			\draw(0,2)--(1,1);
			\draw(0,1)--(0.5,0.5); 
                \draw[opacity=0.2](0.5,0.5)--(1,0); 
                \draw[opacity=0.2](1,1)--(2,0); 
                \draw[opacity=0.2](1.5,1.5)--(3,0); 
                \draw[opacity=0.2](2,2)--(4,0);
                \draw[opacity=0.2](2.5,2.5)--(4,1);
                \draw[opacity=0.2](3,3)--(4,2);
                \draw[opacity=0.2](3.5,3.5)--(4,3);
                \draw[opacity=0.2](3.5,4.5)--(4,5); 
                \draw[opacity=0.2](3,5)--(4,6);
                \draw[opacity=0.2](2.5,5.5)--(4,7);
                \draw[opacity=0.2](2,6)--(4,8);
                \draw[opacity=0.2](1.5,6.5)--(3,8);
                \draw[opacity=0.2](1,7)--(2,8);
                \draw[opacity=0.2](0.5,7.5)--(1,8);
                \draw[opacity=0.2](1,0)--(4,3);
                \draw[opacity=0.2](2,0)--(4,2);
                \draw[opacity=0.2](3,0)--(4,1);
                \draw[opacity=0.2](4,5)--(1,8);
                \draw[opacity=0.2](4,6)--(2,8);
                \draw[opacity=0.2](4,7)--(3,8);
			\node at (0.5,0.5) {\tiny$\bullet$};
                \node at (1,1) {\tiny$\bullet$};
			\node at (1.5,1.5) {\tiny$\bullet$};
                \node at (2,2) {\tiny$\bullet$};
			\node at (2.5,2.5) {\tiny$\bullet$};
                \node at (3,3) {\tiny$\bullet$};
			\node at (3.5,3.5) {\tiny$\bullet$};
                \node at (0.5,1.5) {\tiny$\bullet$};
                \node at (1,2) {\tiny$\bullet$};
			\node at (1.5,2.5) {\tiny$\bullet$};
                \node at (2,3) {\tiny$\bullet$};
			\node at (2.5,3.5) {\tiny$\bullet$};
                \node at (3,4) {\tiny$\bullet$};
			\node at (3.5,4.5) {\tiny$\bullet$};
                \node at (0.5,2.5) {\tiny$\bullet$};
                \node at (1,3) {\tiny$\bullet$};
			\node at (1.5,3.5) {\tiny$\bullet$};
                \node at (2,4) {\tiny$\bullet$};
			\node at (2.5,4.5) {\tiny$\bullet$};
                \node at (3,5) {\tiny$\bullet$};
                \node at (0.5,3.5) {\tiny$\bullet$};
                \node at (1,4) {\tiny$\bullet$};
			\node at (1.5,4.5) {\tiny$\bullet$};
                \node at (2,5) {\tiny$\bullet$};
			\node at (2.5,5.5) {\tiny$\bullet$};
                \node at (0.5,4.5) {\tiny$\bullet$};
                \node at (1,5) {\tiny$\bullet$};
			\node at (1.5,5.5) {\tiny$\bullet$};
                \node at (2,6) {\tiny$\bullet$};
                \node at (0.5,5.5) {\tiny$\bullet$};
                \node at (1,6) {\tiny$\bullet$};
			\node at (1.5,6.5) {\tiny$\bullet$};
                \node at (0.5,6.5) {\tiny$\bullet$};
                \node at (1,7) {\tiny$\bullet$};
                \node at (0.5,7.5) {\tiny$\bullet$};
                \fill[color=orange,opacity=0.2] (0,2)--(1.5,3.5)--(0,5);
                \fill[color=orange,opacity=0.2] (4,1)--(1.5,3.5)--(4,6);
                \node at (1.5,3.5) {\color{red}\tiny$\bullet$};
                \node[left] at (0,0) {\scriptsize $1$};
                \node[left] at (0,1) {\scriptsize $2$};
                \node[left] at (0,2) {\scriptsize $3$};
                \node[left] at (0,3) {\scriptsize $4$};
                \node[left] at (0,4) {\scriptsize $5$};
                \node[left] at (0,5) {\scriptsize $6$};
                \node[left] at (0,6) {\scriptsize $7$};
                \node[left] at (0,7) {\scriptsize $8$};
                \node[left] at (0,8) {\scriptsize $9$};
                \node[right] at (4,0) {\scriptsize $6$};
                \node[right] at (4,1) {\scriptsize $7$};
                \node[right] at (4,2) {\scriptsize $8$};
                \node[right] at (4,3) {\scriptsize $9$};
                \node[right] at (4,4) {\scriptsize $10$};
                \node[right] at (4,5) {\scriptsize $1$};
                \node[right] at (4,6) {\scriptsize $2$};
                \node[right] at (4,7) {\scriptsize $3$};
                \node[right] at (4,8) {\scriptsize $4$};
		      \end{tikzpicture}
                \caption{}
                \label{fig:Fact}
            \end{subfigure}
    \caption{(a) a rectangular region with $X_T=X_{49}$, $X_B=X_{26}$, $X_L=X_{46}$ and $X_R=X_{29}$; using the rectangle rule the sum of the $c_{ij}$ with $i\in\{2,3,4\}$ and $j\in \{6,7,8,9\}$ is equal to $X_{49} + X_{26}-X_{46}-X_{29}$. (b) on the pole $X_{37}$ the amplitude $A_{10}$ factors into the product of $A_5[3,4,5,6,7]$ (the left \color{orange}\textbf{orange} \color{black} triangle) and $A_7[7,8,9,10,1,2,3]$ (the right \color{orange}\textbf{orange} \color{black} triangle). }
    \label{fig:mesh2}
\end{figure}
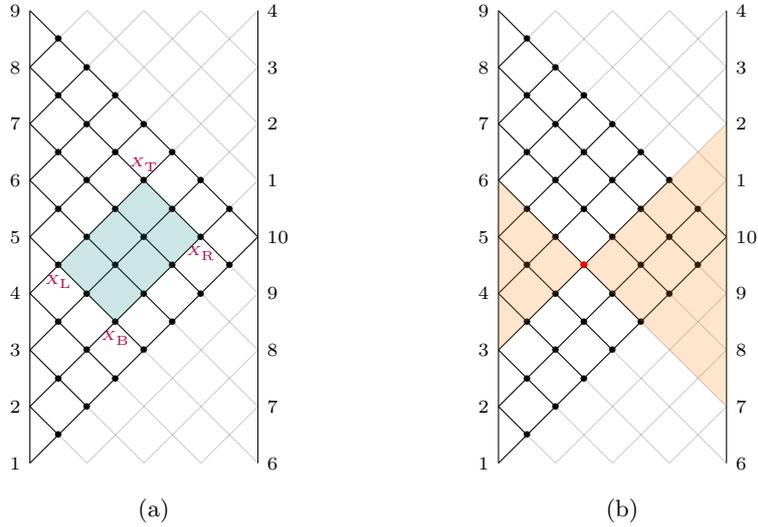

    \item \textbf{Hidden zeros:} We define a \textit{maximal rectangle} as a rectangular region of the mesh with $X_{L}$ and $X_R$ on the corresponding boundaries. For the models considered in this paper (Tr$\phi^3$, YMS, NLSM...) if all of the $c_{ij}$ on the interior of a maximal rectangle are set to zero, then the amplitude will vanish. Since each zero is defined by a choice of maximal rectangle, which is in turn defined by a choice of $X_B$. We will denote the zero using the notation
    \begin{equation}
        A_n\left[1,2,...,n\right] \xrightarrow{\mathcal{Z}\left(X_B\right)} 0.
    \end{equation}
    More formally, the zero $\mathcal{Z}(X_{ij})$ is defined by 
    \begin{equation}
        c_{kl}=0, \hspace{5mm} k=i,...,j-2 , \hspace{5mm} l =j,..,i-2.
    \end{equation}

    \item \textbf{Smooth splitting:} If one of the conditions defining a hidden zero is \textit{relaxed}, meaning for a single plaquette on the interior of the chosen maximal rectangle $c_*\neq 0$, then the amplitude will \textit{split}. We can sub-divide the splitting into two cases. 

    \begin{figure}
    \centering
            \begin{subfigure}[t]{0.4\textwidth}
                \centering
                    		\begin{tikzpicture}[scale=0.75]
                \draw (0,0)--(4,4);
			\draw (0,1)--(3.5,4.5);
			\draw(0,2)--(3,5);
			\draw(0,3)--(2.5,5.5);
			\draw(0,4)--(2,6);
			\draw (0,5)--(1.5,6.5);
			\draw(0,6)--(1,7);
			\draw(0,7)--(0.5,7.5);
			\draw(0,0)--(0,8);
			\draw (4,0)--(4,8);
			\draw(0,8)--(4,4);
			\draw(0,7)--(3.5,3.5);
			\draw(0,6)--(3,3);
			\draw(0,5)--(2.5,2.5);
			\draw(0,4)--(2,2);
			\draw(0,3)--(1.5,1.5);
			\draw(0,2)--(1,1);
			\draw(0,1)--(0.5,0.5); 
                \draw[opacity=0.2](0.5,0.5)--(1,0); 
                \draw[opacity=0.2](1,1)--(2,0); 
                \draw[opacity=0.2](1.5,1.5)--(3,0); 
                \draw[opacity=0.2](2,2)--(4,0);
                \draw[opacity=0.2](2.5,2.5)--(4,1);
                \draw[opacity=0.2](3,3)--(4,2);
                \draw[opacity=0.2](3.5,3.5)--(4,3);
                \draw[opacity=0.2](3.5,4.5)--(4,5); 
                \draw[opacity=0.2](3,5)--(4,6);
                \draw[opacity=0.2](2.5,5.5)--(4,7);
                \draw[opacity=0.2](2,6)--(4,8);
                \draw[opacity=0.2](1.5,6.5)--(3,8);
                \draw[opacity=0.2](1,7)--(2,8);
                \draw[opacity=0.2](0.5,7.5)--(1,8);
                \draw[opacity=0.2](1,0)--(4,3);
                \draw[opacity=0.2](2,0)--(4,2);
                \draw[opacity=0.2](3,0)--(4,1);
                \draw[opacity=0.2](4,5)--(1,8);
                \draw[opacity=0.2](4,6)--(2,8);
                \draw[opacity=0.2](4,7)--(3,8);
			\node at (0.5,0.5) {\tiny$\bullet$};
                \node at (1,1) {\tiny$\bullet$};
			\node at (1.5,1.5) {\tiny$\bullet$};
                \node at (2,2) {\tiny$\bullet$};
			\node at (2.5,2.5) {\tiny$\bullet$};
                \node at (3,3) {\tiny$\bullet$};
			\node at (3.5,3.5) {\tiny$\bullet$};
                \node at (0.5,1.5) {\tiny$\bullet$};
                \node at (1,2) {\tiny$\bullet$};
			\node at (1.5,2.5) {\tiny$\bullet$};
                \node at (2,3) {\tiny$\bullet$};
			\node at (2.5,3.5) {\tiny$\bullet$};
                \node at (3,4) {\tiny$\bullet$};
			\node at (3.5,4.5) {\tiny$\bullet$};
                \node at (0.5,2.5) {\tiny$\bullet$};
                \node at (1,3) {\tiny$\bullet$};
			\node at (1.5,3.5) {\tiny$\bullet$};
                \node at (2,4) {\tiny$\bullet$};
			\node at (2.5,4.5) {\tiny$\bullet$};
                \node at (3,5) {\tiny$\bullet$};
                \node at (0.5,3.5) {\tiny$\bullet$};
                \node at (1,4) {\tiny$\bullet$};
			\node at (1.5,4.5) {\tiny$\bullet$};
                \node at (2,5) {\tiny$\bullet$};
			\node at (2.5,5.5) {\tiny$\bullet$};
                \node at (0.5,4.5) {\tiny$\bullet$};
                \node at (1,5) {\tiny$\bullet$};
			\node at (1.5,5.5) {\tiny$\bullet$};
                \node at (2,6) {\tiny$\bullet$};
                \node at (0.5,5.5) {\tiny$\bullet$};
                \node at (1,6) {\tiny$\bullet$};
			\node at (1.5,6.5) {\tiny$\bullet$};
                \node at (0.5,6.5) {\tiny$\bullet$};
                \node at (1,7) {\tiny$\bullet$};
                \node at (0.5,7.5) {\tiny$\bullet$};
                \node at (3.5,4) {\color{black}\tiny $c_{19}$};
                \node[left] at (0,0) {\scriptsize $1$};
                \node[left] at (0,1) {\scriptsize $2$};
                \node[left] at (0,2) {\scriptsize $3$};
                \node[left] at (0,3) {\scriptsize $4$};
                \node[left] at (0,4) {\scriptsize $5$};
                \node[left] at (0,5) {\scriptsize $6$};
                \node[left] at (0,6) {\scriptsize $7$};
                \node[left] at (0,7) {\scriptsize $8$};
                \node[left] at (0,8) {\scriptsize $9$};
                \node[right] at (4,0) {\scriptsize $6$};
                \node[right] at (4,1) {\scriptsize $7$};
                \node[right] at (4,2) {\scriptsize $8$};
                \node[right] at (4,3) {\scriptsize $9$};
                \node[right] at (4,4) {\scriptsize $10$};
                \node[right] at (4,5) {\scriptsize $1$};
                \node[right] at (4,6) {\scriptsize $2$};
                \node[right] at (4,7) {\scriptsize $3$};
                \node[right] at (4,8) {\scriptsize $4$};
                \fill[color=teal,opacity=0.2] (0,3)--(1.5,1.5)--(3.5,3.5)--(2,5);
                \fill[color=teal,opacity=0.2] (2,5)--(2.5,5.5)--(3.5,4.5)--(3,4);
                \node[above] at (2.5,5.5) {\color{purple}\tiny $X_{4,10}$};
                \node at (2.5,5.5) {\color{purple}\tiny$\bullet$};
                \node[below] at (1.5,1.5) {\color{purple}\tiny $X_{15}$};
                \node at (1.5,1.5) {\color{purple}\tiny$\bullet$};
                \fill[color=orange,opacity=0.2] (0,0)--(1.5,1.5)--(0,3);
                \fill[color=orange,opacity=0.2] (0,3)--(2.5,5.5)--(0,8);
		      \end{tikzpicture}
                \caption{}
                \label{fig:CornerPrimer}
            \end{subfigure}
            \begin{subfigure}[t]{0.4\textwidth}
                \centering
                    		\begin{tikzpicture}[scale=0.75]
                \draw (0,0)--(4,4);
			\draw (0,1)--(3.5,4.5);
			\draw(0,2)--(3,5);
			\draw(0,3)--(2.5,5.5);
			\draw(0,4)--(2,6);
			\draw (0,5)--(1.5,6.5);
			\draw(0,6)--(1,7);
			\draw(0,7)--(0.5,7.5);
			\draw(0,0)--(0,8);
			\draw (4,0)--(4,8);
			\draw(0,8)--(4,4);
			\draw(0,7)--(3.5,3.5);
			\draw(0,6)--(3,3);
			\draw(0,5)--(2.5,2.5);
			\draw(0,4)--(2,2);
			\draw(0,3)--(1.5,1.5);
			\draw(0,2)--(1,1);
			\draw(0,1)--(0.5,0.5); 
                \draw[opacity=0.2](0.5,0.5)--(1,0); 
                \draw[opacity=0.2](1,1)--(2,0); 
                \draw[opacity=0.2](1.5,1.5)--(3,0); 
                \draw[opacity=0.2](2,2)--(4,0);
                \draw[opacity=0.2](2.5,2.5)--(4,1);
                \draw[opacity=0.2](3,3)--(4,2);
                \draw[opacity=0.2](3.5,3.5)--(4,3);
                \draw[opacity=0.2](3.5,4.5)--(4,5); 
                \draw[opacity=0.2](3,5)--(4,6);
                \draw[opacity=0.2](2.5,5.5)--(4,7);
                \draw[opacity=0.2](2,6)--(4,8);
                \draw[opacity=0.2](1.5,6.5)--(3,8);
                \draw[opacity=0.2](1,7)--(2,8);
                \draw[opacity=0.2](0.5,7.5)--(1,8);
                \draw[opacity=0.2](1,0)--(4,3);
                \draw[opacity=0.2](2,0)--(4,2);
                \draw[opacity=0.2](3,0)--(4,1);
                \draw[opacity=0.2](4,5)--(1,8);
                \draw[opacity=0.2](4,6)--(2,8);
                \draw[opacity=0.2](4,7)--(3,8);
			\node at (0.5,0.5) {\tiny$\bullet$};
                \node at (1,1) {\tiny$\bullet$};
			\node at (1.5,1.5) {\tiny$\bullet$};
                \node at (2,2) {\tiny$\bullet$};
			\node at (2.5,2.5) {\tiny$\bullet$};
                \node at (3,3) {\tiny$\bullet$};
			\node at (3.5,3.5) {\tiny$\bullet$};
                \node at (0.5,1.5) {\tiny$\bullet$};
                \node at (1,2) {\tiny$\bullet$};
			\node at (1.5,2.5) {\tiny$\bullet$};
                \node at (2,3) {\tiny$\bullet$};
			\node at (2.5,3.5) {\tiny$\bullet$};
                \node at (3,4) {\tiny$\bullet$};
			\node at (3.5,4.5) {\tiny$\bullet$};
                \node at (0.5,2.5) {\tiny$\bullet$};
                \node at (1,3) {\tiny$\bullet$};
			\node at (1.5,3.5) {\tiny$\bullet$};
                \node at (2,4) {\tiny$\bullet$};
			\node at (2.5,4.5) {\tiny$\bullet$};
                \node at (3,5) {\tiny$\bullet$};
                \node at (0.5,3.5) {\tiny$\bullet$};
                \node at (1,4) {\tiny$\bullet$};
			\node at (1.5,4.5) {\tiny$\bullet$};
                \node at (2,5) {\tiny$\bullet$};
			\node at (2.5,5.5) {\tiny$\bullet$};
                \node at (0.5,4.5) {\tiny$\bullet$};
                \node at (1,5) {\tiny$\bullet$};
			\node at (1.5,5.5) {\tiny$\bullet$};
                \node at (2,6) {\tiny$\bullet$};
                \node at (0.5,5.5) {\tiny$\bullet$};
                \node at (1,6) {\tiny$\bullet$};
			\node at (1.5,6.5) {\tiny$\bullet$};
                \node at (0.5,6.5) {\tiny$\bullet$};
                \node at (1,7) {\tiny$\bullet$};
                \node at (0.5,7.5) {\tiny$\bullet$};
                \node at (1.5,3) {\color{black}\tiny $c_{26}$};
                \node[left] at (0,0) {\scriptsize $1$};
                \node[left] at (0,1) {\scriptsize $2$};
                \node[left] at (0,2) {\scriptsize $3$};
                \node[left] at (0,3) {\scriptsize $4$};
                \node[left] at (0,4) {\scriptsize $5$};
                \node[left] at (0,5) {\scriptsize $6$};
                \node[left] at (0,6) {\scriptsize $7$};
                \node[left] at (0,7) {\scriptsize $8$};
                \node[left] at (0,8) {\scriptsize $9$};
                \node[right] at (4,0) {\scriptsize $6$};
                \node[right] at (4,1) {\scriptsize $7$};
                \node[right] at (4,2) {\scriptsize $8$};
                \node[right] at (4,3) {\scriptsize $9$};
                \node[right] at (4,4) {\scriptsize $10$};
                \node[right] at (4,5) {\scriptsize $1$};
                \node[right] at (4,6) {\scriptsize $2$};
                \node[right] at (4,7) {\scriptsize $3$};
                \node[right] at (4,8) {\scriptsize $4$};
                \fill[color=teal,opacity=0.2] (0,3)--(0.5,3.5)--(2,2)--(1.5,1.5);
                \fill[color=teal,opacity=0.2] (0.5,3.5)--(1,4)--(1.5,3.5)--(1,3);
                \fill[color=teal,opacity=0.2] (1.5,2.5)--(2,3)--(2.5,2.5)--(2,2);
                \fill[color=teal,opacity=0.2] (1,4)--(2.5,5.5)--(4,4)--(2.5,2.5);
                \node[above] at (2.5,5.5) {\color{purple}\tiny $X_{4,10}$};
                \node at (2.5,5.5) {\color{purple}\tiny$\bullet$};
                \node[below] at (1.5,1.5) {\color{purple}\tiny $X_{15}$};
                \node at (1.5,1.5) {\color{purple}\tiny$\bullet$};
                \fill[color=orange,opacity=0.2] (0,0)--(1.5,1.5)--(0,3);
                \fill[color=orange,opacity=0.2] (0,3)--(2.5,5.5)--(0,8);
                \node[above] at (3.5,4.5) {\color{red}\tiny $X_{2,10}$};
                \node at (3.5,4.5) {\color{red}\tiny$\bullet$};
                \draw[color=red,->,thick=0.1] (3.5,4.5)--(1,2);
                \node[below] at (2.5,2.5) {\color{red}\tiny $X_{17}$};
                \node at (2.5,2.5) {\color{red}\tiny$\bullet$};
                \draw[color=red,->,thick=0.1] (2.5,2.5)--(1,4);
                \node[below] at (3,3) {\color{red}\tiny $X_{18}$};
                \node at (3,3) {\color{red}\tiny$\bullet$};
                \draw[color=red,->,thick=0.1] (3,3)--(1.5,4.5);
                \node[below] at (3.5,3.5) {\color{red}\tiny $X_{19}$};
                \node at (3.5,3.5) {\color{red}\tiny$\bullet$};
                \draw[color=red,->,thick=0.1] (3.5,3.5)--(2,5);
		      \end{tikzpicture}
                \caption{}
                \label{fig:NonCornerPrimer}
            \end{subfigure}
    \caption{Splitting rule at 10-point for a near-zero with $X_B=X_{15}$. (a) For this configuration the corner case corresponds to relaxing $c_* = c_{19}\neq0$; the sub-amplitudes that appear on the split ($A_5[1,2,3,4,5]$ and $A_7[4,5,6,7,8,9,10]$) correspond to the smaller triangular regions to the left of the maximal rectangle. (b) A non-corner split for the choice $c_* = c_{26}\neq0$; the sub-amplitudes are the same as the corner case, but with a remapping of kinematic variables depicted with the red arrows.      }
    \label{fig:mesh3}
\end{figure}
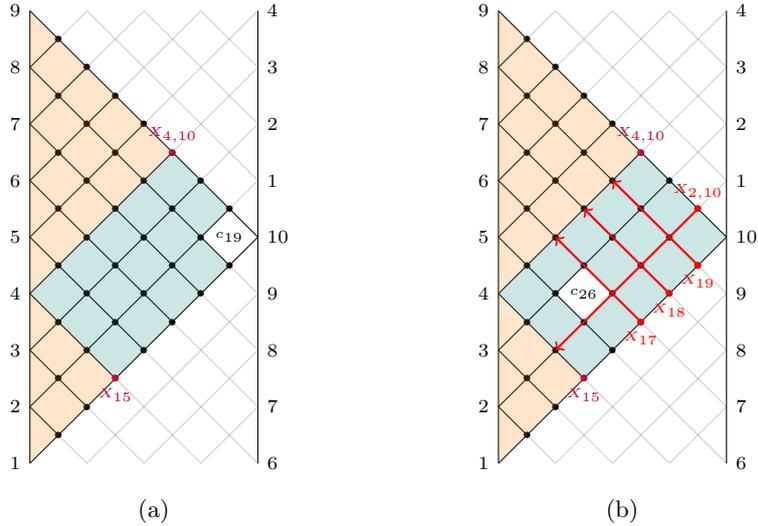
    
    \textit{Corner case}: if $X_{B}=X_{ij}$ (therefore $X_T = X_{j-1,i-1}$) and $c_*=c_{i,i-2}$ the amplitude splits according to the formula
    \begin{equation}
        \label{cornersplit}
        A_n\left[1,2,...,n\right] \xrightarrow[\{c_{i,i-2}\}\neq 0]{\mathcal{Z}\left(X_{ij}\right)} \left(\frac{1}{X_T}+\frac{1}{X_B}\right)A\left[i,i+1,...,j-1,j\right] A\left[j-1,j,...,i-2,i-1\right].
    \end{equation}
    As the name suggests, this case corresponds to choosing $c_*$ in the right-most corner of the maximal rectangle, adjacent to the edge of the mesh. We will use the notation $\xrightarrow[\{c_*\}\neq 0]{\mathcal{Z}\left(X_B\right)}$ to denote the kinematic limit defined by taking the zero $\mathcal{Z}\left(X_B\right)$ and relaxing $c_*\neq 0$.\\
    \\
    \textit{Generic case}: if $c_* = c_{k,l}$ for $i\leq k\leq j-2$ and $j\leq l \leq i-2$, then there is an additional mapping of kinematic variables in the sub-amplitudes on the split
    \begin{equation}
        \label{genericsplit}
        A_n\left[1,2,...,n\right] \xrightarrow[\{c_*\}\neq 0]{\mathcal{Z}\left(X_B\right)}  \left(\frac{1}{X_T}+\frac{1}{X_B}\right)\times A_B \times A_T,
    \end{equation}
    where
    \begin{align}
        A_B &= A\left[i,i+1,...,j-1,j\right]\biggr\vert_{X_{a,j}\rightarrow X_{a,i-1}}, \hspace{10mm} a=i+1,...,k \nonumber\\
        A_T &= A\left[j-1,j,...,i-2,i-1\right]\biggr\vert_{X_{j-1,b}\rightarrow X_{i,b}}, \hspace{5mm} b=l+1,...,i-2.
    \end{align}
    The generic and corner cases are depicted in Figures \ref{fig:CornerPrimer} and \ref{fig:NonCornerPrimer} respectively.\\
    \\
    To see that the splitting formula (\ref{genericsplit}) reduces to the hidden zero in the limit $c_*\rightarrow 0$ is a simple application of the rectangle rule. Since the rectangle that defines the split is maximal $X_L=X_R=0$, together with the assumption that all $c_{ij}$ on the interior except $c_*$ are set to zero the rectangle rule gives
    \begin{equation}
        X_T + X_B = c_*.
    \end{equation}
    The prefactor in the splitting formula (\ref{genericsplit}) then vanishes when $c_*\rightarrow 0$; in this sense the hidden zero is a trivial corollary of the more complicated splitting property. 
    
\end{itemize}

\subsection{Shifting planar variables}
\label{sec:shifts}

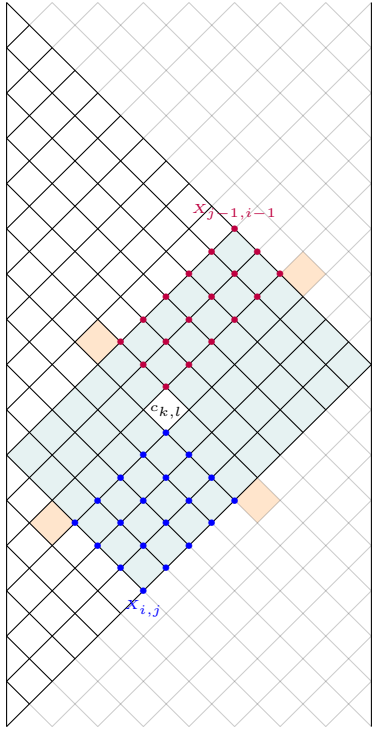
\begin{figure}
    \centering
        \begin{center}
		\begin{tikzpicture}[scale=1.2]
            \fill[color=teal, opacity=0.1] (1.5,1.5)--(4,4)--(2.5,5.5)--(0,3);
            \fill[color=white] (1.75,3.25)--(2,3.5)--(1.75,3.75)--(1.5,3.5);
            \fill[color=orange, opacity=0.2] (2.5,2.5)--(2.75,2.75)--(3,2.5)--(2.75,2.25);
            \fill[color=orange, opacity=0.2] (0.75,2.25)--(0.5,2.5)--(0.25,2.25)--(0.5,2);
            \fill[color=orange, opacity=0.2] (3,5)--(3.25,5.25)--(3.5,5)--(3.25,4.75);
            \fill[color=orange, opacity=0.2] (1.25,4.25)--(1,4.5)--(0.75,4.25)--(1,4);
                \draw(0,0)--(0,8);
			\draw (4,0)--(4,8);
                \draw (0,0)--(4,4);
                \draw (0,0.5)--(3.75,4.25);
			\draw (0,1)--(3.5,4.5);
                \draw (0,1.5)--(3.25,4.75);
			\draw(0,2)--(3,5);
                \draw (0,2.5)--(2.75,5.25);
			\draw(0,3)--(2.5,5.5);
                \draw (0,3.5)--(2.25,5.75);
			\draw(0,4)--(2,6);
                \draw (0,4.5)--(1.75,6.25);
			\draw (0,5)--(1.5,6.5);
                \draw (0,5.5)--(1.25,6.75);
			\draw(0,6)--(1,7);
                \draw (0,6.5)--(0.75,7.25);
			\draw(0,7)--(0.5,7.5);
                \draw (0,7.5)--(0.25,7.75);
			\draw(0,8)--(4,4);
                \draw(0,7.5)--(3.75,3.75);
			\draw(0,7)--(3.5,3.5);
                \draw(0,6.5)--(3.25,3.25);
			\draw(0,6)--(3,3);
                \draw(0,5.5)--(2.75,2.75);
			\draw(0,5)--(2.5,2.5);
                \draw(0,4.5)--(2.25,2.25);
			\draw(0,4)--(2,2);
                \draw(0,3.5)--(1.75,1.75);
			\draw(0,3)--(1.5,1.5);
                \draw(0,2.5)--(1.25,1.25);
			\draw(0,2)--(1,1);
                \draw(0,1.5)--(0.75,0.75);
			\draw(0,1)--(0.5,0.5); 
                \draw(0,0.5)--(0.25,0.25);
                \draw[opacity=0.2](0.5,0.5)--(1,0); 
                \draw[opacity=0.2](0.25,0.25)--(0.5,0); 
                \draw[opacity=0.2](1,1)--(2,0); 
                \draw[opacity=0.2](0.75,0.75)--(1.5,0); 
                \draw[opacity=0.2](1.5,1.5)--(3,0);
                \draw[opacity=0.2](1.25,1.25)--(2.5,0); 
                \draw[opacity=0.2](2,2)--(4,0);
                \draw[opacity=0.2](1.75,1.75)--(3.5,0); 
                \draw[opacity=0.2](2.5,2.5)--(4,1);
                \draw[opacity=0.2](2.25,2.25)--(4,0.5); 
                \draw[opacity=0.2](3,3)--(4,2);
                \draw[opacity=0.2](2.75,2.75)--(4,1.5); 
                \draw[opacity=0.2](3.5,3.5)--(4,3);
                \draw[opacity=0.2](3.25,3.25)--(4,2.5); 
                \draw[opacity=0.2](3.75,3.75)--(4,3.5); 
                \draw[opacity=0.2](3.5,4.5)--(4,5);
                \draw[opacity=0.2](3.75,4.25)--(4,4.5); 
                \draw[opacity=0.2](3,5)--(4,6);
                \draw[opacity=0.2](3.25,4.75)--(4,5.5); 
                \draw[opacity=0.2](2.5,5.5)--(4,7);
                \draw[opacity=0.2](2.75,5.25)--(4,6.5);
                \draw[opacity=0.2](2,6)--(4,8);
                \draw[opacity=0.2](2.25,5.75)--(4,7.5);
                \draw[opacity=0.2](1.5,6.5)--(3,8);
                \draw[opacity=0.2](1.75,6.25)--(3.5,8);
                \draw[opacity=0.2](1,7)--(2,8);
                \draw[opacity=0.2](1.25,6.75)--(2.5,8);
                \draw[opacity=0.2](0.5,7.5)--(1,8);
                \draw[opacity=0.2](0.75,7.25)--(1.5,8);
                \draw[opacity=0.2](0.25,7.75)--(0.5,8);
                \draw[opacity=0.2](0.5,8)--(4,4.5);
                \draw[opacity=0.2](1.5,8)--(4,5.5);
                \draw[opacity=0.2](2.5,8)--(4,6.5);
                \draw[opacity=0.2](3.5,8)--(4,7.5);
                \draw[opacity=0.2](0.5,0)--(4,3.5);
                \draw[opacity=0.2](1,0)--(4,3);
                \draw[opacity=0.2](1.5,0)--(4,2.5);
                \draw[opacity=0.2](2,0)--(4,2);
                \draw[opacity=0.2](2.5,0)--(4,1.5);
                \draw[opacity=0.2](3,0)--(4,1);
                \draw[opacity=0.2](3.5,0)--(4,0.5);
                \draw[opacity=0.2](4,5)--(1,8);
                \draw[opacity=0.2](4,6)--(2,8);
                \draw[opacity=0.2](4,7)--(3,8);
                \node[blue] at (1.5,1.5) {\tiny$\bullet$};
                \node[blue] at (1.75,1.75) {\tiny$\bullet$};
                \node[blue] at (2,2) {\tiny$\bullet$};
                \node[blue] at (2.25,2.25) {\tiny$\bullet$};
                \node[blue] at (2.5,2.5) {\tiny$\bullet$};
                \node[blue] at (1.5,2) {\tiny$\bullet$};
                \node[blue] at (1.75,2.25) {\tiny$\bullet$};
                \node[blue] at (2,2.5) {\tiny$\bullet$};
                \node[blue] at (2.25,2.75) {\tiny$\bullet$};
                \node[blue] at (1.25,1.75) {\tiny$\bullet$};
                \node[blue] at (1.5,2.5) {\tiny$\bullet$};
                \node[blue] at (1.75,2.75) {\tiny$\bullet$};
                \node[blue] at (2,3) {\tiny$\bullet$};
                \node[blue] at (1,2) {\tiny$\bullet$};
                \node[blue] at (1.25,2.25) {\tiny$\bullet$};
                \node[blue] at (1.5,3) {\tiny$\bullet$};
                \node[blue] at (1.75,3.25) {\tiny$\bullet$};
                \node[blue] at (0.75,2.25) {\tiny$\bullet$};
                \node[blue] at (1,2.5) {\tiny$\bullet$};
                \node[blue] at (1.25,2.75) {\tiny$\bullet$};
                \node[purple] at (1.75,3.75) {\tiny$\bullet$};
                \node[purple] at (2,4) {\tiny$\bullet$};
                \node[purple] at (2.25,4.25) {\tiny$\bullet$};
                \node[purple] at (2.5,4.5) {\tiny$\bullet$};
                \node[purple] at (2.75,4.75) {\tiny$\bullet$};
                \node[purple] at (3,5) {\tiny$\bullet$};
                \node[purple] at (1.75,4.25) {\tiny$\bullet$};
                \node[purple] at (2,4.5) {\tiny$\bullet$};
                \node[purple] at (2.25,4.75) {\tiny$\bullet$};
                \node[purple] at (2.5,5) {\tiny$\bullet$};
                \node[purple] at (2.75,5.25) {\tiny$\bullet$};
                \node[purple] at (1.5,4) {\tiny$\bullet$};
                \node[purple] at (1.75,4.75) {\tiny$\bullet$};
                \node[purple] at (2,5) {\tiny$\bullet$};
                \node[purple] at (2.25,5.25) {\tiny$\bullet$};
                \node[purple] at (2.5,5.5) {\tiny$\bullet$};
                \node[purple] at (1.25,4.25) {\tiny$\bullet$};
                \node[purple] at (1.5,4.5) {\tiny$\bullet$};
                \node[above] at (1.75,3.3) {\tiny $c_{k,l}$};
                \node[below,blue] at (1.5,1.5) {\tiny $X_{i,j}$};
                \node[above,purple] at (2.5,5.5) {\tiny $X_{j-1,i-1}$};
		\end{tikzpicture}
	\end{center}
    \caption{The $(X_{i,j},c_{k,l})$-shift. The $X$-variables shifted as $X\rightarrow X-z$ are denoted in \color{blue}\textbf{blue}\color{black}, and as $X\rightarrow X+z$ denoted in \color{purple}\textbf{purple}\color{black}. Also shown in \color{orange}\textbf{orange} \color{black}are the $c$-variables that shift, all of which lie outside the chosen maximal rectangle. Not shown are the cyclic images of any of these features. }
    \label{fig:shiftdef}
\end{figure}

The main tool we introduce in this paper is a family of kinematic shifts with properties tailored to understanding zeros and splitting; in particular we make repeated use of what we will call an $(X_{i,j},c_{k,l})$-shift. This is defined, for $n$-particle scattering, by choosing a maximal rectangle with $X_B=X_{i,j}$ together with a single plaquette $c_{k,l}$ on the interior of this rectangle. We then shift a subset of the $X$-variables as follows
\begin{equation}
    \label{Xcshift}
    \hat{X}_{a,b} \equiv \begin{cases}
        X_{a,b}-z & a=i,...,k, \hspace{11.5mm} b=j,...,l,\\
        X_{a,b}+z & a=k+1...j-1, \hspace{2mm} b=l+1,...,i-1,\\
        X_{a,b} & \text{otherwise}.
    \end{cases}
\end{equation}
The shifted $X$-variables are shown in Figure \ref{fig:shiftdef}. This shift preserves almost all of the non-cyclic $c$-variables, in particular only
\begin{equation}
    c_{i{-}1,k}, \;\; c_{j{-}1,k}, \;\; c_{i{-}1,l} \;\;\text{and}\;\; c_{j{-}1,l},
\end{equation}
are shifted; we note that all of these lie outside the maximal rectangle defined by $X_B=X_{i,j}$. This means that when we impose kinematic conditions on the $c$-variables on the interior of the rectangle, they are not deformed by the action of the shift. 

Applying an $(X_{i,j},c_{k,l})$-shift to an amplitude defines a deformed or shifted amplitude $\hat{A}_n(z)$. In the derivation of on-shell recursion relations from contour integration, an important criterion for their validity is the absence of a residue at $z=\infty$ of the function $\frac{\hat{A}_n(z)}{z}$. We therefore record here the large-$z$ fall-off behavior of amplitudes in the various considered models under different kinematic conditions. 

\begin{itemize}
    \item $\text{Tr}\phi^3$: For all $(X,c)$-shifts, in unconstrained kinematics the amplitudes scale as $\hat{A}_n^{\phi^3}(z)\sim z^{-2}$. We prove this in Section \ref{sec:gvector} by relating these to $g$-vector shifts and using properties of the \textit{surfaceology} construction of $\text{Tr}\phi^3$.
    \item NLSM: This model has only even-multiplicity interactions, and so not every $X$-variable corresponds to a pole. In particular only $X_{eo}$ corresponds to a factorization pole while $X_{ee}$ and $X_{oo}$ have zero residue, where $e=$ even and $o=$ odd. In generic kinematics, for all choices of $(X,c)$-shift the amplitude scales as $\hat{A}_n^{\text{NLSM}}(z)\sim  z^0$, and therefore does not define a valid recursion relation. When special kinematic restrictions are imposed the scaling is sometimes enhanced. A non-exhaustive list of the cases relevant for the discussion in this paper are:
    \begin{itemize}
        \item On even splitting kinematics ($\mathcal{Z}(X_B)$ with $c_*\neq 0$ and $X_B=X_{eo}$) the scaling of an $(X_B,c_*)$-shift is enhanced to $\sim z^{-2}$. This is proven in Section \ref{sec:gvector} using the $\delta$-shift relation to $\text{Tr}\phi^3$ \cite{Arkani-Hamed:2023swr, Arkani-Hamed:2024nhp}.
        \item On odd splitting kinematics ($\mathcal{Z}(X_B)$ with $c_*\neq 0$ and $X_B=X_{ee}$ or $X_{oo}$) the scaling of the $(X_B,c_*)$-shift is not enhanced, but remains $\sim z^0$.
        \item On even \textit{higher-order} splitting kinematics where each of the relaxed $c^{(a)}_*\neq 0$ are in the \textit{top or bottom row}, e.g. $X_B = X_{ij}$ and $c_*^{(a)}=c_{i,a}$ or $c_*^{(a)}=c_{a,j-1}$, the scaling of an $(X_B,c_*^{(a)})$-shift for any $c_*^{(a)}$, is enhanced to $\sim z^{-1}$. This case is important to the discussion of generalized splitting in Section \ref{sec:highersplits}. 
        \item On higher-order splitting kinematics, if the $c$-variables are not in the same row, the scaling is the same as the unconstrained case $\sim z^0$. 
    \end{itemize}
    \item YMS\footnote{In this context, Yang-Mills-Scalar is defined as the dimensional reduction of pure Yang-Mills from $d+2n$-dimensions to $d$-dimensions, producing a model of $d$-dimensional gluons coupled to $n$ complex, massless adjoint scalars $(\phi_i,\bar{\phi}_i)$ for $i=1,...,n$, with a specific quartic potential. The YMS amplitudes for which the splitting and zero properties hold are of the special form $A_{2n}\left[1^{\phi_1}2^{\bar{\phi}_1}3^{\phi_2}4^{\bar{\phi}_2}...\right]$, see \cite{Arkani-Hamed:2023swr, Arkani-Hamed:2023jry} for more details.  }: In this model $X_{eo}$ corresponds to scalar factorization, $X_{oo}$ corresponds to gluon factorization and $X_{ee}$ has zero residue.  In unconstrained kinematics we have observed that the scaling behavior of the shifted amplitude is always either $z^{-1}$ or $z^0$ depending on the shift. For example for $A_8^{\text{YMS}}\left[1^{\phi_1}2^{\overline{\phi}_1}3^{\phi_2}4^{\overline{\phi}_2}5^{\phi_3}6^{\overline{\phi}_3}7^{\phi_4}8^{\overline{\phi}_4}\right]$, under an $(X_{14},c_{25})$-shift the amplitude scales as $z^0$, but for an $(X_{14},c_{17})$-shift it is enhanced to $z^{-1}$ and therefore gives a valid recursion relation. We have neither a systematic understanding of when the scaling is enhanced nor a first-principles derivation of this fact. Empirically we have observed the following in numerous explicit cases and will conjecture that they are general in sequel:
    \begin{itemize}
        \item On scalar splitting kinematics ($\mathcal{Z}(X_B)$ with $c_*\neq 0$ and $X_B=X_{eo}$) the scaling of an $(X_B,c_*)$-shift is enhanced to $\sim z^{-2}$. If one \textit{assumes} the splitting formula (\ref{genericsplit}) then this scaling follows; to avoid a circular argument it would be preferable to have an independent understanding of this fact.
        \item On higher-order scalar splitting kinematics where each of the relaxed $c^{(a)}_*\neq 0$ are in the top or bottom row, and that row contains only scalar poles, the scaling of an $(X_B,c_*^{(a)})$-shift for any $c_*^{(a)}$, is enhanced to $\sim z^{-1}$. This is the same behavior as NLSM and will be discussed briefly in Section \ref{sec:highersplits}. 
    \end{itemize}
    \item Special Galileon: As an uncolored model with only even-multiplicity interactions, the Galileon has poles at $X_{oe}$ and no poles at $X_{ee}$ and $X_{oo}$. It also has many other non-planar poles located at $s_{i\cdots i+k}=0$ for $k$ even. While we do not have a fundamental reason to expect special Galileon amplitudes to scale in any particular way, we have empirically observed the following pattern:
    \begin{itemize}
        \item On unconstrained kinematics, 6- and 8-point special Galileon amplitudes behave as $z^2$ at infinity, under a $(X_{oe},c_{kl})$ shift. This is much lower than the naive power-counting estimates which are $z^5$ and $z^7$ at 6- and 8-point respectively. This extreme reduction in large $z$ fall-off encourages us to conjecture that this $z^2$ behavior continues for all $n$-point amplitudes.
        \item On split kinematics, i.e. $\mathcal{Z}(X_{oe})$ with $c_*\neq 0$, the 6- and 8-point amplitudes behave as $z^{{-}2}$ at infinity. Again, this $z^{{-}4}$ improvement in behavior on split kinematics vs. generic kinematics might be indicative that this behavior is shared at all multiplicity.
    \end{itemize}
    In this work, we only consider $(X_{oe},c_{kl})$ shifts in the context of the special Galileon model, leaving odd-splits involving soft-extended theories \cite{Cachazo:2016njl} to future work.
\end{itemize}

\subsection{\texorpdfstring{$g$}{g}-vector shifts and surfaceology}
\label{sec:gvector}

\paragraph{$\text{Tr}\phi^3$:} In the case of Tr$\phi^3$ theory, (\ref{Xcshift}) is actually a special case of a larger class of shifts known as $g$-vector shifts. These shifts arise naturally in the surface description of Tr$\phi^3$ and are special in that they preserve the combinatorial structure of these amplitudes even under large $g$-vector deformations \cite{Yang:2019esm, He:2018svj, Paranjape:2025wjk}. Let us understand why these are relevant to our discussion of the large $z$ behavior of Tr$\phi^3$.

Amplitudes in Tr$\phi^3$ have an alternative description as the canonical form on a positive geometry called the associahedron. This canonical form is known to be projectively invariant i.e. is preserved under the transformation $X_{ij}\to\Lambda(X) X_{ij}$. In \cite{He:2018svj, Yang:2019esm, Arkani-Hamed:2017mur}, projective invariance was shown to guarantee the absence of a pole at infinity. In terms of the amplitude itself, this is the statement that $g$-vector shifts of Tr$\phi^3$ amplitudes fall off as $z^{-2}$ or faster as $z\to\infty$ \cite{Paranjape:2025wjk}. This means that proving that Tr$\phi^3$ amplitudes behave as $z^{-2}$ at infinity is equivalent to demonstrating that the shift (\ref{Xcshift}) is a $g$-vector shift. 

Under the shift (\ref{Xcshift}), every Feynman diagram has at least one shifted propagator, this gives rise to the naive expectation of a $z^{-1}$ fall-off. For the fall-off to enhance to $z^{-2}$ there must be a cancellation between diagrams. It is instructive to see how this happens. At 5-point for example, performing an $(X_{13},c_{14})$-shift leaves us with two classes of diagrams: those with one shifted propagator and those with two. Grouping these in pairs by unshifted propagators, we can write the amplitude in the form
\begin{align}
\label{eq:5ptmanifestscaling}
    A_5^{\phi^3}\left[12345\right] &= \frac{1}{X_{24}}A_4^{\phi^3}\left[1245\right] + \frac{1}{X_{35}}A_4^{\phi^3}\left[1235\right] + \frac{1}{X_{13}X_{14}}.
\end{align}
Under the shift, each of the 4-point amplitudes that appears satisfies
\begin{align}
    \hat{A}^{\phi^3}_4[1234] = \frac{1}{\hat{X}_{13}}+\frac{1}{\hat{X}_{14}} = \frac{1}{{X}_{13}+z}+\frac{1}{{X}_{14}-z} \overset{z\to\infty}{\longrightarrow} -\frac{c_{13}}{z^2}.
\end{align}
Thus the organization in \eqref{eq:5ptmanifestscaling} manifests the enhanced $z^{-2}$ scaling.
Similarly at 6-point for an $(X_{14},c_{15})$-shift we can write the amplitude in the form
\begin{align}
    A_6^{\phi^3}\left[123456\right] &= \frac{1}{X_{13}X_{46}}A_4^{\phi^3}\left[1346\right] + \frac{1}{X_{13}X_{35}}A_4^{\phi^3}\left[1356\right]+ \frac{1}{X_{24}X_{46}}A_4^{\phi^3}\left[1246\right] \nonumber\\
    &\hspace{5mm} + \frac{1}{X_{24}X_{25}}A_4^{\phi^3}\left[1256\right]+ \frac{1}{X_{24}X_{35}}A_4^{\phi^3}\left[1256\right] \nonumber\\
    &\hspace{5mm }+ \frac{1}{X_{13}X_{14}X_{15}} + \frac{1}{X_{14}X_{15}X_{24}} + \frac{1}{X_{26}X_{35}X_{36}}+ \frac{1}{X_{26}X_{36}X_{46}}.
\end{align}
Again each of the 4-point amplitudes on the right-hand-side scale as $z^{-2}$ under the shift. The elementary observation in these examples is that the individual Feynman diagrams that scale as $z^{-1}$ can always be combined pairwise into an $A_4$ that scales as $z^{-2}$. From the Feynman diagram expansion, it is not at all obvious that this will continue at higher multiplicity, or for all choices of $(X,c)$-shift. The more formal argument below based on the surfaceology construction establishes concretely that this pattern does continue.

\begin{figure}
    \centering
    \begin{subfigure}[t]{0.55\textwidth}
                \centering
            \begin{tikzpicture}[scale=1.2, baseline={(0,-5)cm}]
    \begin{scope}[xshift=1.88cm,yshift=0.68cm]
    \fill[color=orange, opacity=0.2] (0,0) (20:2)--(100:2) (0,0) arc (20:100:2) ;
    \end{scope}
    \begin{scope}[xshift=-0.35cm,yshift=1.97cm]
    \fill[color=orange, opacity=0.2] (0,0) (100:2)--(160:2) (0,0) arc (100:160:2);
    \end{scope}
    \begin{scope}[xshift=-1.88cm,yshift=0.68cm]
    \fill[color=orange, opacity=0.2] (0,0) (160:2)--(260:2) (0,0) arc (160:260:2);
    \end{scope}
    \begin{scope}[xshift=-0.35cm,yshift=-1.97cm]
    \fill[color=orange, opacity=0.2] (0,0) (-100:2)--(20:2) (0,0) arc (-100:20:2);
    \end{scope}
    \draw (0,0) circle (2);
    \foreach \a in {1,2,3,4,5,6,7,8,9,10,11,12,13,14,15,16,17,18}{
    \node at (20*\a:2) {$\bullet$};
    \node at (20*\a:2.2) {\a};}
    \draw (20:2)--(100:2);
    \draw (100:2)--(160:2);
    \draw (160:2)--(260:2);
    \draw (260:2)--(20:2);
    \draw (20:2)--(60:2);
    \draw (20:2)--(80:2);
    \draw[red] (20:2)--(160:2);
    \draw (20:2)--(260:2);
    \draw (20:2)--(280:2);
    \draw (20:2)--(300:2);
    \draw (20:2)--(320:2);
    \draw (20:2)--(340:2);
    \draw (160:2)--(120:2);
    \draw (160:2)--(200:2);
    \draw (160:2)--(220:2);
    \draw (160:2)--(240:2);
\end{tikzpicture}
                \caption{}
                \label{fig:subsurfaces1}
            \end{subfigure}
            \begin{subfigure}[t]{0.44\textwidth}
                \centering
    \input{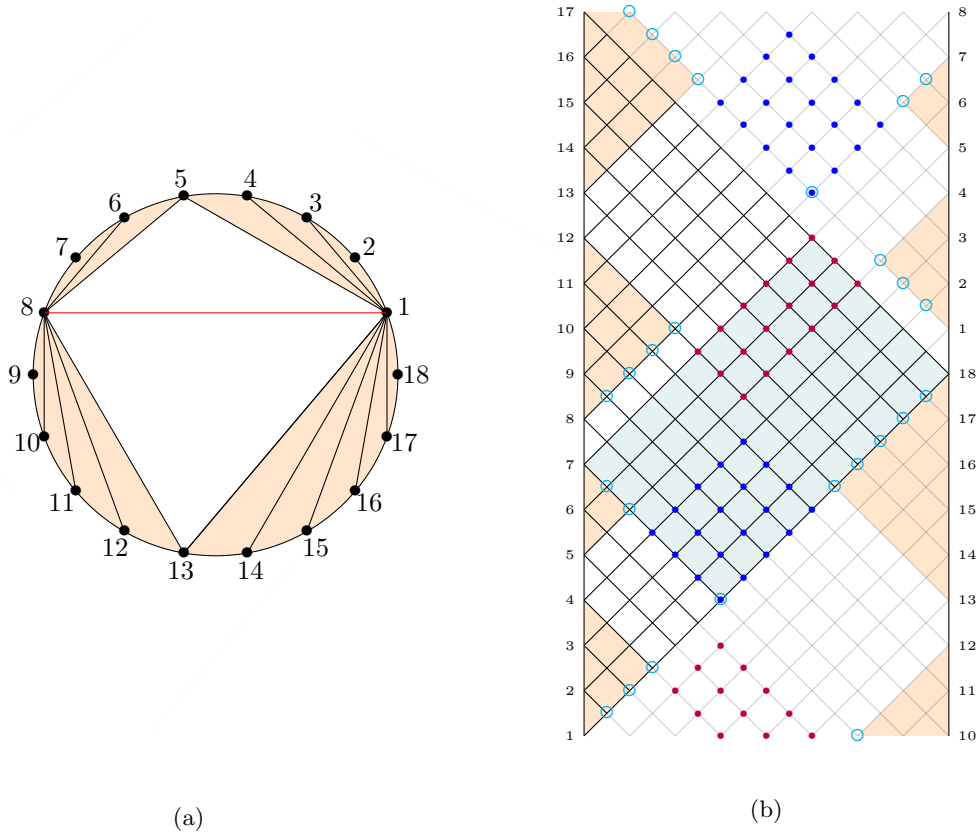}
                \caption{}
           \label{fig:subsurfaces2}
            \end{subfigure}
    \caption{Figure (a) the chords in triangulation $\mathcal{T}$ for an 18-point surface, the shifted chord for a $(X_{18},c_{4,12})$ shift in \textcolor{red}{\textbf{red}} and the four sub-surfaces in \textcolor{orange}{\textbf{orange}}. Figure (b) 18-point mesh under a $(X_{18},c_{4,12})$ shift. The \textcolor{blue}{\textbf{blue}} vertices indicate $X$'s that have been shifted negatively i.e. $\hat{X}=X-z$, while the \textcolor{purple}{\textbf{purple}} vertices indicate positively shifted $X$'s. The four orange regions denote the four sub-surfaces that $(X_{ik{+}1},X_{jk{+}1},X_{il{+}1},X_{jl{+}1})$ divides the $n$-point surface into. The vertices encircled in \textcolor{cyan}{\textbf{cyan}} denote all $X_{ij}\in\mathcal{T}$. The $c_{ij}$'s not in the basis are the ones directly below the encircled $X$'s.}
    \label{fig:subsurfaces}
\end{figure}

 Now we provide a brief introduction to $g$-vector shifts and subsequently describe which $g$-vector shift is equivalent to (\ref{Xcshift}) and thus prove the large $z$ behavior of Tr$\phi^3$ amplitudes. Begin by considering a kinematic basis consisting of $n{-}3$ planar variables $X_{ij}$ and $(n{-}2)(n{-}3)/2$ non-planar variables $c_{kl}$. These form a basis if
\begin{align}
    \{X_{ij}\}= \mathcal{T} \ \forall\ X_{ij}\in\text{basis,\  and }\  X_{i{+}1j{+}1}\notin\mathcal{T}\ \forall\ c_{ij}\in\text{basis}\,.
\end{align}
Here $\mathcal{T}$ is a triangulation of the $n$-point surface i.e. the chords $(ij)$ associated to the $X_{ij}$'s do not cross. See Figure \ref{fig:subsurfaces1} for an example of a triangulation of a generic surface. In other words, the set $\{X_{ij}\}$ in the chosen kinematic basis must be a valid set of propagators in a single Feynman diagram of Tr$\phi^3$. The set $\{c_{kl}\}$ in the basis are all of the non-planar variables, excluding the ones that lie directly below $\{X_{ij}\}$ in the kinematic mesh.

A $g$-vector shift\footnote{These shifts can also be written as 
\begin{align}
    \hat{X}_{ij}=X_{ij} + \vec{g}_{ij}.\vec{t} z
\end{align}
for every planar variable, including those not in the basis. The vector $g$ is the Feynman fan vector associated to $X_{ij}$, which is also the vector normal to the $X_{ij}=0$ facet of the associahedron. This is why these shifts are also known as $g$-vector shifts.} can now be defined as one that only affects the planar variables in the basis, leaving all the non-planar variables unshifted. Thus a $g$-vector shift is fully specified by a triangulation $\mathcal{T}$ and a direction $\vec{t}\in\mathbb{R}^{n{-}3}$. It is then given by
\begin{align}
    \hat{\vec{X}} = \vec{X} + z\,\vec{t}\,, 
\end{align}
where $\vec{X}=(X_{ij}\in \mathcal{T})$.

Let us look at an example. A possible 6-point basis can be constructed by starting with a triangulation $\{X_{24},X_{46},X_{26}\}$. Next, we add all $c_{ij}$ not directly below these $X_{ij}$ in the mesh i.e. $\{c_{14},c_{24},c_{25},c_{26},c_{36},c_{46}\}$. There is now a set of $g$-vector shifts available to us parametrized by a 3d vector $\vec{t}$. Taking for example $\vec{t}=(0,0,1)$, the corresponding shifts for the basis elements are
\begin{align}
    \hat{X}_{24}= X_{24}\,,&&\hat{X}_{46} =X_{46}\,,&&\hat{X}_{26}=X_{26}+z\,, && \hat{c}_{ij}=c_{ij}\ \forall\ c_{ij}\in\text{basis}\,.
\end{align}
Solving for the other $X$- and $c$-variables in terms of this basis, then tells us how these other variables shift
\begin{align}
    \hat{X}_{13}=&-X_{24}+c_{24}+c_{25}+c_{26} = X_{13}\,,\nonumber\\
    \hat{X}_{14}=&-\hat{X}_{26}+ X_{46}+c_{26}+c_{36} = X_{14}-z\,,\nonumber\\
    \hat{X}_{15}=&-\hat{X}_{26}+c_{26}+c_{36}+c_{46} = X_{15}-z\,,\nonumber\\
    \hat{X}_{25}=&-X_{46}+ X_{24}+c_{14}+c_{46} = X_{25}\,,\nonumber\\
    \hat{X}_{35}=&-X_{46}+c_{14}+c_{24}+c_{46} = X_{35}\,,\nonumber\\
    \hat{X}_{36}=&-X_{24}+\hat{X}_{26}+c_{24}+c_{25} = X_{36}+z\,,\nonumber\\
    \hat{c}_{13}=&X_{13}+X_{24}-\hat{X}_{14} = c_{13}+z\,,\nonumber\\
    \hat{c}_{15}=&\hat{X}_{15}+\hat{X}_{26}-X_{25}=c_{15}\,,\nonumber\\
    \hat{c}_{35}=&X_{35}+X_{46}-\hat{X}_{36}=c_{35}-z\,.
\end{align}
Note that this is exactly the $(X_{14},c_{15})$-shift defined in Section \ref{sec:shifts}.

One can now ask whether there exists a $g$-vector shift $(\mathcal{T},\vec{t})$ that coincides with \textit{any} choice of $(X_{ij},c_{kl})$-shift (\ref{Xcshift}). The answer surprisingly is yes! The triangulation of the associated $g$-vector shift can be found as follows:
\begin{itemize}
    \item For generic $(X_{ij},c_{kl})$, there are four $c_{ij}$'s that shift. These are $c_{i{-}1k}$, $c_{j{-}1k}$, $c_{i{-}1l}$ and $c_{j{-}1l}$. These cannot be in the basis i.e. $\{X_{ik{+}1},X_{jk{+}1},X_{il{+}1},X_{jl{+}1}\}\in\mathcal{T}$. Further, we include the diagonal of this quadrilateral $X_{ij}$ in $\mathcal{T}$ (see Figure \ref{fig:subsurfaces1}).
    \item The other elements of $\mathcal{T}$ come from triangulations of the four sub-surfaces that the quadrilateral $(i,k{+}1,j,l{+}1)$ divides the $n$-point surface into (see Figure \ref{fig:subsurfaces1}). We choose these to be
    \begin{align}
        \{X_{ii{+}2},\cdots,X_{ik}\}= &\mathcal{T}_{(i\cdots k{+}1)}\,,\nonumber\\
        \{X_{il{+}1},\cdots,X_{ii{-}2}\}= &\mathcal{T}_{(l{+}1\cdots i)}\,,\nonumber\\
        \{X_{jj{+}2},\cdots,X_{jl}\}= &\mathcal{T}_{(j\cdots l{+}1)}\,,\nonumber\\
        \{X_{jk{+}1},\cdots,X_{jj{-}2}\}= &\mathcal{T}_{(k{+}1\cdots j)}\,.
    \end{align}
\end{itemize}

The full triangulation is then
\begin{align}
    \mathcal{T} = \{X_{ik{+}1},X_{jk{+}1},X_{il{+}1},X_{jl{+}1},\mathcal{T}_{(i\cdots k{+}1)},\mathcal{T}_{(k{+}1\cdots j)},\mathcal{T}_{(j\cdots l{+}1)},\mathcal{T}_{(l{+}1\cdots i)},X_{ij}\},.
\end{align}
These are highlighted in Figure \ref{fig:subsurfaces2}, along with the four relevant sub-surfaces. The last ingredient to map the shift (\ref{Xcshift}) to a $g$-vector shift is the vector $\vec{t}\in\mathbb{R}^{n{-}3}$. This is
\begin{align}
    \vec{t}= (\underbrace{0,0,\cdots,0,0}_{n{-}4},1)\,.
\end{align}
The behavior of the amplitude when $z\to\infty$ is then determined to be \cite{Paranjape:2025wjk, Yang:2019esm, He:2018svj}
\begin{align}
    \hat{A}_n^{\phi^3}(z) \sim z^{-2} \Rightarrow\ \text{Tr$\phi^3$ has no pole at infinity}.
\end{align}

\paragraph{NLSM:} Remarkably, in \cite{Arkani-Hamed:2023swr} it was noted that pion amplitudes in the NLSM can be obtained from those in Tr$\phi^3$ via a so-called $\delta$-shift. This is a shift on the $X$-variables defined by
\begin{align}
    \hat{X}_{oo}=X_{oo}+\delta\,,\ \hat{X}_{ee}=X_\text{ee}-\delta,
\end{align}
where $e=$ even and $o=$ odd. Applying this to a Tr$\phi^3$ amplitude and taking $\delta\to\infty$ then gives
\begin{equation}
    \label{deltatrphi3}
    \lim_{\delta\rightarrow \infty}\delta^{2-2n} \left(A_{2n}^{\phi^3}\left[1,2,...,2n-1,2n\right]\biggr\vert_{\substack{X_{ee}\rightarrow X_{ee}-\delta \\ X_{oo}\rightarrow X_{oo}+\delta}}\right) = A_{2n}^{\text{NLSM}}\left[1,2,...,2n-1,2n\right].
\end{equation}
Indeed this $\delta$-shift is also a special case of the larger class of $g$-vector shifts, and the behavior of Tr$\phi^3$ at infinity under $g$-vector shifts was recently studied in \cite{Paranjape:2025wjk} where it was shown that the result of $A_{2n}^{\text{Tr}\phi^3}$ at infinity under two $g$-vector shifts is independent of the order in which they were carried out. In other words, the large $\delta$ limit that gives NLSM commutes with the large $z$ limit. This allows us to understand the behavior of NLSM at large $z$ from the behavior of Tr$\phi^3$.

On even split kinematics for example,
\begin{align}
    A_{2n}^{\phi^3} \xrightarrow[\{c_{k,l}\}\neq 0]{\mathcal{Z}\left(X_B\right)} \left(\frac{1}{X_B}+\frac{1}{X_T}\right) A_{n_B}^{\phi^3} A_{n_T}^{\phi^3}.
\end{align}
Under a $(X_B,c_{kl})$ shift, none of the $X$'s in the sub-amplitudes $A_{n_i}$ shift. This leaves only the prefactor that gives
\begin{align}
    \hat{A}_{2n}^{\phi^3} =&\left(\frac{1}{X_B+z}+\frac{1}{X_T-z}\right) A_{n_B}^{\phi^3} A_{n_T}^{\phi^3}\nonumber\\
      &\overset{z\to\infty}{\longrightarrow}\frac1{z^2} \frac{c_{kl}}{X_BX_T} A_{n_B}^{\phi^3} A_{n_T}^{\phi^3}\,.
\end{align}
Performing a $\delta$-shift next gives,
\begin{align}
\label{eq:NLSMfalloff1}
    \lim_{\delta\to\infty}\lim_{z\to\infty} \hat{A}_{2n}^{\phi^3} = \frac1{z^2\delta^{2n-2}} \frac{c_{kl}}{X_BX_T} A_{n_B}^\text{NLSM} A_{n_T}^\text{NLSM}\,,
\end{align}
where we have used the delta scaling of the sub-amplitudes and the fact that $n_B+n_T=n+2$.

Commutativity of $g$-vector shifts then tells us that this is equivalent to first doing a large $\delta$-shift (which gives us NLSM) and then doing a large $z$-shift. Thus, the fall-off of NLSM amplitudes at infinity,
\begin{align}
\label{eq:NLSMfalloff2}
    \hat{A}_n^\text{NLSM}(z) \sim z^{-k},
\end{align}
satisfies the condition that the fall-offs must match:
\begin{align}
    2 + 2n -2 = k + 2n-2 \Rightarrow k=2\,.
\end{align}
On the LHS above is the fall-off coefficient of $\delta\sim z\to\infty$ in \eqref{eq:NLSMfalloff1} while on the RHS we have the fall-off coefficient in \eqref{eq:NLSMfalloff2}. This gives us the result we are after,
\begin{align}
    \hat{A}_n^\text{NLSM}(z) \sim z^{-2}\ \Rightarrow\ \text{ NLSM has no pole at infinity on split kinematics.}
\end{align}
In Section \ref{sec:highersplits}, we also discuss other types of higher-order splits. The large $z$ behavior of NLSM on such kinematics can also be read off of the Tr$\phi^3$ higher-order splitting theorem in a similar manner.

Finally, we note that on generic kinematics NLSM scales as $\sim z^0$. This is because, as shown in \cite{Paranjape:2025wjk}, the products of a large $g$-vector deformation (such as NLSM) have a ``$c$-expansion'' i.e. the amplitude can be written in a form that contains only $c$'s in the numerator and $X$'s in the denominator, where $c$ and $X$ form a kinematic basis. This means that under a subsequent $g$-vector shift, the worst behavior one can obtain is $\sim z^0$.

\section{Hidden Zeros and Generalized Splitting}
\label{sec:zeros}

\subsection{Recursive proof of zeros and splitting}
\label{sec:proofofzeros}

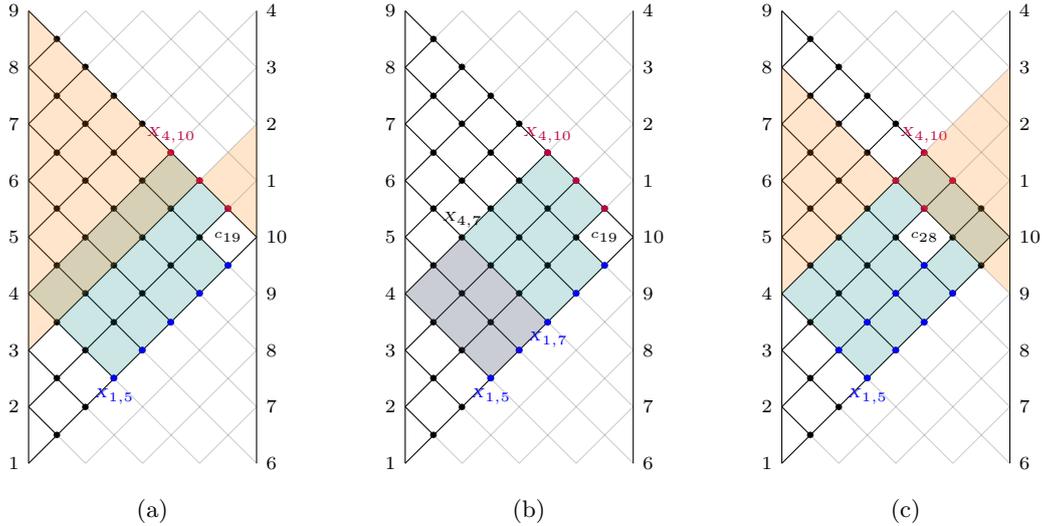
\begin{figure}
    \centering
            \begin{subfigure}[t]{0.32\textwidth}
                \centering
                    		\begin{tikzpicture}[scale=0.75]
                \draw (0,0)--(4,4);
			\draw (0,1)--(3.5,4.5);
			\draw(0,2)--(3,5);
			\draw(0,3)--(2.5,5.5);
			\draw(0,4)--(2,6);
			\draw (0,5)--(1.5,6.5);
			\draw(0,6)--(1,7);
			\draw(0,7)--(0.5,7.5);
			\draw(0,0)--(0,8);
			\draw (4,0)--(4,8);
			\draw(0,8)--(4,4);
			\draw(0,7)--(3.5,3.5);
			\draw(0,6)--(3,3);
			\draw(0,5)--(2.5,2.5);
			\draw(0,4)--(2,2);
			\draw(0,3)--(1.5,1.5);
			\draw(0,2)--(1,1);
			\draw(0,1)--(0.5,0.5); 
                \draw[opacity=0.2](0.5,0.5)--(1,0); 
                \draw[opacity=0.2](1,1)--(2,0); 
                \draw[opacity=0.2](1.5,1.5)--(3,0); 
                \draw[opacity=0.2](2,2)--(4,0);
                \draw[opacity=0.2](2.5,2.5)--(4,1);
                \draw[opacity=0.2](3,3)--(4,2);
                \draw[opacity=0.2](3.5,3.5)--(4,3);
                \draw[opacity=0.2](3.5,4.5)--(4,5); 
                \draw[opacity=0.2](3,5)--(4,6);
                \draw[opacity=0.2](2.5,5.5)--(4,7);
                \draw[opacity=0.2](2,6)--(4,8);
                \draw[opacity=0.2](1.5,6.5)--(3,8);
                \draw[opacity=0.2](1,7)--(2,8);
                \draw[opacity=0.2](0.5,7.5)--(1,8);
                \draw[opacity=0.2](1,0)--(4,3);
                \draw[opacity=0.2](2,0)--(4,2);
                \draw[opacity=0.2](3,0)--(4,1);
                \draw[opacity=0.2](4,5)--(1,8);
                \draw[opacity=0.2](4,6)--(2,8);
                \draw[opacity=0.2](4,7)--(3,8);
			\node at (0.5,0.5) {\tiny$\bullet$};
                \node at (1,1) {\tiny$\bullet$};
			\node at (1.5,1.5) {\tiny$\bullet$};
                \node at (2,2) {\tiny$\bullet$};
			\node at (2.5,2.5) {\tiny$\bullet$};
                \node at (3,3) {\tiny$\bullet$};
			\node at (3.5,3.5) {\tiny$\bullet$};
                \node at (0.5,1.5) {\tiny$\bullet$};
                \node at (1,2) {\tiny$\bullet$};
			\node at (1.5,2.5) {\tiny$\bullet$};
                \node at (2,3) {\tiny$\bullet$};
			\node at (2.5,3.5) {\tiny$\bullet$};
                \node at (3,4) {\tiny$\bullet$};
			\node at (3.5,4.5) {\tiny$\bullet$};
                \node at (0.5,2.5) {\tiny$\bullet$};
                \node at (1,3) {\tiny$\bullet$};
			\node at (1.5,3.5) {\tiny$\bullet$};
                \node at (2,4) {\tiny$\bullet$};
			\node at (2.5,4.5) {\tiny$\bullet$};
                \node at (3,5) {\tiny$\bullet$};
                \node at (0.5,3.5) {\tiny$\bullet$};
                \node at (1,4) {\tiny$\bullet$};
			\node at (1.5,4.5) {\tiny$\bullet$};
                \node at (2,5) {\tiny$\bullet$};
			\node at (2.5,5.5) {\tiny$\bullet$};
                \node at (0.5,4.5) {\tiny$\bullet$};
                \node at (1,5) {\tiny$\bullet$};
			\node at (1.5,5.5) {\tiny$\bullet$};
                \node at (2,6) {\tiny$\bullet$};
                \node at (0.5,5.5) {\tiny$\bullet$};
                \node at (1,6) {\tiny$\bullet$};
			\node at (1.5,6.5) {\tiny$\bullet$};
                \node at (0.5,6.5) {\tiny$\bullet$};
                \node at (1,7) {\tiny$\bullet$};
                \node at (0.5,7.5) {\tiny$\bullet$};
                \node at (3.5,4) {\color{black}\tiny $c_{19}$};
                \node[left] at (0,0) {\scriptsize $1$};
                \node[left] at (0,1) {\scriptsize $2$};
                \node[left] at (0,2) {\scriptsize $3$};
                \node[left] at (0,3) {\scriptsize $4$};
                \node[left] at (0,4) {\scriptsize $5$};
                \node[left] at (0,5) {\scriptsize $6$};
                \node[left] at (0,6) {\scriptsize $7$};
                \node[left] at (0,7) {\scriptsize $8$};
                \node[left] at (0,8) {\scriptsize $9$};
                \node[right] at (4,0) {\scriptsize $6$};
                \node[right] at (4,1) {\scriptsize $7$};
                \node[right] at (4,2) {\scriptsize $8$};
                \node[right] at (4,3) {\scriptsize $9$};
                \node[right] at (4,4) {\scriptsize $10$};
                \node[right] at (4,5) {\scriptsize $1$};
                \node[right] at (4,6) {\scriptsize $2$};
                \node[right] at (4,7) {\scriptsize $3$};
                \node[right] at (4,8) {\scriptsize $4$};
                \fill[color=teal,opacity=0.2] (0,3)--(1.5,1.5)--(3.5,3.5)--(2,5);
                \fill[color=teal,opacity=0.2] (2,5)--(2.5,5.5)--(3.5,4.5)--(3,4);
                \node[above] at (2.5,5.5) {\color{purple}\tiny $X_{4,10}$};
                \node at (2.5,5.5) {\color{purple}\tiny$\bullet$};
                \node at (3,5) {\color{purple}\tiny$\bullet$};
                \node at (3.5,4.5) {\color{purple}\tiny$\bullet$};
                \node[below] at (1.5,1.5) {\color{blue}\tiny $X_{1,5}$};
                \node at (1.5,1.5) {\color{blue}\tiny$\bullet$};
                \node at (2,2) {\color{blue}\tiny$\bullet$};
                \node at (2.5,2.5) {\color{blue}\tiny$\bullet$};
                \node at (3,3) {\color{blue}\tiny$\bullet$};
                \node at (3.5,3.5) {\color{blue}\tiny$\bullet$};
                \fill[color=orange,opacity=0.2] (0,2)--(3,5)--(0,8);
                \fill[color=orange,opacity=0.2] (4,4)--(3,5)--(4,6);
		      \end{tikzpicture}
                \caption{}
                \label{fig:Proof1}
            \end{subfigure}
            \begin{subfigure}[t]{0.32\textwidth}
            \centering
                		\begin{tikzpicture}[scale=0.75]
                \draw (0,0)--(4,4);
			\draw (0,1)--(3.5,4.5);
			\draw(0,2)--(3,5);
			\draw(0,3)--(2.5,5.5);
			\draw(0,4)--(2,6);
			\draw (0,5)--(1.5,6.5);
			\draw(0,6)--(1,7);
			\draw(0,7)--(0.5,7.5);
			\draw(0,0)--(0,8);
			\draw (4,0)--(4,8);
			\draw(0,8)--(4,4);
			\draw(0,7)--(3.5,3.5);
			\draw(0,6)--(3,3);
			\draw(0,5)--(2.5,2.5);
			\draw(0,4)--(2,2);
			\draw(0,3)--(1.5,1.5);
			\draw(0,2)--(1,1);
			\draw(0,1)--(0.5,0.5); 
                \draw[opacity=0.2](0.5,0.5)--(1,0); 
                \draw[opacity=0.2](1,1)--(2,0); 
                \draw[opacity=0.2](1.5,1.5)--(3,0); 
                \draw[opacity=0.2](2,2)--(4,0);
                \draw[opacity=0.2](2.5,2.5)--(4,1);
                \draw[opacity=0.2](3,3)--(4,2);
                \draw[opacity=0.2](3.5,3.5)--(4,3);
                \draw[opacity=0.2](3.5,4.5)--(4,5); 
                \draw[opacity=0.2](3,5)--(4,6);
                \draw[opacity=0.2](2.5,5.5)--(4,7);
                \draw[opacity=0.2](2,6)--(4,8);
                \draw[opacity=0.2](1.5,6.5)--(3,8);
                \draw[opacity=0.2](1,7)--(2,8);
                \draw[opacity=0.2](0.5,7.5)--(1,8);
                \draw[opacity=0.2](1,0)--(4,3);
                \draw[opacity=0.2](2,0)--(4,2);
                \draw[opacity=0.2](3,0)--(4,1);
                \draw[opacity=0.2](4,5)--(1,8);
                \draw[opacity=0.2](4,6)--(2,8);
                \draw[opacity=0.2](4,7)--(3,8);
			\node at (0.5,0.5) {\tiny$\bullet$};
                \node at (1,1) {\tiny$\bullet$};
			\node at (1.5,1.5) {\tiny$\bullet$};
                \node at (2,2) {\tiny$\bullet$};
			\node at (2.5,2.5) {\tiny$\bullet$};
                \node at (3,3) {\tiny$\bullet$};
			\node at (3.5,3.5) {\tiny$\bullet$};
                \node at (0.5,1.5) {\tiny$\bullet$};
                \node at (1,2) {\tiny$\bullet$};
			\node at (1.5,2.5) {\tiny$\bullet$};
                \node at (2,3) {\tiny$\bullet$};
			\node at (2.5,3.5) {\tiny$\bullet$};
                \node at (3,4) {\tiny$\bullet$};
			\node at (3.5,4.5) {\tiny$\bullet$};
                \node at (0.5,2.5) {\tiny$\bullet$};
                \node at (1,3) {\tiny$\bullet$};
			\node at (1.5,3.5) {\tiny$\bullet$};
                \node at (2,4) {\tiny$\bullet$};
			\node at (2.5,4.5) {\tiny$\bullet$};
                \node at (3,5) {\tiny$\bullet$};
                \node at (0.5,3.5) {\tiny$\bullet$};
                \node at (1,4) {\tiny$\bullet$};
			\node at (1.5,4.5) {\tiny$\bullet$};
                \node at (2,5) {\tiny$\bullet$};
			\node at (2.5,5.5) {\tiny$\bullet$};
                \node at (0.5,4.5) {\tiny$\bullet$};
                \node at (1,5) {\tiny$\bullet$};
			\node at (1.5,5.5) {\tiny$\bullet$};
                \node at (2,6) {\tiny$\bullet$};
                \node at (0.5,5.5) {\tiny$\bullet$};
                \node at (1,6) {\tiny$\bullet$};
			\node at (1.5,6.5) {\tiny$\bullet$};
                \node at (0.5,6.5) {\tiny$\bullet$};
                \node at (1,7) {\tiny$\bullet$};
                \node at (0.5,7.5) {\tiny$\bullet$};
                \node at (3.5,4) {\color{black}\tiny $c_{19}$};
                \node[left] at (0,0) {\scriptsize $1$};
                \node[left] at (0,1) {\scriptsize $2$};
                \node[left] at (0,2) {\scriptsize $3$};
                \node[left] at (0,3) {\scriptsize $4$};
                \node[left] at (0,4) {\scriptsize $5$};
                \node[left] at (0,5) {\scriptsize $6$};
                \node[left] at (0,6) {\scriptsize $7$};
                \node[left] at (0,7) {\scriptsize $8$};
                \node[left] at (0,8) {\scriptsize $9$};
                \node[right] at (4,0) {\scriptsize $6$};
                \node[right] at (4,1) {\scriptsize $7$};
                \node[right] at (4,2) {\scriptsize $8$};
                \node[right] at (4,3) {\scriptsize $9$};
                \node[right] at (4,4) {\scriptsize $10$};
                \node[right] at (4,5) {\scriptsize $1$};
                \node[right] at (4,6) {\scriptsize $2$};
                \node[right] at (4,7) {\scriptsize $3$};
                \node[right] at (4,8) {\scriptsize $4$};
                \fill[color=teal,opacity=0.2] (0,3)--(1.5,1.5)--(3.5,3.5)--(2,5);
                \fill[color=teal,opacity=0.2] (2,5)--(2.5,5.5)--(3.5,4.5)--(3,4);
                \node[above] at (2.5,5.5) {\color{purple}\tiny $X_{4,10}$};
                \node[above] at (1,4) {\tiny $X_{4,7}$};
                \node at (2.5,5.5) {\color{purple}\tiny$\bullet$};
                \node at (3,5) {\color{purple}\tiny$\bullet$};
                \node at (3.5,4.5) {\color{purple}\tiny$\bullet$};
                \node[below] at (1.5,1.5) {\color{blue}\tiny $X_{1,5}$};
                \node[below] at (2.5,2.5) {\color{blue}\tiny $X_{1,7}$};
                \node at (1.5,1.5) {\color{blue}\tiny$\bullet$};
                \node at (2,2) {\color{blue}\tiny$\bullet$};
                \node at (2.5,2.5) {\color{blue}\tiny$\bullet$};
                \node at (3,3) {\color{blue}\tiny$\bullet$};
                \node at (3.5,3.5) {\color{blue}\tiny$\bullet$};
                \fill[color=purple,opacity=0.1] (1.5,1.5)--(2.5,2.5)--(1,4)--(0,3);
		      \end{tikzpicture}
            \caption{}
            \label{fig:Proof2}
            \end{subfigure}
            \begin{subfigure}[t]{0.32\textwidth}
            \centering
                		\begin{tikzpicture}[scale=0.75]
                \draw (0,0)--(4,4);
			\draw (0,1)--(3.5,4.5);
			\draw(0,2)--(3,5);
			\draw(0,3)--(2.5,5.5);
			\draw(0,4)--(2,6);
			\draw (0,5)--(1.5,6.5);
			\draw(0,6)--(1,7);
			\draw(0,7)--(0.5,7.5);
			\draw(0,0)--(0,8);
			\draw (4,0)--(4,8);
			\draw(0,8)--(4,4);
			\draw(0,7)--(3.5,3.5);
			\draw(0,6)--(3,3);
			\draw(0,5)--(2.5,2.5);
			\draw(0,4)--(2,2);
			\draw(0,3)--(1.5,1.5);
			\draw(0,2)--(1,1);
			\draw(0,1)--(0.5,0.5); 
                \draw[opacity=0.2](0.5,0.5)--(1,0); 
                \draw[opacity=0.2](1,1)--(2,0); 
                \draw[opacity=0.2](1.5,1.5)--(3,0); 
                \draw[opacity=0.2](2,2)--(4,0);
                \draw[opacity=0.2](2.5,2.5)--(4,1);
                \draw[opacity=0.2](3,3)--(4,2);
                \draw[opacity=0.2](3.5,3.5)--(4,3);
                \draw[opacity=0.2](3.5,4.5)--(4,5); 
                \draw[opacity=0.2](3,5)--(4,6);
                \draw[opacity=0.2](2.5,5.5)--(4,7);
                \draw[opacity=0.2](2,6)--(4,8);
                \draw[opacity=0.2](1.5,6.5)--(3,8);
                \draw[opacity=0.2](1,7)--(2,8);
                \draw[opacity=0.2](0.5,7.5)--(1,8);
                \draw[opacity=0.2](1,0)--(4,3);
                \draw[opacity=0.2](2,0)--(4,2);
                \draw[opacity=0.2](3,0)--(4,1);
                \draw[opacity=0.2](4,5)--(1,8);
                \draw[opacity=0.2](4,6)--(2,8);
                \draw[opacity=0.2](4,7)--(3,8);
			\node at (0.5,0.5) {\tiny$\bullet$};
                \node at (1,1) {\tiny$\bullet$};
			\node at (1.5,1.5) {\tiny$\bullet$};
                \node at (2,2) {\tiny$\bullet$};
			\node at (2.5,2.5) {\tiny$\bullet$};
                \node at (3,3) {\tiny$\bullet$};
			\node at (3.5,3.5) {\tiny$\bullet$};
                \node at (0.5,1.5) {\tiny$\bullet$};
                \node at (1,2) {\tiny$\bullet$};
			\node at (1.5,2.5) {\tiny$\bullet$};
                \node at (2,3) {\tiny$\bullet$};
			\node at (2.5,3.5) {\tiny$\bullet$};
                \node at (3,4) {\tiny$\bullet$};
			\node at (3.5,4.5) {\tiny$\bullet$};
                \node at (0.5,2.5) {\tiny$\bullet$};
                \node at (1,3) {\tiny$\bullet$};
			\node at (1.5,3.5) {\tiny$\bullet$};
                \node at (2,4) {\tiny$\bullet$};
			\node at (2.5,4.5) {\tiny$\bullet$};
                \node at (3,5) {\tiny$\bullet$};
                \node at (0.5,3.5) {\tiny$\bullet$};
                \node at (1,4) {\tiny$\bullet$};
			\node at (1.5,4.5) {\tiny$\bullet$};
                \node at (2,5) {\tiny$\bullet$};
			\node at (2.5,5.5) {\tiny$\bullet$};
                \node at (0.5,4.5) {\tiny$\bullet$};
                \node at (1,5) {\tiny$\bullet$};
			\node at (1.5,5.5) {\tiny$\bullet$};
                \node at (2,6) {\tiny$\bullet$};
                \node at (0.5,5.5) {\tiny$\bullet$};
                \node at (1,6) {\tiny$\bullet$};
			\node at (1.5,6.5) {\tiny$\bullet$};
                \node at (0.5,6.5) {\tiny$\bullet$};
                \node at (1,7) {\tiny$\bullet$};
                \node at (0.5,7.5) {\tiny$\bullet$};
                \node at (2.5,4) {\color{black}\tiny $c_{28}$};
                \node[left] at (0,0) {\scriptsize $1$};
                \node[left] at (0,1) {\scriptsize $2$};
                \node[left] at (0,2) {\scriptsize $3$};
                \node[left] at (0,3) {\scriptsize $4$};
                \node[left] at (0,4) {\scriptsize $5$};
                \node[left] at (0,5) {\scriptsize $6$};
                \node[left] at (0,6) {\scriptsize $7$};
                \node[left] at (0,7) {\scriptsize $8$};
                \node[left] at (0,8) {\scriptsize $9$};
                \node[right] at (4,0) {\scriptsize $6$};
                \node[right] at (4,1) {\scriptsize $7$};
                \node[right] at (4,2) {\scriptsize $8$};
                \node[right] at (4,3) {\scriptsize $9$};
                \node[right] at (4,4) {\scriptsize $10$};
                \node[right] at (4,5) {\scriptsize $1$};
                \node[right] at (4,6) {\scriptsize $2$};
                \node[right] at (4,7) {\scriptsize $3$};
                \node[right] at (4,8) {\scriptsize $4$};
                \fill[color=teal,opacity=0.2] (0,3)--(1.5,1.5)--(3,3)--(1.5,4.5);
                \fill[color=teal,opacity=0.2] (2,5)--(2.5,5.5)--(4,4)--(3.5,3.5);
                \fill[color=teal,opacity=0.2] (3.5,3.5)--(3,3)--(2.5,3.5)--(3,4);
                \fill[color=teal,opacity=0.2] (1.5,4.5)--(2,5)--(2.5,4.5)--(2,4);
                \node[above] at (2.5,5.5) {\color{purple}\tiny $X_{4,10}$};
                \node at (2.5,5.5) {\color{purple}\tiny$\bullet$};
                \node at (3,5) {\color{purple}\tiny$\bullet$};
                \node at (2.5,4.5) {\color{purple}\tiny$\bullet$};
                \node at (2,5) {\color{purple}\tiny$\bullet$};
                \node[below] at (1.5,1.5) {\color{blue}\tiny $X_{1,5}$};
                \node at (1.5,1.5) {\color{blue}\tiny$\bullet$};
                \node at (2,2) {\color{blue}\tiny$\bullet$};
                \node at (2.5,2.5) {\color{blue}\tiny$\bullet$};
                \node at (3,3) {\color{blue}\tiny$\bullet$};
                \node at (2.5,3.5) {\color{blue}\tiny$\bullet$};
                \node at (2,3) {\color{blue}\tiny$\bullet$};
                \node at (1.5,2.5) {\color{blue}\tiny$\bullet$};
                \node at (1,2) {\color{blue}\tiny$\bullet$};
                \fill[color=orange,opacity=0.2] (0,3)--(2,5)--(0,7);
                \fill[color=orange,opacity=0.2] (4,3)--(2,5)--(4,7);
		      \end{tikzpicture}
            \caption{}
            \label{fig:Proof3}
            \end{subfigure}
    \caption{(a) and (c): at least one of the sub-amplitudes on each of the residues of the shifted poles (except $X_T$ and $X_B$) is evaluated on lower-point zero kinematics and therefore vanishes. (b): an example of the rectangle rule argument leading to the kinematic remapping; in this case the application of the rule to the rectangle shown gives $X_{4,7}+\hat{X}_{1,5}-\hat{X}_{1,7}=0$. When the relaxed plaquette is not in the corner, this rectangle rule argument will be modified leading to the non-trivial kinematic remapping in the generic case.}
    \label{fig:meshproof}
\end{figure}

In this section we present a new proof of the smooth splitting relations (\ref{genericsplit}), and as a corollary the hidden zeros \cite{Arkani-Hamed:2023swr}. The idea is to use an $(X_{ij},c_{kl})$-shift to reconstruct the amplitude as a contour integral, with the external kinematics of the unshifted amplitude taken to be near the zero $\mathcal{Z}(X_{ij})$ with $c_*=c_{kl}\neq 0$. We will first give the argument in the simpler corner case and then extend it to the generic case. Without loss of generality we will take $i=1$. We present the argument for $\text{Tr}\phi^3$; for NLSM and YMS the argument is identical except for the restrictions on the multiplicity $n$ and choice of $X_B$ to ensure that the resulting formula describes splitting on scalar channels. We will comment on non-scalar splitting below but otherwise leave this for future work.\\
\\
\textbf{Corner case:} We consider an $n$-particle amplitude on the special kinematics defined near the zero $\mathcal{Z}(X_{1j})$ with the corner non-cyclic invariant relaxed, $c_{*}=c_{1,n-1}\neq 0$. The style of argument is essentially identical to the derivation of BCFW recursion relations \cite{Britto:2005fq}. We define a deformed amplitude $\hat{A}_n(z)$ by applying an $(X_{1j},c_{1,n-1})$-shift. We can then reconstruct the unshifted amplitude by a contour integral 
\begin{equation}
    A_n\left[1,...,n\right] = \oint_\mathcal{C} \frac{\text{d}z }{2\pi i} \frac{\hat{A}_n(z)}{z},
\end{equation}
where $\mathcal{C}$ is a small circular contour surrounding $z=0$ and no other poles. Deforming the contour and using the fact that, as discussed in Section \ref{sec:shifts}, for this deformation there is no residue at $z=\infty$, we can write the amplitude as
\begin{equation}
    A_n\left[1,...,n\right] = -\sum_{i} \text{Res}\left[\frac{\hat{A}_n(z)}{z};z=z_i\right],
\end{equation}
where $z_i$ are the non-zero poles of the deformed amplitude. For the assumed shift these poles are located at $z=X_{1,m}$ for $m=j,j+1,...,n-1$ and $z=-X_{m',n}$ for $m'=2,3,...,j-1$. Now is the key step of the argument. As discussed in Section \ref{sec:shifts}, under an $(X,c)$-shift, the non-cyclic $c$-variables inside the chosen maximal rectangle do not shift. Therefore for all values of the deformation parameter $z$ the amplitude is evaluated on near-zero kinematics. On all of the residues, \textit{except} $z=-X_T$ and $z=X_B$, one of the sub-amplitudes is evaluated on kinematics corresponding to a hidden zero, an illustrative example is given in Figure \ref{fig:meshproof}. If we assume that for amplitudes with fewer than $n$ external particles the hidden zeros have been proven, then the reconstructed amplitude is the sum of two terms 
\begin{equation}
    A_n\left[1,...,n\right] = -\frac{1}{X_B}\text{Res}\left[\hat{A}_n(z);z=X_B\right] +\frac{1}{X_T}\text{Res}\left[\hat{A}_n(z);z=-X_T\right].
\end{equation}
Next we use the fact, proven in Section \ref{sec:gvector}, that the deformed amplitude scales like $\sim z^{-2}$ as $z\rightarrow \infty$, and therefore we have a bonus relation\footnote{An alternative proof is not to make use of the bonus relation, but instead to assume only that the scaling of the deformed amplitude is \textit{at least} $\sim z^{-1}$ as $z\rightarrow \infty$, the behavior of each individual Feynman diagram in $\text{Tr}\phi^3$. The reconstructed amplitude is then given as the sum of two residues, proceeding as in the main text one rediscovers the equality (\ref{proofbonus}) after relabeling states on both the $z=-X_T$ and $z=X_B$ residues and hence the enhanced $z^{-2}$ scaling.  }
\begin{equation}
    \label{proofbonus}
    \text{Res}\left[\hat{A}_n(z);z=X_B\right] + \text{Res}\left[\hat{A}_n(z);z=-X_T\right] =0.
\end{equation}
Inserting this into the above gives 
\begin{equation}
    A_n\left[1,...,n\right] = -\left(\frac{1}{X_B}+\frac{1}{X_T}\right)\text{Res}\left[\hat{A}_n(z);z=X_B\right]. 
\end{equation}
The final step is to evaluate the residue at $z=X_B=X_{1,j}$ as a product of sub-amplitudes, this is given by
\begin{align}
    \label{proofres}
    \text{Res}\left[\hat{A}_n(z);z=X_B\right] &= -A\left[1,2,...j-1,j\right]\hat{A}\left[j,j+1,...,n,1\right] \nonumber\\
    &= -A\left[1,2,...j-1,j\right]\left(A\left[j-1,j,j+1,...,n\right]\biggr\vert_{X_{j-1,m}\rightarrow \hat{X}_{1,m}}\right),
\end{align}
for $m=j+1,...,n-1$. Note that the second line is a trivial relabeling of the first, together with cyclicity of the labels. By a simple application of the rectangle rule
\begin{equation}
    \hat{X}_{1,m} = X_{j-1,m},
\end{equation}
using the fact that the rectangle bordered by $(X_{L'},X_{R'},X_{B'},X_{T'})=(X_{j-1,j},\hat{X}_{1,k},X_{1,j},X_{j-1,k})$, encloses only vanishing $c$-variables. Therefore we find
\begin{equation}
    A_n\left[1,...,n\right] \xrightarrow[\{c_{1,n-1}\}\neq 0]{\mathcal{Z}\left(X_{1,j}\right)} \left(\frac{1}{X_B}+\frac{1}{X_T}\right)A\left[1,2,...j-1,j\right]A\left[j-1,j,...,n\right],
\end{equation}
which is exactly the corner splitting formula (\ref{cornersplit}).

We can now make an inductive proof of the corner splitting: if we assume that the $n=4$ case of the zero has been verified explicitly, then this argument applied at $n=5$ establishes the 5-point splitting formula. Since the hidden zero is a trivial corollary of splitting from setting the remaining corner $c$-variable to zero this also establishes the 5-point zero. Applying the same argument again at 6-point then gives the 6-point splitting formula and so on. \\
\\
\textbf{Generic case:} The proof in the generic case is almost identical, in this case we choose $c_* = c_{k,l}\neq c_{1,n-1}$. The argument is exactly the same up to (\ref{proofres}), except in this case both sub-amplitudes are evaluated on shifted kinematics 
\begin{align}
    &\text{Res}\left[\hat{A}_n(z);z=X_B\right]\nonumber\\
    &= -\hat{A}\left[1,2,...j-1,j\right]\hat{A}\left[j,j+1,...,n,1\right] \nonumber\\
    &= -\left(A\left[1,2,...j-1,j\right]\biggr\vert_{X_{m,j}\rightarrow \hat{X}_{m,j}}\right)\left(A\left[j-1,j,j+1,...,n\right]\biggr\vert_{X_{j-1,m'}\rightarrow \hat{X}_{1,m'}}\right),
\end{align}
for $m=1,...,j-2$ and $m'=j,...,n-1$. For $A[j-1,...,n]$ the only difference is now that only for $m'=l+1,...,n-1$ are the invariants shifted. On the residue $z=X_{1,j}$ the shifted invariants that appear on the second line of (\ref{proofres}) are given explicitly by 
\begin{equation}
    \hat{X}_{1,m'} = 
    \begin{cases}
     X_{j-1,m'}, & m'=j+1,..,l\\
     X_{1,m'}, & m'=l+1,...,n-1.
    \end{cases}
\end{equation}
Similarly for $A[1,...,j]$, the shifted invariants are given by
\begin{equation}
    \hat{X}_{m,j} = 
    \begin{cases}
     X_{m,n}, & m=2,...,k\\
     X_{m,j}, & m=k+1,...,j-2.
    \end{cases}
\end{equation}
Therefore the residue is given by 
\begin{align}
    &\text{Res}\left[\hat{A}_n(z);z=X_B\right]\nonumber\\
    &= -\left(A\left[1,2,...j-1,j\right]\biggr\vert_{X_{m,j}\rightarrow X_{m,n}}\right)\left(A\left[j-1,j,j+1,...,n\right]\biggr\vert_{X_{j-1,m'}\rightarrow X_{j-1,m'}}\right),
\end{align}
for $m=2,...,k$ and $m'=j+1,...,n-1$. This gives exactly the generic case of the splitting formula (\ref{genericsplit}).

Given the inductive logic of the proof we can now better understand the origin of the shifts introduced in Section \ref{sec:shifts}. One can discover these shifts by the following argument. Choose near-zero kinematics $\mathcal{Z}\left(X_B\right)$ and $c_{*}\neq 0$, and consider a generic linear shift of the $X$-variables 
\begin{equation}
    X_{ij}\rightarrow X_{ij}+a_{ij}z,
\end{equation}
for some constants $a_{ij}$ to be determined. We impose that this shift accomplishes two things: (\textit{i}) the $c$-variables that have been set to zero should not shift, and (\textit{ii}) in addition to $X_T$ and $X_B$, only those $X$-variables with residues given as products of sub-amplitudes, at least one of which is evaluated on lower-point zero kinematics, are allowed to shift. These two conditions have a unique solution given by (\ref{Xcshift}). 

A final comment about the application of the proof to YMS. In this case the proof is not completely inductive since it only applies to the case of splitting in scalar channels. To prove the $n$-point splitting formula we need to assume that the $m$-point hidden zeros for $m<n$ have been established. For YMS, some of these $m$-point zeros are related to \textit{gluon splitting}, meaning one of $X_T$ or $X_B$ corresponds to a factorization channel with gluon exchange. We have not discussed this case, and empirically its splitting behavior has a significantly different structure than scalar splits. Nonetheless, the gluon channel zeros are still present in the usual form, a fact that can be established by assuming the amplitudes satisfy the fundamental BCJ relations \cite{Bartsch:2024amu}. With this additional input the proof presented above goes through.

\subsection{Generalized splitting}
\label{sec:highersplits}

The smooth splitting formula, for which we relax one of the hidden zero kinematic conditions, corresponds to a special limit in which the structure of the amplitude simplifies dramatically. Intuitively, if we relax further kinematic conditions the expression will become progressively more complicated until we relax all of the conditions, recovering the original amplitude in generic kinematics. The intermediate cases, where more than one but fewer than all of the kinematic conditions are relaxed may be of some interest in exposing hidden structures of the amplitudes. In this section, we use the recursive approach introduced in this paper to provide a clear and systematic way to write down these \textit{higher-order splitting formulae}.

\subsubsection{Tr\texorpdfstring{$\phi^3$}{phi3}}

The general procedure to derive a higher-order splitting formula is straightforward: we choose a maximal rectangle defined by $X_B=X_{ij}$ and relax some set of interior non-cyclic invariants $\{c_*^{(a)}\neq 0,a=1,..,k\}$; for any valid $(X_{ij},c_*^{(b)})$-shift, the associated contour integral can be used to reconstruct the amplitude giving a $k$-th order splitting formula. 

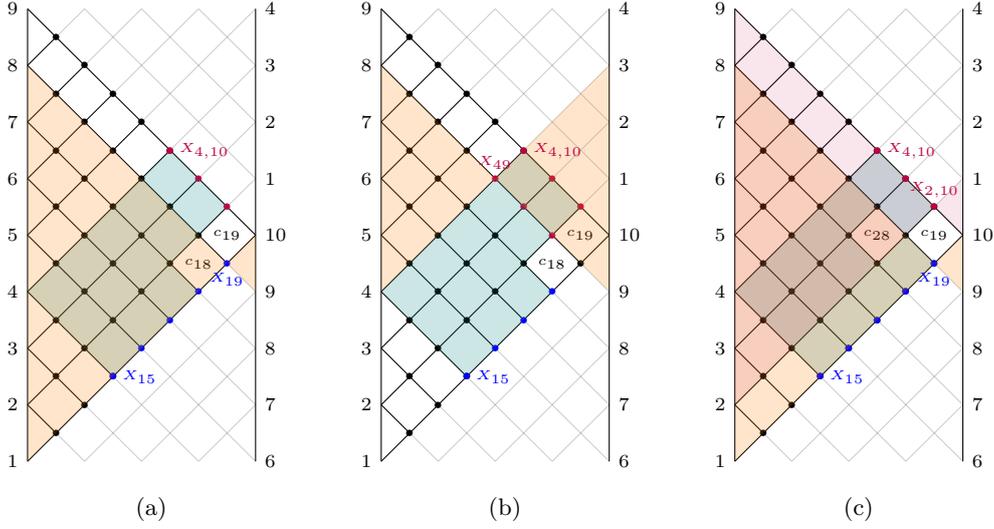
\begin{figure}
    \centering
            \begin{subfigure}[t]{0.3\textwidth}
                \centering
                		\begin{tikzpicture}[scale=0.75]
                \draw (0,0)--(4,4);
			\draw (0,1)--(3.5,4.5);
			\draw(0,2)--(3,5);
			\draw(0,3)--(2.5,5.5);
			\draw(0,4)--(2,6);
			\draw (0,5)--(1.5,6.5);
			\draw(0,6)--(1,7);
			\draw(0,7)--(0.5,7.5);
			\draw(0,0)--(0,8);
			\draw (4,0)--(4,8);
			\draw(0,8)--(4,4);
			\draw(0,7)--(3.5,3.5);
			\draw(0,6)--(3,3);
			\draw(0,5)--(2.5,2.5);
			\draw(0,4)--(2,2);
			\draw(0,3)--(1.5,1.5);
			\draw(0,2)--(1,1);
			\draw(0,1)--(0.5,0.5); 
                \draw[opacity=0.2](0.5,0.5)--(1,0); 
                \draw[opacity=0.2](1,1)--(2,0); 
                \draw[opacity=0.2](1.5,1.5)--(3,0); 
                \draw[opacity=0.2](2,2)--(4,0);
                \draw[opacity=0.2](2.5,2.5)--(4,1);
                \draw[opacity=0.2](3,3)--(4,2);
                \draw[opacity=0.2](3.5,3.5)--(4,3);
                \draw[opacity=0.2](3.5,4.5)--(4,5); 
                \draw[opacity=0.2](3,5)--(4,6);
                \draw[opacity=0.2](2.5,5.5)--(4,7);
                \draw[opacity=0.2](2,6)--(4,8);
                \draw[opacity=0.2](1.5,6.5)--(3,8);
                \draw[opacity=0.2](1,7)--(2,8);
                \draw[opacity=0.2](0.5,7.5)--(1,8);
                \draw[opacity=0.2](1,0)--(4,3);
                \draw[opacity=0.2](2,0)--(4,2);
                \draw[opacity=0.2](3,0)--(4,1);
                \draw[opacity=0.2](4,5)--(1,8);
                \draw[opacity=0.2](4,6)--(2,8);
                \draw[opacity=0.2](4,7)--(3,8);
			\node at (0.5,0.5) {\tiny$\bullet$};
                \node at (1,1) {\tiny$\bullet$};
			\node at (1.5,1.5) {\tiny$\bullet$};
                \node at (2,2) {\color{blue}\tiny$\bullet$};
			\node at (2.5,2.5) {\color{blue}\tiny$\bullet$};
                \node at (3,3) {\color{blue}\tiny$\bullet$};
			\node at (3.5,3.5) {\color{blue}\tiny$\bullet$};
                \node at (0.5,1.5) {\tiny$\bullet$};
                \node at (1,2) {\tiny$\bullet$};
			\node at (1.5,2.5) {\tiny$\bullet$};
                \node at (2,3) {\tiny$\bullet$};
			\node at (2.5,3.5) {\tiny$\bullet$};
                \node at (3,4) {\tiny$\bullet$};
			\node at (3.5,4.5) {\color{purple}\tiny$\bullet$};
                \node at (0.5,2.5) {\tiny$\bullet$};
                \node at (1,3) {\tiny$\bullet$};
			\node at (1.5,3.5) {\tiny$\bullet$};
                \node at (2,4) {\tiny$\bullet$};
			\node at (2.5,4.5) {\tiny$\bullet$};
                \node at (3,5) {\color{purple}\tiny$\bullet$};
                \node at (0.5,3.5) {\tiny$\bullet$};
                \node at (1,4) {\tiny$\bullet$};
			\node at (1.5,4.5) {\tiny$\bullet$};
                \node at (2,5) {\tiny$\bullet$};
			\node at (2.5,5.5) {\tiny$\bullet$};
                \node at (0.5,4.5) {\tiny$\bullet$};
                \node at (1,5) {\tiny$\bullet$};
			\node at (1.5,5.5) {\tiny$\bullet$};
                \node at (2,6) {\tiny$\bullet$};
                \node at (0.5,5.5) {\tiny$\bullet$};
                \node at (1,6) {\tiny$\bullet$};
			\node at (1.5,6.5) {\tiny$\bullet$};
                \node at (0.5,6.5) {\tiny$\bullet$};
                \node at (1,7) {\tiny$\bullet$};
                \node at (0.5,7.5) {\tiny$\bullet$};
			\node at (3,3.5) {\color{black}\tiny $c_{18}$};
			\node at (3.5,4) {\color{black}\tiny $c_{19}$};
                \node[left] at (0,0) {\scriptsize $1$};
                \node[left] at (0,1) {\scriptsize $2$};
                \node[left] at (0,2) {\scriptsize $3$};
                \node[left] at (0,3) {\scriptsize $4$};
                \node[left] at (0,4) {\scriptsize $5$};
                \node[left] at (0,5) {\scriptsize $6$};
                \node[left] at (0,6) {\scriptsize $7$};
                \node[left] at (0,7) {\scriptsize $8$};
                \node[left] at (0,8) {\scriptsize $9$};
                \node[right] at (4,0) {\scriptsize $6$};
                \node[right] at (4,1) {\scriptsize $7$};
                \node[right] at (4,2) {\scriptsize $8$};
                \node[right] at (4,3) {\scriptsize $9$};
                \node[right] at (4,4) {\scriptsize $10$};
                \node[right] at (4,5) {\scriptsize $1$};
                \node[right] at (4,6) {\scriptsize $2$};
                \node[right] at (4,7) {\scriptsize $3$};
                \node[right] at (4,8) {\scriptsize $4$};
                \fill[color=teal,opacity=0.2] (1,2)--(1.5,1.5)--(3,3)--(2.5,3.5);
                \fill[color=teal,opacity=0.2] (0,3)--(1,2)--(3.5,4.5)--(2.5,5.5);
                \node[right] at (2.5,5.5) {\color{purple}\tiny $X_{4,10}$};
                \node at (2.5,5.5) {\color{purple}\tiny$\bullet$};
                \node[right] at (1.5,1.5) {\color{blue}\tiny $X_{15}$};
                \node at (1.5,1.5) {\color{blue}\tiny$\bullet$};
                \node[below] at (3.5,3.5) {\color{blue}\tiny $X_{19}$};
                \fill[color=orange,opacity=0.2] (3.5,3.5)--(4,3)--(4,4);
                \fill[color=orange,opacity=0.2] (3.5,3.5)--(0,0)--(0,7);
		\end{tikzpicture}
                \caption{}
                \label{fig:HigherOrderTrphi3Mesh1}
            \end{subfigure}
            \begin{subfigure}[t]{0.3\textwidth}
                \centering
                		\begin{tikzpicture}[scale=0.75]
                \draw (0,0)--(4,4);
			\draw (0,1)--(3.5,4.5);
			\draw(0,2)--(3,5);
			\draw(0,3)--(2.5,5.5);
			\draw(0,4)--(2,6);
			\draw (0,5)--(1.5,6.5);
			\draw(0,6)--(1,7);
			\draw(0,7)--(0.5,7.5);
			\draw(0,0)--(0,8);
			\draw (4,0)--(4,8);
			\draw(0,8)--(4,4);
			\draw(0,7)--(3.5,3.5);
			\draw(0,6)--(3,3);
			\draw(0,5)--(2.5,2.5);
			\draw(0,4)--(2,2);
			\draw(0,3)--(1.5,1.5);
			\draw(0,2)--(1,1);
			\draw(0,1)--(0.5,0.5); 
                \draw[opacity=0.2](0.5,0.5)--(1,0); 
                \draw[opacity=0.2](1,1)--(2,0); 
                \draw[opacity=0.2](1.5,1.5)--(3,0); 
                \draw[opacity=0.2](2,2)--(4,0);
                \draw[opacity=0.2](2.5,2.5)--(4,1);
                \draw[opacity=0.2](3,3)--(4,2);
                \draw[opacity=0.2](3.5,3.5)--(4,3);
                \draw[opacity=0.2](3.5,4.5)--(4,5); 
                \draw[opacity=0.2](3,5)--(4,6);
                \draw[opacity=0.2](2.5,5.5)--(4,7);
                \draw[opacity=0.2](2,6)--(4,8);
                \draw[opacity=0.2](1.5,6.5)--(3,8);
                \draw[opacity=0.2](1,7)--(2,8);
                \draw[opacity=0.2](0.5,7.5)--(1,8);
                \draw[opacity=0.2](1,0)--(4,3);
                \draw[opacity=0.2](2,0)--(4,2);
                \draw[opacity=0.2](3,0)--(4,1);
                \draw[opacity=0.2](4,5)--(1,8);
                \draw[opacity=0.2](4,6)--(2,8);
                \draw[opacity=0.2](4,7)--(3,8);
			\node at (0.5,0.5) {\tiny$\bullet$};
                \node at (1,1) {\tiny$\bullet$};
			\node at (1.5,1.5) {\tiny$\bullet$};
                \node at (2,2) {\color{blue}\tiny$\bullet$};
			\node at (2.5,2.5) {\color{blue}\tiny$\bullet$};
                \node at (3,3) {\color{blue}\tiny$\bullet$};
			\node at (3.5,3.5) {\tiny$\bullet$};
                \node at (0.5,1.5) {\tiny$\bullet$};
                \node at (1,2) {\tiny$\bullet$};
			\node at (1.5,2.5) {\tiny$\bullet$};
                \node at (2,3) {\tiny$\bullet$};
			\node at (2.5,3.5) {\tiny$\bullet$};
                \node at (3,4) {\color{purple}\tiny$\bullet$};
			\node at (3.5,4.5) {\color{purple}\tiny$\bullet$};
                \node at (0.5,2.5) {\tiny$\bullet$};
                \node at (1,3) {\tiny$\bullet$};
			\node at (1.5,3.5) {\tiny$\bullet$};
                \node at (2,4) {\tiny$\bullet$};
			\node at (2.5,4.5) {\color{purple}\tiny$\bullet$};
                \node at (3,5) {\color{purple}\tiny$\bullet$};
                \node at (0.5,3.5) {\tiny$\bullet$};
                \node at (1,4) {\tiny$\bullet$};
			\node at (1.5,4.5) {\tiny$\bullet$};
                \node at (2,5) {\color{purple}\tiny$\bullet$};
			\node at (2.5,5.5) {\tiny$\bullet$};
                \node at (0.5,4.5) {\tiny$\bullet$};
                \node at (1,5) {\tiny$\bullet$};
			\node at (1.5,5.5) {\tiny$\bullet$};
                \node at (2,6) {\tiny$\bullet$};
                \node at (0.5,5.5) {\tiny$\bullet$};
                \node at (1,6) {\tiny$\bullet$};
			\node at (1.5,6.5) {\tiny$\bullet$};
                \node at (0.5,6.5) {\tiny$\bullet$};
                \node at (1,7) {\tiny$\bullet$};
                \node at (0.5,7.5) {\tiny$\bullet$};
			\node at (3,3.5) {\color{black}\tiny $c_{18}$};
			\node at (3.5,4) {\color{black}\tiny $c_{19}$};
                \node[left] at (0,0) {\scriptsize $1$};
                \node[left] at (0,1) {\scriptsize $2$};
                \node[left] at (0,2) {\scriptsize $3$};
                \node[left] at (0,3) {\scriptsize $4$};
                \node[left] at (0,4) {\scriptsize $5$};
                \node[left] at (0,5) {\scriptsize $6$};
                \node[left] at (0,6) {\scriptsize $7$};
                \node[left] at (0,7) {\scriptsize $8$};
                \node[left] at (0,8) {\scriptsize $9$};
                \node[right] at (4,0) {\scriptsize $6$};
                \node[right] at (4,1) {\scriptsize $7$};
                \node[right] at (4,2) {\scriptsize $8$};
                \node[right] at (4,3) {\scriptsize $9$};
                \node[right] at (4,4) {\scriptsize $10$};
                \node[right] at (4,5) {\scriptsize $1$};
                \node[right] at (4,6) {\scriptsize $2$};
                \node[right] at (4,7) {\scriptsize $3$};
                \node[right] at (4,8) {\scriptsize $4$};
                \fill[color=teal,opacity=0.2] (1,2)--(1.5,1.5)--(3,3)--(2.5,3.5);
                \fill[color=teal,opacity=0.2] (0,3)--(1,2)--(3.5,4.5)--(2.5,5.5);
                \fill[color=orange,opacity=0.2] (2,5)--(0,3)--(0,7);
                \fill[color=orange,opacity=0.2] (2,5)--(4,3)--(4,7);
                \node[right] at (2.5,5.5) {\color{purple}\tiny $X_{4,10}$};
                \node[above] at (2,5) {\color{purple}\tiny $X_{49}$};
                \node at (2.5,5.5) {\color{purple}\tiny$\bullet$};
                \node[right] at (1.5,1.5) {\color{blue}\tiny $X_{15}$};
                \node at (1.5,1.5) {\color{blue}\tiny$\bullet$};
		\end{tikzpicture}
                \caption{}
                \label{fig:HigherOrderTrphi3Mesh2}
            \end{subfigure}
            \begin{subfigure}[t]{0.3\textwidth}
                \centering
                		\begin{tikzpicture}[scale=0.75]
                \draw (0,0)--(4,4);
			\draw (0,1)--(3.5,4.5);
			\draw(0,2)--(3,5);
			\draw(0,3)--(2.5,5.5);
			\draw(0,4)--(2,6);
			\draw (0,5)--(1.5,6.5);
			\draw(0,6)--(1,7);
			\draw(0,7)--(0.5,7.5);
			\draw(0,0)--(0,8);
			\draw (4,0)--(4,8);
			\draw(0,8)--(4,4);
			\draw(0,7)--(3.5,3.5);
			\draw(0,6)--(3,3);
			\draw(0,5)--(2.5,2.5);
			\draw(0,4)--(2,2);
			\draw(0,3)--(1.5,1.5);
			\draw(0,2)--(1,1);
			\draw(0,1)--(0.5,0.5); 
                \draw[opacity=0.2](0.5,0.5)--(1,0); 
                \draw[opacity=0.2](1,1)--(2,0); 
                \draw[opacity=0.2](1.5,1.5)--(3,0); 
                \draw[opacity=0.2](2,2)--(4,0);
                \draw[opacity=0.2](2.5,2.5)--(4,1);
                \draw[opacity=0.2](3,3)--(4,2);
                \draw[opacity=0.2](3.5,3.5)--(4,3);
                \draw[opacity=0.2](3.5,4.5)--(4,5); 
                \draw[opacity=0.2](3,5)--(4,6);
                \draw[opacity=0.2](2.5,5.5)--(4,7);
                \draw[opacity=0.2](2,6)--(4,8);
                \draw[opacity=0.2](1.5,6.5)--(3,8);
                \draw[opacity=0.2](1,7)--(2,8);
                \draw[opacity=0.2](0.5,7.5)--(1,8);
                \draw[opacity=0.2](1,0)--(4,3);
                \draw[opacity=0.2](2,0)--(4,2);
                \draw[opacity=0.2](3,0)--(4,1);
                \draw[opacity=0.2](4,5)--(1,8);
                \draw[opacity=0.2](4,6)--(2,8);
                \draw[opacity=0.2](4,7)--(3,8);
			\node at (0.5,0.5) {\tiny$\bullet$};
                \node at (1,1) {\tiny$\bullet$};
			\node at (1.5,1.5) {\tiny$\bullet$};
                \node at (2,2) {\tiny$\bullet$};
			\node at (2.5,2.5) {\tiny$\bullet$};
                \node at (3,3) {\tiny$\bullet$};
			\node at (3.5,3.5) {\tiny$\bullet$};
                \node at (0.5,1.5) {\tiny$\bullet$};
                \node at (1,2) {\tiny$\bullet$};
			\node at (1.5,2.5) {\tiny$\bullet$};
                \node at (2,3) {\tiny$\bullet$};
			\node at (2.5,3.5) {\tiny$\bullet$};
                \node at (3,4) {\tiny$\bullet$};
			\node at (3.5,4.5) {\tiny$\bullet$};
                \node at (0.5,2.5) {\tiny$\bullet$};
                \node at (1,3) {\tiny$\bullet$};
			\node at (1.5,3.5) {\tiny$\bullet$};
                \node at (2,4) {\tiny$\bullet$};
			\node at (2.5,4.5) {\tiny$\bullet$};
                \node at (3,5) {\tiny$\bullet$};
                \node at (0.5,3.5) {\tiny$\bullet$};
                \node at (1,4) {\tiny$\bullet$};
			\node at (1.5,4.5) {\tiny$\bullet$};
                \node at (2,5) {\tiny$\bullet$};
			\node at (2.5,5.5) {\tiny$\bullet$};
                \node at (0.5,4.5) {\tiny$\bullet$};
                \node at (1,5) {\tiny$\bullet$};
			\node at (1.5,5.5) {\tiny$\bullet$};
                \node at (2,6) {\tiny$\bullet$};
                \node at (0.5,5.5) {\tiny$\bullet$};
                \node at (1,6) {\tiny$\bullet$};
			\node at (1.5,6.5) {\tiny$\bullet$};
                \node at (0.5,6.5) {\tiny$\bullet$};
                \node at (1,7) {\tiny$\bullet$};
                \node at (0.5,7.5) {\tiny$\bullet$};
			\node at (3.5,4) {\color{black}\tiny $c_{19}$};
			\node at (2.5,4) {\color{black}\tiny $c_{28}$};
                \node[left] at (0,0) {\scriptsize $1$};
                \node[left] at (0,1) {\scriptsize $2$};
                \node[left] at (0,2) {\scriptsize $3$};
                \node[left] at (0,3) {\scriptsize $4$};
                \node[left] at (0,4) {\scriptsize $5$};
                \node[left] at (0,5) {\scriptsize $6$};
                \node[left] at (0,6) {\scriptsize $7$};
                \node[left] at (0,7) {\scriptsize $8$};
                \node[left] at (0,8) {\scriptsize $9$};
                \node[right] at (4,0) {\scriptsize $6$};
                \node[right] at (4,1) {\scriptsize $7$};
                \node[right] at (4,2) {\scriptsize $8$};
                \node[right] at (4,3) {\scriptsize $9$};
                \node[right] at (4,4) {\scriptsize $10$};
                \node[right] at (4,5) {\scriptsize $1$};
                \node[right] at (4,6) {\scriptsize $2$};
                \node[right] at (4,7) {\scriptsize $3$};
                \node[right] at (4,8) {\scriptsize $4$};
                \fill[color=teal,opacity=0.2] (0,3)--(1.5,1.5)--(3,3)--(1.5,4.5);
                \fill[color=teal,opacity=0.2] (3,3)--(3.5,3.5)--(3,4)--(2.5,3.5);
                \fill[color=teal,opacity=0.2] (1.5,4.5)--(2.5,5.5)--(3,5)--(2,4);
                \fill[color=teal,opacity=0.2] (2.5,4.5)--(3,5)--(3.5,4.5)--(3,4);
                \node[right] at (2.5,5.5) {\color{purple}\tiny $X_{4,10}$};
                \node at (2.5,5.5) {\color{purple}\tiny$\bullet$};
                \node[right] at (1.5,1.5) {\color{blue}\tiny $X_{15}$};
                \node at (1.5,1.5) {\color{blue}\tiny$\bullet$};
                \node at (2,2) {\color{blue}\tiny$\bullet$};
                \node at (2.5,2.5) {\color{blue}\tiny$\bullet$};
                \node at (3,3) {\color{blue}\tiny$\bullet$};
                \node at (3.5,3.5) {\color{blue}\tiny$\bullet$};
                \node at (3.5,4.5) {\color{purple}\tiny$\bullet$};
                \node at (3,5) {\color{purple}\tiny$\bullet$};
                \fill[color=orange,opacity=0.2] (3.5,3.5)--(4,3)--(4,4);
                \fill[color=orange,opacity=0.2] (3.5,3.5)--(0,0)--(0,7);
                \fill[color=purple,opacity=0.1] (3.5,4.5)--(4,4)--(4,5);
                \fill[color=purple,opacity=0.1] (3.5,4.5)--(0,1)--(0,8);
                \node[below] at (3.5,3.5) {\color{blue}\tiny $X_{19}$};
                \node[above] at (3.5,4.5) {\color{purple}\tiny $X_{2,10}$};
		\end{tikzpicture}
                \caption{}
                \label{fig:HigherOrderTrphi3Mesh3}
            \end{subfigure}
    \caption{Configurations of second-order splitting. Diagrams (a) and (b) are evaluated on the same kinematics but are calculated with different shifts; in each case a third non-zero residue is present which contains a sub-amplitude on first-order splitting kinematics. In (c) a non-adjacent second-order split is considered, in this case there are two additional non-zero residues. }
    \label{fig:HOsplitTrphi3}
\end{figure}

A simple example to illustrate this in detail is 10-point scattering amplitude in $\text{Tr}\phi^3$ with $X_B = X_{15}$ on the second-order splitting kinematics $\{c_*^{(1)},c_*^{(2)}\}= \{c_{19},c_{18}\}$ as shown in Figure \ref{fig:HigherOrderTrphi3Mesh1}. If we make an $(X_{15},c_{19})$-shift, similar to the derivation of the usual (first-order) splitting formula, the shifted amplitude has poles as $z=X_{15}$ and $z=-X_{4,10}$. However since $c_{18}\neq 0$, there is an additional pole at $z=X_{19}$ with non-zero residue. As described in Section \ref{sec:shifts}, for $\text{Tr}\phi^3$ this shift will fall off like $\sim z^{-2}$ as $z\rightarrow \infty$ for any kinematics, this means the contour integral can be used to derive a bonus relation of the form 
\begin{equation}
    \text{Res}\left[\hat{A}^{\phi^3}_{10}(z);z=-X_{4,10}\right] + \text{Res}\left[\hat{A}^{\phi^3}_{10}(z);z=X_{15}\right] + \text{Res}\left[\hat{A}^{\phi^3}_{10}(z);z=-X_{49}\right] = 0.
\end{equation}
Using this we can remove one of the residues in the recursive formula. A choice that generalizes naturally to splitting at higher-orders is to choose this to be $z=X_{B}=X_{15}$, giving 
\begin{align}
    A^{\phi^3}_{10}\left[1,2,3,4,5,6,7,8,9,10\right] \xrightarrow[\{c_{18},c_{19}\}\neq 0]{\mathcal{Z}\left(X_{15}\right)} &\left(\frac{1}{X_{4,10}}+\frac{1}{X_{15}}\right)\text{Res}\left[\hat{A}^{\phi^3}_{10}(z);z=-X_{4,10}\right] \nonumber\\
    &-\left(\frac{1}{X_{19}}-\frac{1}{X_{15}}\right)\text{Res}\left[\hat{A}^{\phi^3}_{10}(z);z=X_{19}\right].
\end{align}
The residue at $z=-X_{4,10}$ takes the same form as the first-order splitting formula
\begin{align}
    \text{Res}\left[\hat{A}^{\phi^3}_{10}(z);z=-X_{4,10}\right] =  A^{\phi^3}_5\left[1,2,3,4,5\right]A^{\phi^3}_7\left[4,5,6,7,8,9,10\right].
\end{align}
The residue at $z=X_{19}$ is more interesting, as shown in Figure \ref{fig:HigherOrderTrphi3Mesh1} this factors into a product $\hat{A}^{\phi^3}_9[1,2,3,4,5,6,7,8,9] \times A^{\phi^3}_3[9,10,1]$, where $\hat{A}^{\phi^3}_9$ is evaluated on first-order splitting kinematics. We therefore proceed iteratively and use the known first-order splitting formula to simplify this reside
\begin{align}
    \text{Res}\left[\hat{A}^{\phi^3}_{10}(z);z=X_{19}\right] = -\left(\frac{1}{X_{49}}+\frac{1}{X_{15}-X_{19}}\right)A^{\phi^3}_5\left[1,2,3,4,5\right]A^{\phi^3}_6\left[4,5,6,7,8,9\right],
\end{align}
where we have also used $A^{\phi^3}_3=1$. Putting this together the result simplifies to the form 
\begin{align}
    \label{secondorderphi310ptv1}
    &A^{\phi^3}_{10}\left[1,2,3,4,5,6,7,8,9,10\right] \xrightarrow[\{c_{18},c_{19}\}\neq 0]{\mathcal{Z}\left(X_{15}\right)} \nonumber\\
    &A^{\phi^3}_5\left[1,2,3,4,5\right]\left\{\frac{c_{19}+c_{18}}{X_{4,10} X_{15}}A^{\phi^3}_7\left[4,5,6,7,8,9,10\right]+\frac{c_{18}}{X_{15}X_{19}X_{49}}A^{\phi^3}_6\left[4,5,6,7,8,9\right]\right\}.
\end{align}
Despite being more complicated than the first-order splitting formula, we see that this second-order example retains the feature that the amplitude is a product, in this case the sub-amplitude $A^{\phi^3}_5[1,2,3,4,5]$ is a common factor. Additionally, we see that this expression manifestly reduces to the expected first-order expression in the limit $c_{18}\rightarrow 0$. 

In the derivation of the 10-point second order split, we made a choice to use an $(X_{15},c_{19})$-shift since $c_{19}\neq 0$. But since we have relaxed two conditions, it seems equally reasonable to derive a second-order splitting formula using a $(X_{15},c_{18})$-shift as shown in Figure \ref{fig:HigherOrderTrphi3Mesh2}. The analysis in this case is similar, the shifted amplitude has 3 non-zero residues at $z=-X_{4,10},X_{15}$ and $-X_{49}$, and we use the associated bonus relation to remove the residue at $z=X_{15}$. The final result has quite a different form 
\begin{align}
    \label{secondorderphi310ptv2}
    &A^{\phi^3}_{10}\left[1,2,3,4,5,6,7,8,9,10\right] \xrightarrow[\{c_{18},c_{19}\}\neq 0]{\mathcal{Z}\left(X_{15}\right)} \nonumber\\
    &A^{\phi^3}_5\left[1,2,3,4,5\right]\left\{\left(\frac{1}{X_{4,10}}+\frac{1}{X_{15}}\right)\left(A^{\phi^3}_7\left[4,5,6,7,8,9,10\right]\biggr\vert_{X_{49}\rightarrow X_{49}-X_{4,10}}\right)\right.\nonumber\\
    &\hspace{2.5cm} \left.+\left(\frac{1}{X_{49}}+\frac{1}{X_{15}}\right)\left(\frac{1}{X_{19}}+\frac{1}{X_{4,10}-X_{49}}\right)A^{\phi^3}_6\left[4,5,6,7,8,9\right]\right\}.
\end{align}
Despite appearances (\ref{secondorderphi310ptv1}) and (\ref{secondorderphi310ptv2}) are the same amplitude in the same kinematics, a fact that can be verified explicitly. This second version is still a product, but the fact that it reduces to the first-order splitting formula is now non-manifest. Moreover we have the appearance of a \textit{spurious pole} at $X_{49}=X_{4,10}$. This is non-physical and must cancel in the sum of the two terms in the second factor. 

In retrospect, it is somewhat remarkable that (\ref{secondorderphi310ptv1}) has no spurious poles; in most examples we have studied they are present for any choice of shift. If we consider the same 10-point amplitude with the same maximal rectangle ($X_B=X_{15}$), but we relax $\{c_*^{(1)},c_*^{(2)}\} = \{c_{19},c_{28}\}$, using the $(X_{15},c_{19})$-shift we find 
\begin{align}
    &A^{\phi^3}_{10}\left[1,2,3,4,5,6,7,8,9,10\right]\xrightarrow[\{c_{19},c_{28}\}\neq 0]{\mathcal{Z}\left(X_{15}\right)}\nonumber\\
    &\left(\frac{1}{X_{4,10}}+\frac{1}{X_{15}}\right)\left(A^{\phi^3}_5\left[1,2,3,4,5\right]\biggr\vert_{\substack{X_{25}\rightarrow X_{2,10}-X_{4,10}}}\right)A^{\phi^3}_7\left[4,5,6,7,8,9,10\right]\nonumber\\
    &+\left(\frac{1}{X_{2,10}}+\frac{1}{X_{15}}\right)\left(\frac{1}{X_{25}}-\frac{1}{X_{2,10}-X_{4,10}}\right)A^{\phi^3}_4\left[2,3,4,5\right]\left(A^{\phi^3}_7\left[4,5,6,7,8,9,10\right]\biggr\vert_{X_{49}\rightarrow X_{29}}\right) \nonumber\\
    &+ \frac{c_{28}}{X_{15}X_{19}X_{49}}\left(A^{\phi^3}_5\left[1,2,3,4,5\right]\biggr\vert_{X_{25}\rightarrow X_{29}}\right)A^{\phi^3}_6\left[4,5,6,7,8,9\right],
\end{align}
this configuration is shown in Figure \ref{fig:HigherOrderTrphi3Mesh3}. Unlike the previous examples, this expression does not factor into a product; this appears to be a generic feature of relaxing invariants in different ``rows" of the maximal rectangle.

Proceeding to higher orders along the bottom row of the maximal rectangle; on the \textit{third-order} splitting kinematics $\{c_*^{(1)},c_*^{(2)},c_*^{(3)}\}=\{c_{19},c_{18},c_{17}\}$, under the $(X_{15},c_{19})$-shift, we find 
\begin{align}
    \label{thirdorderphi310pt}
    &A^{\phi^3}_{10}\left[1,2,3,4,5,6,7,8,9,10\right] \xrightarrow[\{c_{17},c_{18},c_{19}\}\neq 0]{\mathcal{Z}\left(X_{15}\right)} \nonumber\\
    &A^{\phi^3}_5\left[1,2,3,4,5\right]\left\{\frac{c_{19}+c_{18}+c_{17}}{X_{4,10}X_{15}}A^{\phi^3}_7\left[4,5,6,7,8,9,10\right]\right.\nonumber\\
    &\hspace{2.5cm}+\frac{c_{18}+c_{17}}{X_{15}X_{19}X_{49}}A^{\phi^3}_6\left[4,5,6,7,8,9\right] \nonumber\\
    &\hspace{2.5cm}\left.+\frac{c_{17}}{X_{15}  X_{18} X_{48}}A^{\phi^3}_5\left[4,5,6,7,8\right]A^{\phi^3}_4\left[8,9,10,1\right]\right\}.
\end{align}
Comparing (\ref{secondorderphi310ptv1}) and (\ref{thirdorderphi310pt}) the pattern is clear, if we relax the entire row
\begin{align}
    \label{rowsplit10}
    &A^{\phi^3}_{10}\left[1,2,3,4,5,6,7,8,9,10\right] \xrightarrow[\{c_{15},c_{16},c_{17},c_{18},c_{19}\}\neq 0]{\mathcal{Z}\left(X_{15}\right)} \nonumber\\
    &A^{\phi^3}_5\left[1,2,3,4,5\right]\left\{\frac{c_{19}+c_{18}+c_{17}+c_{16}+c_{15}}{X_{15}X_{4,10}}A^{\phi^3}_7\left[4,5,6,7,8,9,10\right]\right.\nonumber\\
    &\hspace{2.5cm}+\frac{c_{18}+c_{17}+c_{16}+c_{15}}{X_{15}X_{19}X_{49}}A^{\phi^3}_6\left[4,5,6,7,8,9\right]A_3^{\phi^3}\left[9,10,1\right] \nonumber\\
    &\hspace{2.5cm}\left.+\frac{c_{17}+c_{16}+c_{15}}{X_{15}  X_{18} X_{48}}A^{\phi^3}_5\left[4,5,6,7,8\right]A^{\phi^3}_4\left[8,9,10,1\right]\right. \nonumber\\
    &\hspace{2.5cm}\left.+\frac{c_{16}+c_{15}}{X_{15}  X_{17} X_{47}}A^{\phi^3}_4\left[4,5,6,7\right]A^{\phi^3}_5\left[7,8,9,10,1\right]\right. \nonumber\\
    &\hspace{2.5cm}\left.+\frac{c_{15}}{X_{15}  X_{16} X_{46}}A_3^{\phi^3}\left[4,5,6\right] A^{\phi^3}_6\left[6,7,8,9,10,1\right]\right\}.
\end{align}
It is straightforward to verify from explicit expressions that this generalization is correct. This formula has a remarkably simple nested structure; progressively setting $c_{15}=0$ then $c_{16}=0$ and so on gives a sequence of similar formulae with fewer terms.  

The generalization of the above formula to any zero at any multiplicity can be immediately written down. Without loss of generality, for the maximal rectangle with $X_B=X_{1,j}$ when the entire bottom row is relaxed $\{c_*^{(a)}\} = \{c_{1,k};\;j\leq k \leq n-1\}$ the amplitude satisfies the following \textit{higher-order splitting formula}
\begin{empheq}[box=\fbox]{align}
    \label{HOsplittingphi3}
  &A^{\phi^3}_n\left[1,...,n\right] \xrightarrow[\{c_{1,j},c_{1,j+1},...,c_{1,n-1}\}\neq 0]{\mathcal{Z}\left(X_{1,j}\right)} \nonumber\\
    &A^{\phi^3}\left[1,...,j\right]\biggr\{\sum_{k=j}^{n-1}\frac{c_{1,k}}{X_{1,j}X_{j-1,n}}A^{\phi^3}\left[j-1,...,n\right] \nonumber\\
    &\hspace{2cm}+ \sum_{k=j}^{n-1}\sum_{l=j}^k\frac{ c_{1,l}}{X_{1,j} X_{1,k+1}X_{j-1,k+1}}A^{\phi^3}\left[j-1,...,k+1\right]A^{\phi^3}\left[k+1,...,n,1\right]\biggr\}.
\end{empheq}
In addition to recovering lower-order splitting formulae, from this general result we observe that there are other interesting kinematic limits. For instance if we take (\ref{rowsplit10}) in the limit
\begin{equation}
    c_{15}=c_{16}=c_{19}=0, \hspace{10mm} \text{and} \hspace{10mm} c_{17}+c_{18}=0,
\end{equation}
we find the amplitude splits into a product of \textit{three} sub-amplitudes
\begin{align}
    A^{\phi^3}_{10}\left[1,2,3,4,5,6,7,8,9,10\right] \rightarrow \frac{c_{17}}{X_{15}X_{18}X_{48}}A^{\phi^3}_{5}\left[1,2,3,4,5\right]A^{\phi^3}_{5}\left[4,5,6,7,8\right]A^{\phi^3}_{4}\left[8,9,10,1\right]. 
\end{align}
More generally, for any maximal rectangle with $X_B=X_{1,j}$, if we relax the pair of invariants $\{c_{1,k},c_{1,k+1}\}\neq 0$ with $j\leq k\leq n-2$, but still set the \textit{sum} to zero,  $c_{1,k}+c_{1,k+1}=0$, then we have a \textit{triple-splitting} formula 
\begin{empheq}[box=\fbox]{align}
    \label{triplesplitting}
  A^{\phi^3}_{n}\left[1,...,n\right] \rightarrow \frac{c_{1,k}}{X_{1,j}X_{1,k+1}X_{j-1,k}}A^{\phi^3}\left[1,..,j-1,j\right]A^{\phi^3}\left[j-1,...,k+1\right]A^{\phi^3}\left[k+1,...,n,1\right].
\end{empheq}
There may be further interesting kinematic limits and simple closed formulae for other patterns of relaxation, and we expect that these on-shell recursion relations provide a systematic approach to exploring these questions. For now, we will move on to higher-order splitting in other models.

\subsubsection{Non-Linear Sigma Model}

Like $\text{Tr}\phi^3$, the NLSM is a model of colored scalars, but has significantly different pole structure since it is a model with only even-point interactions. For the purposes of using on-shell recursion, this means that some of the shifted $X$-variables do not correspond to poles of the amplitude; in particular:
\begin{itemize}
    \item $X_{eo}$ is a factorization channel,
    \item $X_{ee}$ and $X_{oo}$ have zero residue,
\end{itemize}
where $e=\text{even}$ and $o=\text{odd}$. As discussed in Section \ref{sec:shifts}, on near-zero kinematics if we relax multiple $c_{ij}$ in the same row the $\sim z^{-2}$ scaling decreases to $\sim z^{-1}$. This means we no longer have a bonus relation and have to sum over all of the residues. However, due to the restriction on which poles can appear discussed above, there are still typically fewer residues in a given contour integral compared to the same kinematics and multiplicity for $\text{Tr}\phi^3$.

For example, consider the 10-point amplitude on the near-zero kinematics defined by the maximal rectangle $X_B=X_{16}$ and the third-order splitting conditions, $\{c_*^{(1)},c_*^{(2)},c_*^{(3)}\}=\{c_{19},c_{29},c_{39}\}$. Using a $(X_{16},c_{19})$-shift we find 
\begin{align}   
    \label{NLSM10ptExample}
    & A_{10}^{\text{NLSM}}\left[1,2,3,4,5,6,7,8,9,10\right] \xrightarrow[\{c_{19},c_{29},c_{39}\}\neq 0]{\mathcal{Z}\left(X_{16}\right)}  \nonumber\\
    &\biggr\{\frac{1}{X_{3,10}}\left(\frac{1}{X_{36}}-\frac{1}{X_{36}-c_{39}}\right)\left(A_4^{\text{NLSM}}\left[1,2,3,10\right]\biggr\vert_{\substack{X_{2,10}\rightarrow X_{2,10}-X_{3,10} }}\right)A_4^{\text{NLSM}}\left[3,4,5,6\right]\nonumber\\
    &\hspace{0.25cm} +\frac{1}{X_{5,10}}\left(A_6^{\text{NLSM}}\left[1,2,3,4,5,6\right]\biggr\vert_{\substack{X_{26}\rightarrow X_{26}-c_{29}-c_{39}\\X_{36}\rightarrow X_{36}-c_{39}}}\right) \nonumber\\
    &\hspace{0.25cm} + \frac{1}{X_{16}}A_6^{\text{NLSM}}\left[1,2,3,4,5,6\right]\biggr\}A_6^{\text{NLSM}}\left[5,6,7,8,9,10\right].
\end{align}
In this expression the first line corresponds to a new residue at $z=-X_{3,10}$ which is evaluated by an iterative application of the first-order splitting formula. Note that despite the various shifted invariants, only $X_{36}-c_{39}$ is an actual spurious singularity that cancels in the sum of two residues.

An alternative approach to obtain all-multiplicity higher-order splitting formulae for the NLSM is to begin with the formulae (\ref{HOsplittingphi3}) for $\text{Tr}\phi^3$ and apply the $\delta$-deformation \cite{Arkani-Hamed:2023swr} described in (\ref{deltatrphi3}). The general formula (\ref{HOsplittingphi3}), applicable when relaxing an entire row of conditions defining a maximal rectangle, contains sub-amplitudes with both even and odd numbers of external particles. The latter do not exist in NLSM (though they do exist in so-called \textit{extended} theories \cite{Cachazo:2016njl,Arkani-Hamed:2023swr}), and so we impose that certain linear combinations of relaxed $c_{ij}$-variables vanish. This removes all terms with odd-multiplicity amplitudes and leads to a version of the higher-order splitting formula for the NLSM. On a maximal rectangle with $X_B=X_{1,j}$ where $j\in 2\mathds{Z}_{\geq0}$, when \textit{almost} the entire bottom row is relaxed $\{c_*^{(a)}\} = \{c_{1,k};\;j< k \leq n-1\}$, and we additionally impose the linear constraints $c_{1,l}+c_{1,l+1}=0$ for $l=j+1,...,n-3$, the amplitude satisfies
\begin{empheq}[box=\fbox]{align}
    \label{HOsplittingNLSM}
  &A_{n}^{\text{NLSM}}\left[1,...,n\right] \rightarrow \nonumber\\
    &A^{\text{NLSM}}\left[1,...,j\right]\biggr\{\frac{c_{1,n-1}}{X_{1,j}X_{j-1,n}}A^{\text{NLSM}}\left[j-1,j,...,n\right] \nonumber \\
    &\hspace{0.25cm}+ \sum_{k=1}^{\frac{1}{2}(n-j-2)} \frac{c_{1,n-1-2k}}{X_{1,j}X_{j-1,n-2k} X_{1,n-2k}}A^{\text{NLSM}}\left[j-1,...,n-2k\right]A^{\text{NLSM}}\left[n-2k,...,n,1\right]\biggr\}. 
\end{empheq}
As a representative example, consider the 12-point amplitude near the maximal rectangle with $X_B=X_{16}$ with $\{c_{17},c_{18},c_{19},c_{1,10},c_{1,11}\}\neq 0$ together with the additional linear constraints $c_{17}+c_{18}=0$ and $c_{19}+c_{1,10}=0$, the above formula gives 
\begin{align}
    &A_{12}^{\text{NLSM}}\left[1,2,3,4,5,6,7,8,9,10,11,12\right] \rightarrow \nonumber\\
    &A_6^{\text{NLSM}}\left[1,2,3,4,5,6\right] \biggr\{\frac{c_{1,11}}{X_{16}X_{5,12}}A_8^{\text{NLSM}}\left[5,6,7,8,9,10,11,12\right] \nonumber\\
    &\hspace{3.25cm}+\frac{c_{19}}{X_{16}X_{5,10}X_{1,10}} A_6^{\text{NLSM}}\left[5,6,7,8,9,10\right]A_4^{\text{NLSM}}\left[10,11,12,1\right] \nonumber\\
    &\hspace{3.25cm}+\frac{c_{17}}{X_{16}X_{58}X_{18}} A_4^{\text{NLSM}}\left[5,6,7,8\right]A_6^{\text{NLSM}}\left[8,9,10,11,12,1\right]\biggr\}.
\end{align}
Likewise we can relax fewer $c_{ij}$ and isolate a single term in this sum, giving a triple splitting formula for NLSM. This has the same form as (\ref{triplesplitting}) except that we must choose $j,k\in 2\mathds{Z}_{\geq 0}$. For example, for the 12-point amplitude we again consider kinematics near the zero defined by $X_B=X_{16}$ but now we relax $c_{19}\neq0$ and $c_{1,10}\neq 0$ subject to the linear constraint $c_{19}+c_{1,10}=0$, this gives 
\begin{align}
    &A_{12}^{\text{NLSM}}\left[1,2,3,4,5,6,7,8,9,10,11,12\right] \rightarrow \nonumber\\
    &\frac{c_{1,9}}{X_{16}X_{5,10}X_{1,10}}A_6^{\text{NLSM}}\left[1,2,3,4,5,6\right]A_6^{\text{NLSM}}\left[5,6,7,8,9,10\right]A_4^{\text{NLSM}}\left[10,11,12,1\right].
\end{align}
It should be possible to derive similar formulae for the extended model that appears on odd-point splits.

\subsubsection{Yang-Mills-Scalar}

Finally we consider higher-order splitting in YMS. Without loss of generality we can assume the external states are scalars in a $2n$-point amplitude, where one can recover the $n$-point gluon amplitudes by taking the so-called scaffolding residue \cite{Arkani-Hamed:2023jry}. On the inside of a diagram both scalars and gluons will appear, giving a more complicated classification of singularities: 
\begin{itemize}
    \item $X_{eo}$ is a scalar factorization channel,
    \item $X_{ee}$ has zero residue,
    \item $X_{oo}$ is a gluon factorization channel.
\end{itemize}
Similar to NLSM, on higher-order splitting kinematics, corresponding to relaxing multiple $c_{ij}$ in the same row, the YMS amplitudes scale as $\sim z^{-1}$ and so there is again no bonus relation. Unlike YMS the ``direction" of the row of conditions we are relaxing is important. If we consider the same example as above i.e. $X_B=X_{16}$ and the third-order splitting conditions, $\{c_*^{(1)},c_*^{(2)},c_*^{(3)}\}=\{c_{19},c_{29},c_{39}\}$, the third-order splitting formula we find is formally identical to (\ref{NLSM10ptExample}). If instead we had chosen to relax $\{c_*^{(1)},c_*^{(2)},c_*^{(3)}\}=\{c_{19},c_{18},c_{17}\}$, then we are unable to derive a splitting formula using recursion since the amplitude scales like $z^0$ at infinity. If we were able to circumvent this problem in the latter case, for example by using a once-subtracted contour integral, then we would have a new residue at $z=X_{19}$ which corresponds to a gluon factorization channel.   

It is tempting to guess that simple higher-order splitting formulae like (\ref{HOsplittingphi3}) and (\ref{HOsplittingNLSM}) will also apply to YMS. Unfortunately the simplest guess, that (\ref{HOsplittingphi3}) is correct on the nose, does not make sense. Due to the pattern of labels in each of the ``triple product" terms, one of the products of sub-amplitudes will always correspond to gluon exchange. In this paper we have not discussed splitting on gluon channels, empirically they have a substantially different structure from scalar channel splits. Possibly they are related to some kind of extended theory, similar to the odd-point splits of NLSM.  

Assuming closed form expressions for higher-order YMS splits do exist, we can predict their behavior under a $\delta$-deformation \cite{Arkani-Hamed:2023swr}. Empirically we have observed the following relation 
\begin{equation}
    \lim_{\delta\rightarrow \infty}\delta^{1-n} \left(A_{2n}^{\text{YMS}}\left[1^{\phi_1},2^{\overline{\phi}_1},3^{\phi_2},4^{\overline{\phi}_2},...\right]\biggr\vert_{\substack{X_{ee}\rightarrow X_{ee}-\delta \\ X_{oo}\rightarrow X_{oo}+\delta}}\right) = A_{2n}^{\text{NLSM}}\left[1,2,3,4,...\right],
\end{equation}
mirroring the known relation between $\text{Tr}\phi^3$ and NLSM under the same deformation (\ref{deltatrphi3}). Therefore any conjectural gluon splitting formula should reduce to the corresponding NLSM splitting formula. We leave the investigation of this case to future work.

\subsection{Uncolored models}
\label{sec:galileon}
All the theories we have discussed so far have had some notion of color/flavor, allowing for the amplitude to be broken into partial amplitudes associated to specific orderings. The kinematic mesh is well-suited to describe such theories, since it separates the positions of poles in an ordered amplitude, $X$'s from the non-pole $c$'s. For an uncolored theory, the full permutation symmetric amplitude does not have two such classes of kinematic variables. It is allowed to have poles in both $X$ and $c$, making the kinematic mesh an ill-suited tool. Nonetheless, in this section we see that much of the previous discussion extends to an uncolored scalar model, the special Galileon model.

The special Galileon model is a higher-derivative scalar model with a 6-derivative quartic coupling. It has been shown to have many interesting properties, including a CHY representation \cite{Cachazo:2014xea}, soft recursion relations \cite{Cheung:2014dqa,Cheung:2015ota} and double-copy constructibility \cite{Cachazo:2014xea}. More recently, it was shown to display hidden zeros \cite{Cao:2024gln,Bartsch:2024amu,Li:2024qfp} and its CHY integrand was shown to split near these zeros \cite{Cao:2024gln,Cao:2024qpp}. This makes it a natural question whether smooth splitting in the Galileon theory can be seen as a consequence of the residue theorems we have been discussing. 

Note that the hidden zeros in uncolored theories work similar to the colored case, except that the amplitude is no longer sensitive to the order of the particles. Thus it becomes the statement of splitting all the labels into 3 sets: $\{i{-}1,j{-}1\}$, $A$ and $B$. This then gives the zero:
\begin{align}
    M^\text{sGal}_n(i{-}1, A, j{-}1, B) \to 0 \text{\ \  when\ \  } s_{ab}=0 \ \forall\ a\in A, b\in B\,.
\end{align}

To understand how the splitting theorems are modified, we start by considering the additional poles that are contained in a special Galileon amplitude that are absent in color-ordered amplitudes. Under a generic $(X_{ij},c_{kl})$-shift (\ref{Xcshift}), all but four of the $c$-variables remain unshifted. These variables are $c_{i{-}1k}$, $c_{i{-}1l}$, $c_{j{-}1k}$ and $c_{j{-}1l}$. Thus the possible additional residues that we need to check are at
\begin{align}
    \hat{c}_A=\sum_{\{a,b\}\in A} \hat{c}_{ab}=0
\end{align}
where $A$ must contain either $i{-}1$ or $j{-}1$ and either $l$ or $k$ i.e. it can be a set of the following types:
\begin{enumerate}
    \item $A\ni \{i{-}1,k\}$ and $A\not\ni \{j{-}1,l\}$,
    \item $A\ni \{i{-}1,l\}$ and $A\not\ni \{j{-}1,k\}$,
    \item $A\ni \{j{-}1,k\}$ and $A\not\ni \{i{-}1,l\}$,
    \item $A\ni \{j{-}1,l\}$ and $A\not\ni \{i{-}1,k\}$.
\end{enumerate}
In each of these cases, the residue on the pole is given by factorization of the amplitude into lower-point ones. Since the lowest valence interaction in Galileon amplitudes is 4, there are no poles of the type $(p_a+p_b)^2 = c_{ab}=0$ (or any other even $|A|$ for that matter). In other words, the simplest case is when $A$ contains 3 elements. One can show that on split kinematics, on each of these poles the amplitude factorizes into two lower-point amplitudes, one of which is being evaluated on a zero. Thus the residue on each of these extra poles vanishes on split kinematics.

Let us look at an example of $M^\text{sGal}_6$ under a $(X_{14},c_{15})$ shift on split kinematics $c_{14}=c_{24}=c_{25}=0$. Here the shifted $c$'s are $c_{16}$, $c_{56}$, $c_{13}$, $c_{35}$ giving as additional poles at
\begin{align}
    \{c_{126},c_{146},c_{256},c_{456},c_{123},c_{134},c_{235},c_{345}\}=0\,.
\end{align}
Consider the behavior of the amplitude on the pole $c_{134}=0$,
\begin{align}
    M^\text{sGal}_6\ \overset{c_{134}\to 0}{\longrightarrow}\ M^\text{sGal}_4(1,3,4, P) M^\text{sGal}_4(2,5,6,-P)\,.
\end{align}
The amplitude $M^\text{sGal}_4(2,6,5,-P)=0$ when $c_{25}=0$. Thus the residue on $\hat{c}_{134}$ vanishes on split kinematics. All other residues vanish similarly.

The only residue left is then the one at infinity. For generic kinematics, the special Galileon amplitude scales as
\begin{align}
    M^\text{sGal}_6 \sim z^2, && 
    M^\text{sGal}_8 \sim z^2.
\end{align}
Compared to its $X$ dimension (5 at 6-point and 7 at 8-point), these amplitudes behave much better than expected as $z\rightarrow\infty$. Still, like in the case of NLSM, this is not good enough scaling to construct a recursion relation for general kinematics.

On split kinematics on the other hand, the amplitudes display enhanced fall-off at infinity:
\begin{align}
    M^\text{sGal}_6 \overset{\text{split}}{\sim} z^{-2}, && 
    M^\text{sGal}_8 \overset{\text{split}}{\sim} z^{-2}.
\end{align}
Thus our recursive proof of the existence of zeros and near-zero splitting discussed in Section \ref{sec:proofofzeros} applies, and we see that these properties extend to special Galileon theory. 

In \cite{Bartsch:2024amu, Li:2024qfp}, the existence of hidden zeros in special Galileon theory was implied by its KLT double copy structure,
\begin{align}
    M^{\text{sGal}}_n=\sum_{\alpha,\beta} S[\alpha|\beta] A^{\text{NLSM}}_n[1,m,n,\alpha]A^{\text{NLSM}}_n[1,\beta,m,n],
\end{align}
for a zero associated to the causal diamond based at $X_{1m{+}1}$. It is then natural to ask where the large $z$ behavior of Galileon amplitudes can also be seen as a consequence of KLT. Evaluating the kernel on a $(X_{1m{+}1},c)$ shift introduces no $z$-dependence into the kernel. Thus the only $z$-dependence of $M^{\text{sGal}}_n$ comes from $A^{\text{NLSM}}_n$. Unfortunately, NLSM amplitudes with different orderings do not display enhanced fall-off at infinity and so the fall-off of the Galileon results from cancellations of the leading $z$ behavior between NLSM amplitudes with different orderings. An interesting exception occurs at 6-point where indeed the large $z$ behavior of the Galileon is manifested term by term in the KLT product.

While YMS and NLSM also admitted interesting residue theorems when two of the $c_{ij}=0$ conditions are relaxed, we see that the special Galileon has a pole at infinity when evaluated on such kinematics,
\begin{align}
    M^\text{sGal}_6 \sim z^0, && 
    M^\text{sGal}_8 \sim z^0,
\end{align}
when two adjacent $c$'s are non-zero. This prevents us from accessing similar higher-order splitting theorems in this case. Finally, let us comment on the odd-point splitting theorems. Like in the case of NLSM, the amplitude on odd splitting kinematics has bad large $z$ behavior,
\begin{align}
    M^\text{sGal}_6 \sim z^2, && 
    M^\text{sGal}_8 \sim z^2.
\end{align}
Thus the odd-point splitting theorems (which in this case involve the Galileon-scalar mixed theory) cannot be derived in the same way as the even-point ones.

Other uncolored theories that one might consider are Einstein-Maxwell-Scalar and Dirac-Born-Infeld amplitudes. Neither of these have hidden zeros and both scale poorly as $z\to\infty$,
\begin{align}
    M_6^\text{EMS} \sim z, &&     M_6^\text{DBI} \sim z^2, &&     M_6^\text{DBI}\ \overset{\text{split}}{\sim}\ z\,,
\end{align}
where the last DBI amplitude is evaluated on split kinematics. Thus there are no recursion relations or splitting theorems that we can derive for these theories. Nevertheless, there exist other scalar theories for which zeros and splitting theorems have been reported \cite{Li:2024qfp}. We leave a careful consideration of their zeros and poles to future work.

\section{Four Dimensions}
\label{sec:4d}
To realize splitting and zero properties of the scalar amplitudes, it is necessary to treat the amplitudes as functions of $\frac{n(n-3)}{2}$ independent Mandelstam invariants\footnote{This section can be read independently of the previous sections.}. For an $n$-particle scattering process this is only possible in spacetime dimensions $d>n-2$, otherwise the Mandelstams are further constrained by non-linear Gram determinant identities. In this section, we study the fate of hidden zeros in four dimensions.

\subsection{Dimensional constraints}

The discussion of zeros and splitting in section \ref{sec:zeros} (as well as in all of the previous literature on this subject \cite{Arkani-Hamed:2023swr,Bartsch:2024amu,Li:2024qfp,Feng:2025ofq,Rodina:2024yfc,Cao:2024qpp,Cao:2024gln,Arkani-Hamed:2024fyd}) assumes that the amplitudes are being calculated in $d$-dimensions, where $d$ is assumed to be sufficiently large that there are no additional dimensionality constraints. For a scalar model like $\text{Tr}\phi^3$ and NLSM, if $d<n-1$, where $n$ is the multiplicity of a scattering amplitude, then there are complicated non-linear \textit{Gram determinant} constraints on the $X$-variables\footnote{As a quick reminder, if $d<n-1$ then there are more external momenta $p_i^\mu$ (after solving momentum conservation) than linearly independent vectors. This implies that for any length-$d$ subsets $\rho$ and $\sigma$ of momenta, the Gram matrix $\{p_i\cdot p_j\}$, $i\in\rho$, $j\in\sigma$ must be singular. The vanishing of the determinant of any such Gram matrix then gives a degree-$d$ polynomial constraint on the dot products $p_i\cdot p_j$. }. The proof of the zeros and splitting formulae given in section \ref{sec:zeros} assumes that the amplitude is parametrized by $\frac{n(n-3)}{2}$ Mandelstam variables that can be varied independently. If there are additional dimensionality constraints, then setting some Mandelstams to zero may force others to \textit{accidentally} vanish. Additionally, if those Mandelstams correspond to physical singularities of the amplitude then we may have a $0/0$ cancellation that spoils the zero. 

As an illustrative example, consider the 6-point scattering amplitude of $\text{Tr}\phi^3$ in $d=4$. There is one independent Gram determinant constraint in this case that we can express as $\text{det}\{p_i\cdot p_j\}=0$, where $i,j\in\{1,2,3,4,5\}$. On the support of the ``skinny zero" $c_{13}=c_{14}=c_{15}=0$, this reduces to
\begin{equation}
    \text{det}\{p_i\cdot p_j\} \rightarrow X_{13}^2 X_{35} X_{46} c_{35} = 0.
\end{equation}
Realizing the zero kinematics in $d=4$ therefore requires us to choose one of the factors in the polynomial to ``accidentally" vanish, that is to pick a branch of the constrained kinematic space. Since $\text{Tr}\phi^3$ has poles in every cyclic channel we see that there is a unique choice, $c_{35}=0$, that preserves the zero. As the multiplicity of scattering increases the Gram polynomials become increasingly complicated, and it is far from obvious that there is always a good choice of kinematic branch that realizes the hidden zero in lower dimensions. 

For general multiplicity in $d=4$, the dimensionality constraints can be trivialized by using spinor-helicity variables (we will use the conventions of \cite{Elvang:2013cua}). For each $c_{ij}$ that is set to zero we have a binary choice, either $\langle ij\rangle = 0$ or $[ij] = 0$. The complicated vanishing Gram polynomial constraints are consequences of the more elementary Schouten identities 
\begin{equation}
    |i]_a [jk] + |j]_a [ki] + |k]_a [ij] = 0, \hspace{10mm} |i\rangle^{\dot{a}}\langle jk\rangle + |k\rangle^{\dot{a}}\langle ki\rangle + |k\rangle^{\dot{a}}\langle ij\rangle  = 0,
\end{equation}
for any spinors corresponding to null momenta $p_i,p_j$ and $p_k$. As a simple corollary, if $[ij]=0$ then $|i] \propto |j]$ and also if $\langle ij\rangle =0$ then $|i\rangle \propto |j\rangle$. In this language, the accidental vanishing of additional Mandelstam variables on the hidden zero kinematics is a simple consequence of the transitivity of proportionality: if $|a] \propto |b]$ and $|a] \propto |c]$ then $|b]\propto |c]$. 

Let's repeat the above 6-point example to illustrate the point. Setting $c_{13}=0$ requires us to make a choice, either $\langle 13\rangle=0$ or $[13]=0$; since this is a parity preserving scalar model, without loss of generality we will choose the angle bracket to vanish. For the second kinematic condition $c_{14}=0$ we again have a choice, either $\langle 14\rangle = 0$ or $[14]=0$. If we choose the angle bracket to also vanish in this case then we have the following chain of implications
\begin{equation}
    |1\rangle \propto |3\rangle \;\; \text{and} \;\;|1\rangle \propto |4\rangle \;\; \Rightarrow \;\; |3\rangle \propto |4\rangle \;\;\Rightarrow \;\; \langle 34\rangle = 0 \;\;\Rightarrow \;\; X_{35}=0. 
\end{equation}
This choice was therefore bad since it set a cyclic Mandelstam to zero, so we have to instead choose $[14]=0$. For the final condition $c_{15}=0$, by the same reasoning we find we have to choose $\langle 15\rangle = 0$ to avoid a potential pole at $X_{46}=0$. Since we have chosen $\langle 13\rangle =\langle 15\rangle = 0 $, this forces $c_{35}$ to accidentally vanish (the same conclusion we found from the Gram polynomial) but this is harmless since it is non-cyclic. 

For $\text{Tr}\phi^3$ it is straightforward to generalize this argument to arbitrary multiplicity. When the kinematic conditions that define the zero are organized in the mesh as a rectangle, it is clear that to avoid accidentally vanishing cyclic Mandelstam variables the choice of vanishing angle/square brackets must alternate along each row and column like the squares on a checkerboard. 

\begin{figure}
    \centering
        		\begin{tikzpicture}[scale=0.75]
                \draw (0,0)--(4,4);
			\draw (0,1)--(3.5,4.5);
			\draw(0,2)--(3,5);
			\draw(0,3)--(2.5,5.5);
			\draw(0,4)--(2,6);
			\draw (0,5)--(1.5,6.5);
			\draw(0,6)--(1,7);
			\draw(0,7)--(0.5,7.5);
			\draw(0,0)--(0,8);
			\draw (4,0)--(4,8);
			\draw(0,8)--(4,4);
			\draw(0,7)--(3.5,3.5);
			\draw(0,6)--(3,3);
			\draw(0,5)--(2.5,2.5);
			\draw(0,4)--(2,2);
			\draw(0,3)--(1.5,1.5);
			\draw(0,2)--(1,1);
			\draw(0,1)--(0.5,0.5); 
                \draw[opacity=0.2](0.5,0.5)--(1,0); 
                \draw[opacity=0.2](1,1)--(2,0); 
                \draw[opacity=0.2](1.5,1.5)--(3,0); 
                \draw[opacity=0.2](2,2)--(4,0);
                \draw[opacity=0.2](2.5,2.5)--(4,1);
                \draw[opacity=0.2](3,3)--(4,2);
                \draw[opacity=0.2](3.5,3.5)--(4,3);
                \draw[opacity=0.2](3.5,4.5)--(4,5); 
                \draw[opacity=0.2](3,5)--(4,6);
                \draw[opacity=0.2](2.5,5.5)--(4,7);
                \draw[opacity=0.2](2,6)--(4,8);
                \draw[opacity=0.2](1.5,6.5)--(3,8);
                \draw[opacity=0.2](1,7)--(2,8);
                \draw[opacity=0.2](0.5,7.5)--(1,8);
                \draw[opacity=0.2](1,0)--(4,3);
                \draw[opacity=0.2](2,0)--(4,2);
                \draw[opacity=0.2](3,0)--(4,1);
                \draw[opacity=0.2](4,5)--(1,8);
                \draw[opacity=0.2](4,6)--(2,8);
                \draw[opacity=0.2](4,7)--(3,8);
			\node at (0.5,0.5) {\tiny$\bullet$};
                \node at (1,1) {\tiny$\bullet$};
			\node at (1.5,1.5) {\tiny$\bullet$};
                \node at (2,2) {\tiny$\bullet$};
			\node at (2.5,2.5) {\tiny$\bullet$};
                \node at (3,3) {\tiny$\bullet$};
			\node at (3.5,3.5) {\tiny$\bullet$};
                \node at (0.5,1.5) {\tiny$\bullet$};
                \node at (1,2) {\tiny$\bullet$};
			\node at (1.5,2.5) {\tiny$\bullet$};
                \node at (2,3) {\tiny$\bullet$};
			\node at (2.5,3.5) {\tiny$\bullet$};
                \node at (3,4) {\tiny$\bullet$};
			\node at (3.5,4.5) {\tiny$\bullet$};
                \node at (0.5,2.5) {\tiny$\bullet$};
                \node at (1,3) {\tiny$\bullet$};
			\node at (1.5,3.5) {\tiny$\bullet$};
                \node at (2,4) {\tiny$\bullet$};
			\node at (2.5,4.5) {\tiny$\bullet$};
                \node at (3,5) {\tiny$\bullet$};
                \node at (0.5,3.5) {\tiny$\bullet$};
                \node at (1,4) {\tiny$\bullet$};
			\node at (1.5,4.5) {\tiny$\bullet$};
                \node at (2,5) {\tiny$\bullet$};
			\node at (2.5,5.5) {\tiny$\bullet$};
                \node at (0.5,4.5) {\tiny$\bullet$};
                \node at (1,5) {\tiny$\bullet$};
			\node at (1.5,5.5) {\tiny$\bullet$};
                \node at (2,6) {\tiny$\bullet$};
                \node at (0.5,5.5) {\tiny$\bullet$};
                \node at (1,6) {\tiny$\bullet$};
			\node at (1.5,6.5) {\tiny$\bullet$};
                \node at (0.5,6.5) {\tiny$\bullet$};
                \node at (1,7) {\tiny$\bullet$};
                \node at (0.5,7.5) {\tiny$\bullet$};
			\node at (2,2.5) {\color{black}\tiny $\langle16\rangle$};
			\node at (2.5,3) {\color{black}\tiny $[17]$};
			\node at (3,3.5) {\color{black}\tiny $\langle18\rangle$};
			\node at (3.5,4) {\color{black}\tiny $[19]$};
			\node at (1.5,3) {\color{black}\tiny $[26]$};
			\node at (2,3.5) {\color{black}\tiny $\langle27\rangle$};
			\node at (2.5,4) {\color{black}\tiny $[28]$};
			\node at (3,4.5) {\color{black}\tiny $\langle29\rangle$};
			\node at (1,3.5) {\color{black}\tiny $\langle36\rangle$};
			\node at (1.5,4) {\color{black}\tiny $[37]$};
			\node at (2,4.5) {\color{black}\tiny $\langle38\rangle$};
			\node at (2.5,5) {\color{black}\tiny $[39]$};
			\node at (0.5,4) {\color{black}\tiny $[46]$};
			\node at (1,4.5) {\color{black}\tiny $\langle47\rangle$};
			\node at (1.5,5) {\color{black}\tiny $[48]$};
			\node at (2,5.5) {\color{black}\tiny $\langle49\rangle$};
                \node[left] at (0,0) {\scriptsize $1$};
                \node[left] at (0,1) {\scriptsize $2$};
                \node[left] at (0,2) {\scriptsize $3$};
                \node[left] at (0,3) {\scriptsize $4$};
                \node[left] at (0,4) {\scriptsize $5$};
                \node[left] at (0,5) {\scriptsize $6$};
                \node[left] at (0,6) {\scriptsize $7$};
                \node[left] at (0,7) {\scriptsize $8$};
                \node[left] at (0,8) {\scriptsize $9$};
                \node[right] at (4,0) {\scriptsize $6$};
                \node[right] at (4,1) {\scriptsize $7$};
                \node[right] at (4,2) {\scriptsize $8$};
                \node[right] at (4,3) {\scriptsize $9$};
                \node[right] at (4,4) {\scriptsize $10$};
                \node[right] at (4,5) {\scriptsize $1$};
                \node[right] at (4,6) {\scriptsize $2$};
                \node[right] at (4,7) {\scriptsize $3$};
                \node[right] at (4,8) {\scriptsize $4$};
                \fill[color=teal,opacity=0.2] (0,4)--(2,2)--(4,4)--(2,6);
		\end{tikzpicture}
    \caption{Illustration of a 4d ``checkerboard" zero of $\text{Tr}\phi^3$, the angle/square bracket in each plaquette is chosen to vanish.}
    \label{fig:checkerboard}
\end{figure}
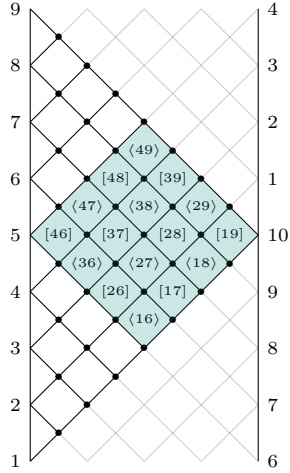

A 10-point example of a ``checkerboard" zero with $X_B=X_{16}$ is shown in Figure \ref{fig:checkerboard}. This example illustrates a general feature of zeros from maximal rectangles with 3 or more rows and columns, the accidental spinor relations in this case are
\begin{align}
    &|1\rangle \propto |3\rangle, \hspace{10mm} |2\rangle \propto |4\rangle, \hspace{10mm}  |6\rangle \propto |8\rangle, \hspace{10mm}  |7\rangle \propto |9\rangle, \nonumber\\
    &|1] \propto |3], \hspace{10.75mm} |2] \propto |4], \hspace{10.75mm}  |6] \propto |8], \hspace{10.75mm}  |7] \propto |9].
\end{align}
Since the same pairs of square and angle spinors are proportional, this is therefore seen to be a multi-collinear limit of non-adjacent pairs of external particles
\begin{equation}
    p_1^\mu \propto p_3^\mu, \hspace{10mm} p_2^\mu \propto p_4^\mu, \hspace{10mm} p_6^\mu \propto p_8^\mu, \hspace{10mm} p_7^\mu \propto p_9^\mu. 
\end{equation}
This also means that without loss of generality the 4d zeros are realizable in \textit{real} kinematics $X_{ij}\in \mathds{R}$. 

We therefore conclude that the hidden zeros of $\text{Tr}\phi^3$ \textit{can} be realized in $d=4$ and moreover the required branch of kinematic space is unique up a trivial parity transformation that corresponds to interchanging all angle and square brackets. 

\subsection{YM and YMS in four dimensions}
\label{sec:4dYM}

For theories with spinning particles, amplitudes are no longer simply a function of $c_{ij}$, but now also depend on other Lorentz invariants involving the polarization vectors. In \cite{Arkani-Hamed:2023swr}, a simple generalization of hidden zeros to particles with spin was presented. These zero loci are still characterized by a maximal rectangle $X_{B}=X_{ij}$, the only difference being that instead of only setting $c_{ab}\in \mathcal{Z}(X_{ij})$ to zero, we also set
\begin{align}
    \varepsilon_a\cdot p_b=0\,, && p_a\cdot \varepsilon_b = 0\,, && \varepsilon_a\cdot\varepsilon_b=0 && \forall\ \ \ \  c_{ab}=0\,.
\end{align}
We will denote this as $\mathcal{Z}^{\text{spin}}(X_{ij})$. Similarly, when particle $a$ has spin 1 but particle $b$ has spin 0, the condition reads
\begin{align}
    \varepsilon_a\cdot p_b =0 &&\forall\ \ \ \ c_{ab}=0\,.
\end{align}
These are a dimension-agnostic representation of these zero conditions i.e. in dimensions higher than $2n{-}2$. In this section, we discuss how hidden zeros manifest in four dimensions in pure YM and YMS amplitudes.

The polarization vectors associated to the two transverse helicity states in four dimensions are
\begin{align}
    \varepsilon^{+\mu}_i \propto \frac{[i|\sigma^\mu|q_i\rangle}{\langle iq_i\rangle}\,, && \varepsilon^{-\mu}_i \propto \frac{\langle i|\overline{\sigma}^\mu|q_i]}{[iq_i]}\,.
\end{align}
The spinors $|q_i\rangle$ and $|q_i]$ are arbitrary, but cannot be proportional to $|i\rangle$ and $|i]$ respectively. If they were it would imply that $\varepsilon_i^\mu \propto p_i^\mu$, and so contradict the assumption that the polarizations are transverse. 

Consider a zero condition that involves one spin 1 particle (for example of positive helicity):
\begin{align}
    p_i\cdot p_j=\sq{ij}\ang{ij} = 0\,, && \varepsilon^+_i\cdot p_j \propto\frac{\sq{ij}\ang{jq_i}}{\ang{iq_i}} = 0\,. 
\end{align}
If $\ang{ij}=0$ and $\sq{ij}\ne0$, then the second condition tells us that $\ang{jq_i}=0$. But,
\begin{align}
    \ang{jq_i} = \ang{ij}=0\ \Rightarrow\ |i\rangle \propto |j\rangle \propto |q_i\rangle\,,
\end{align}
and this is not an allowed choice of reference spinor. Thus, for positive helicity particles the zero condition is $\sq{ij}=0$ while it is $\ang{ij}=0$ for negative helicity particles. This makes it clear that a spinning zero condition, $\{c_{ab},p_a\cdot \varepsilon_b, p_b\cdot \varepsilon_a, \varepsilon_a \cdot \varepsilon_b\}=0$, can never be satisfied in four dimensions unless $\varepsilon_a$ and $\varepsilon_b$ have the same helicity. 

Putting all of this together, we see that unlike in general dimensions, particular zeros $\mathcal{Z}^\text{spin}(X_{ij})$ are related to particular choices of helicity states in four dimensions. Indeed the only type of helicity configurations for all-gluon amplitudes in which a hidden zero can be present is MHV (or anti-MHV). These amplitudes are given by the familiar Parke-Taylor factor, given in a particular cyclic sector as
\begin{align}
\label{eq:YMHV}
    A_n^{\text{YM}}[1^+\cdots i^-\cdots j^-\cdots n^+] = \frac{\ang{ij}^4}{\ang{12}\ang{23}\cdots\ang{n{-}n}\ang{n1}} \,.
\end{align}
This helicity configuration is compatible with the zero conditions $\mathcal{Z}^\text{spin}(X_{i{+}1,j{+}1})$. We can see this by applying the zero condition to the momentum conservation equation
\begin{align}
    &\sq{12}\langle2| + \cdots \sq{1i}\langle i| + \cdots \sq{1j}\langle j|  +\cdots \sq{1n}\langle n| = 0\nonumber\\
    &\Rightarrow [1i]\ang{ij}=0 \Rightarrow \ang{ij}=0\,,
\end{align}
where in the second line we have used the zero condition
\begin{align}
    \sq{1a}=0\ \  \forall \ \ a\ne i,j\ \ \text{ and }\ \ [1i], [1j]\ne 0\,.
\end{align}
We now move on to YMS amplitudes with gluons and some scalars. To begin with, let us consider the case of two scalars. For a zero $\mathcal{Z}^\text{spin}(X_{i{+}1,j{+}1})$, the possible helicity configurations are:
\begin{enumerate}
    \item MHV amplitudes: Replace the two negative helicity particles in \eqref{eq:YMHV} with pairs of scalars e.g. 
    \begin{align}
        A_n^{\text{YMS}} [1^+\cdots i^\phi (i{+}1)^{\bar\phi} \cdots j^\phi (j{+}1)^{\bar\phi} \cdots n^+] = \frac{\ang{ij}^2\ang{i{+}1, j{+}1}^2}{\ang{12}\ang{23}\cdots\ang{n1}} \overset{\mathcal{Z}^\text{spin}(X_{i{+}1,j{+}1})}{\longrightarrow} 0.
    \end{align}
    \item NMHV amplitudes: Replace any pair $a^+ b^+$ in \eqref{eq:YMHV} with $a^\phi b^{\bar\phi}$ e.g.  
    \begin{align}
    \label{eq:NMHV6}
        A_6^{\text{YMS}} [1^+ 2^- 3^+ 4^- 5^\phi 6^{\bar{\phi}}] = &-\frac{\ang{26}^2\sq{35}^2\langle 2|6+1|3]^2}{\ang{12}\sq{34}[45][5|3+4|2\rangle [3|1+2|6\rangle\ang{16}s_{126}}\nonumber\\
        &+ \frac{\sq{15}^2\sq{16}\ang{24}^4}{\sq{56}\langle4|2+3|1] \ang{23}\ang{34} \langle 2|3+4|5] s_{234}}\nonumber\\
        &-\frac{\sq{13}^4\ang{45}\ang{56}}{\sq{12}\sq{23} [3|1+2|6\rangle\langle4|2+3|1] s_{123}}\nonumber\\
        &\overset{\mathcal{Z}^\text{spin}(X_{35})}{\longrightarrow} 0,
    \end{align}
    where $\mathcal{Z}^\text{spin}(X_{24})$ sets $\ang{24}=\ang{25}=\ang{26}=0$ and via Schouten, $\ang{45}=\ang{56}=\ang{46}=0$. 6-point amplitudes such as the one above were calculated in \cite{Badger:2005zh} via the BCFW recursion relations \cite{Britto:2004ap, Britto:2005fq}.
    \item N$^k$ MHV amplitudes: Though one can keep adding $\phi\bar{\phi}$ pairs, generic amplitudes with more than two negative helicity gluons cannot realize the zero conditions on polarization vectors. However, the skinny zero i.e. $\mathcal{Z}^\text{spin}(X_{ii{+}2})$ can be realized in an arbitrary N$^k$MHV amplitude where $i{+}1$ is a scalar, e.g.
    \begin{align}    
    \label{eq:NMHV62}
    A_6^{\text{YMS}}\left[1^\phi 2^+ 3^+ 4^- 5^- 6^{\bar{\phi}}\right]  
    =& \frac{s_{123}\left([64]\langle 41\rangle +[65]\langle 51\rangle \right)^2}{\langle 12\rangle \langle 23\rangle [45][56]  [6|(4+5)|3\rangle [4|(2+3)|1\rangle} \nonumber\\
    &- \frac{[26]^2 [12]\langle 45\rangle^3}{s_{126}\langle 34\rangle [16]\langle 5|(3+4)|2] \langle 3|(4+5)|6]} \nonumber\\
    &-\frac{\langle 56\rangle \langle 15\rangle^2 [23]^3}{ s_{156}\langle 16\rangle [34]\langle 5|(3+4)|2][4|(2+3)|1\rangle} \nonumber\\
    &\overset{\mathcal{Z}^\text{spin}(X_{13})}{\longrightarrow} 0,
\end{align}
    which vanishes when $[13]=\ang{14}=\ang{15}=0$ (leading by Schouten to $\ang{45}=0$ as well). 
\end{enumerate}
These examples make two things clear. The first is that in four dimensions, spinning hidden zeros $\mathcal{Z}^\text{spin}(X_{ij})$ can only be realized in certain helicity configurations. Second we observe that in \eqref{eq:NMHV6} and \eqref{eq:NMHV62}, where the answer is written as a sum of BCFW terms, on the support of the zero each term is vanishing \textit{independently}. This leads to the natural question of whether BCFW recursion relations can prove the existence of the four dimensional counterparts of the $d$-dimensional hidden zeros. Related $d$-dimensional discussions can be found in \cite{Rodina:2024yfc,Feng:2025ofq}.

\subsection{BCFW and helicity zeros}
\label{sec:BCFW}
The BCFW recursion relations can be used to recursively construct tree amplitudes and loop integrands in a variety of theories \cite{Britto:2004ap,Britto:2005fq}. Importantly, both the theories studied in the previous section are BCFW constructible. In this section, we present a proof that each term in the BCFW expansion vanishes on the support of hidden zeros, giving a second kind of recursive proof of the hidden zeros, but in this case one special to four dimensions. 

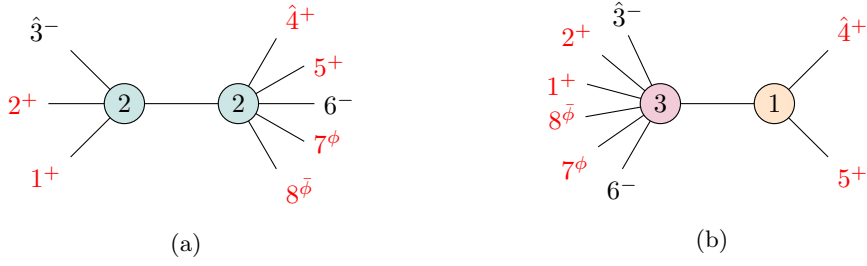
\begin{figure}[t]
    \centering
    \begin{subfigure}[t]{0.45\textwidth}
        \centering
        \begin{tikzpicture}[baseline={(0,0)cm},scale=1]
        \draw (0,0)--(-1,0) node[at end, left, red] {$2^+$};
        \draw (0,0)--(-{cos(45)},{cos(45)}) node[at end, above left] {$\hat{3}^-$};
        \draw (0,0)--(-{cos(45)},-{cos(45)}) node[at end, below left, red] {$1^+$};
        \draw (0,0)--(1.5,0);
        \foreach \a in {4,5,6,7,8} {
        \draw (1.5,0)--({1.5+cos(90-30*(\a-3))},{sin(90-30*(\a-3))});}
        \node[above right, red] at ({1.5+cos(90-30*(4-3))},{sin(90-30*(4-3))}) {$\hat{4}^+$};
        \node[right, red] at ({1.5+cos(90-30*(5-3))},{sin(90-30*(5-3))}) {$5^+$};
        \node[right] at ({1.5+cos(90-30*(6-3))},{sin(90-30*(6-3))}) {$6^-$};
        \node[right, red] at ({1.5+cos(90-30*(7-3))},{sin(90-30*(7-3))}) {$7^\phi$};
        \node[below right, red] at ({1.5+cos(90-30*(8-3))},{sin(90-30*(8-3))}) {$8^{\bar\phi}$};
	\draw[black,fill=white] (1.5,0) circle (1.75ex);
	\draw[black,fill=white] (0,0) circle (1.75ex);
	\draw[black,fill=teal, opacity=0.2] (1.5,0) circle (1.75ex);
	\draw[black,fill=teal, opacity=0.2] (0,0) circle (1.75ex);
        \node at (0,0) {2};
        \node at (1.5,0) {2};
    \end{tikzpicture}
        \caption{}
        \label{fig:MHVtimesMHV}
    \end{subfigure}
    \begin{subfigure}[t]{0.45\textwidth}
        \centering
        \begin{tikzpicture}[baseline={(0,0)cm},scale=1]
        \draw (1.5,0)--({1.5+cos(45)},-{cos(45)}) node[at end, below right, red] {$5^+$};
        \draw (1.5,0)--({1.5+cos(45)},{cos(45)}) node[at end, above right, red] {$\hat{4}^+$};
        \draw (0,0)--(1.5,0);
        \foreach \a in {4,5,6,7,8,9} {
        \draw (0,0)--({0-cos(90-25*(\a-3))},{sin(90-25*(\a-3))});}
        \node[above] at ({0-cos(90-25*(4-3))},{sin(90-25*(4-3))}) {$\hat{3}^-$};
        \node[above left, red] at ({0-cos(90-25*(5-3))},{sin(90-25*(5-3))}) {$2^+$};
        \node[left, red] at ({0-cos(90-25*(6-3))},{sin(90-25*(6-3))}) {$1^+$};
        \node[left, red] at ({0-cos(90-25*(7-3))},{sin(90-25*(7-3))}) {$8^{\bar\phi}$};
        \node[below left, red] at ({0-cos(90-25*(8-3))},{sin(90-25*(8-3))}) {$7^\phi$};
        \node[below] at ({0-cos(90-25*(9-3))},{sin(90-25*(9-3))}) {$6^-$};
	\draw[black,fill=white] (1.5,0) circle (1.75ex);
	\draw[black,fill=white] (0,0) circle (1.75ex);
	\draw[black,fill=orange, opacity=0.2] (1.5,0) circle (1.75ex);
	\draw[black,fill=purple, opacity=0.2] (0,0) circle (1.75ex);
        \node at (0,0) {3};
        \node at (1.5,0) {1};
    \end{tikzpicture} 
        \caption{}
        \label{fig:aMHVtimesNMHV}
    \end{subfigure}
    \caption{The two types of BCFW terms in the $[3,4\rangle$ expansion of $A_8[1^+2^+3^-4^+5^+6^-7^\phi8^{\bar{\phi}}]$ with the states satisfying $[ab]=0$ highlighted in red. The numbers in the blobs refer to N$^{k{-}2}$MHV amplitudes. (a) is an MHV $\times$ MHV term. The term vanishes because the right sub-amplitude is evaluated on a zero. (b) is a anti-MHV $\times$ NMHV term. This vanishes due to 3-point kinematics, provided the lower-point NMHV amplitude has the zero.}
    \label{fig:BCFWofNMHV}
\end{figure}

The BCFW shift chooses two external legs $i, j$ and shifts them as
\begin{align}
    |\hat{i}]=|i]+z|j]\,, &&
    |\hat{j}\rangle = |j\rangle - z |i\rangle\,.
\end{align}
Under such a shift, YM and YMS amplitudes fall off at large $z$, allowing for a recursive construction of its scattering amplitudes. Since these are color-ordered theories, shifting adjacent legs, i.e. $j=i+1$, the recursion relations take the simple form
\begin{align}
    A_n = \sum_{k=3}^{n-1} \text{Res}\left[\frac{\hat{A}_k(z)\hat{A}_{n-k+2}(z)}{zP_I^2(z)};z=z_k\right]\,.
\end{align}
Since the all-gluon case is trivial, we will begin with the case of two scalars. Consider an 8-point example $A^{\text{YMS}}_8[1^+2^+3^-4^+5^+6^-7^\phi8^{\bar{\phi}}]$ on the spinning zero $\mathcal{Z}^{\text{spin}}(X_{47})$. From the discussion above this translates into the following conditions on the spinors
\begin{align}
    |1]\propto|2] \propto |4]\propto|5]\propto|7]\propto|8]\,.
\end{align}
Under a $[3,4\rangle$ shift of this NMHV amplitude, we get two types of terms given in Figure \ref{fig:BCFWofNMHV}. The first type is MHV $\times$ MHV, shown in \ref{fig:MHVtimesMHV}. This vanishes because the right sub-amplitude is evaluated on
\begin{align}
\label{eq:BCFWzero1}
    |4]\propto|5]\propto|7]\propto|8]\,,
\end{align}
which corresponds to the zero $\mathcal{Z}^{\text{spin}}(X_{47})$ of the sub-amplitude $\hat{A}_6^{\text{YMS}}\left[\hat{4}^+5^+6^-7^\phi 8^{\bar{\phi}}\hat{P}^+\right]$. The vanishing of the remaining MHV $\times$ MHV terms is similar. The second type is \ref{fig:aMHVtimesNMHV}. Here the right sub-amplitude is an anti-MHV 3-point amplitude and 3-point kinematics requires that the spinor representation of the intermediate momentum satisfies $|\hat{P}]\propto |\hat{4}] \propto |5]$. The left NMHV amplitude is being evaluated on
\begin{align}
\label{eq:BCFWzero2}
    |1]\propto|2]\propto|\hat{P}]\propto|7]\propto|8]\,,
\end{align}
corresponding to the zero $\mathcal{Z}^{\text{spin}}(X_{P,7})$ of the shifted sub-amplitude $\hat{A}^{\text{YMS}}_7\left[6^-7^\phi 8^{\bar{\phi}} 1^+ 2^+ \hat{3}^- \hat{P}^+\right]$. Thus if the 7-point NMHV amplitude has a zero, so does the 8-point NMHV amplitude.

The proof of the existence of such hidden zeros from BCFW always follows in the same manner. Given an amplitude $A_n[12\cdots i\cdots j\cdots n]$ being evaluated on a zero condition involving all particles except $i$ and $j$. The BCFW expansion under a shift of $i$ and $i{+}1$ has two types of terms: those in which $i$ and $j$ are on different sides and one in which $i$ and $j$ are on the same side. The former contains a sub-amplitude that vanishes manifestly on the zero condition, while the latter requires use of 3-point kinematics in order to see that it vanishes. These arguments prove the existence of 4d hidden zeros via BCFW in all the helicity configurations discussed in Section \ref{sec:4dYM}.

In fact, a small modification of this BCFW argument leads to an interesting new class of zeros that are present also for pure gluon amplitudes and in all helicity sectors: \\
\\
\fbox{
\begin{minipage}{\textwidth}
    \textbf{Helicity zero:} for \textit{any} 4d YM helicity amplitude, either set all of the square spinors of the positive helicity particles proportional \textit{or} set all of the angle spinors of the negative helicity particles proportional, then the amplitude will vanish.
\end{minipage}
}
\\
\vspace{3mm}
\\
Take for example the NMHV gluon amplitude $A^{\text{YM}}_6[1^-2^-3^-4^+5^+6^+]$. This vanishes when 
\begin{align}
    |1\rangle\propto|2\rangle\propto|3\rangle\ \  \text{ or }\ \ |4]\propto|5]\propto|6]\,.
\end{align}
This happens term by term in its BCFW expansion:
\begin{align}
    A^{\text{YM}}_6[1^-2^-3^-4^+5^+6^+] = \frac{\langle 3|1+2|6]^3}{[12]\ang{16}\ang{34}\ang{45}\langle 5|1+6|2]s_{126}}+\frac{\langle 1|5+6|4]^3}{[23][34]\ang{16}\ang{56}\langle 5|1+6|2]s_{156}}\,.
\end{align}
We see this by using momentum conservation and for instance the first condition $|1\rangle\propto|2\rangle\propto|3\rangle$,
\begin{align}
    \langle 1 | 1+ 2+ 3+ 4+ 5+ 6|4] = \langle 1|5+6|4]=0\,.
\end{align}
The proof is essentially identical to the one presented above for hidden zeros in YMS from the BCFW expansion. Term-by-term in the recursion relation, the result vanishes assuming lower-multiplicity helicity zeros have been established. 

The helicity zeros and the 4d realizable hidden zeros are clearly closely related and show that the latter is often over-constrained. For example consider (\ref{eq:NMHV62}), this amplitude vanishes if we set $|1\rangle \propto |4\rangle \propto |5\rangle$, corresponding to a helicity zero where we treat $1^\phi$ as a ``negative helicity" particle.\footnote{This is natural if we recognize that 4d YMS, with a single complex scalar, is the bosonic truncation of pure $\mathcal{N}=2$ super-Yang-Mills; the scalars $\phi$ and $\bar{\phi}$ belonging to CPT conjugate positive- and negative-helicity multiplets respectively. More generally the statement of the helicity zero and the BCFW argument naturally generalizes to arbitrary numbers of complex scalars, where each $\phi$ and $\bar{\phi}$ pair are given opposite helicity assignments.} This means the condition $[13]=0$ in the hidden zero is actually redundant, it can be relaxed and the amplitude still vanishes. We have observed in many other examples that the realizable 4d hidden zeros can be relaxed into a helicity zero in this way. Understanding the full space of zeros of 4d helicity amplitudes clearly deserves further investigation. 

Despite having many more terms, the BCFW expansion of graviton amplitudes satisfies the same helicity selection rules. Thus the discussion above easily extends to gravity, where the zero conditions now involve the polarization tensors
\begin{align}
    \varepsilon^+_{\mu\nu}=\varepsilon_\mu^+ \varepsilon_\nu^+\,, &&\varepsilon^-_{\mu\nu}=\varepsilon_\mu^- \varepsilon_\nu^-\,.
\end{align}
The fact that these ``factorizable'' tensors give \emph{all} graviton polarization tensors is special to 4d. In higher dimensions not all polarization tensors can be written as $\varepsilon^i_\mu\varepsilon^i_\nu$. Thus in four dimensions, we can show the presence of helicity zeros in all helicity configurations and hidden zeros for some N$^k$MHV ones, via the BCFW expansion of graviton amplitudes, analogous to the gluon case .

\section{Discussion}
\label{sec:discussion}

In this work, we introduced a kinematic shift under which Tr$\phi^3$, NLSM, YMS and special Galileon have good large $z$ behavior on split kinematics. This shift is a particular instance of a $g$-vector shift that arises naturally in the study of the surface description of amplitudes in these theories \cite{Arkani-Hamed:2023lbd, Arkani-Hamed:2024vna, Arkani-Hamed:2023mvg}. In some cases, the good behavior extends to generic or near-split kinematics as well, allowing us to derive a large class of smooth splitting theorems in Tr$\phi^3$, NLSM and YMS. One result of these theorems is the proof of the existence of hidden zeros and near-zero splitting in these theories.  
	
For the NLSM, YMS and special Galileon models, we restricted to cases where the split sub-amplitudes are even-point i.e. there is no need to introduce other external states into the amplitude in order to describe the splitting. In the case of YMS, this would require a gluon ``internal state'' whose polarizations would need to be summed over, producing an unfactorized splitting theorem. For NLSM and special Galileon, it is expected that the odd-point amplitudes belong to the soft-extended versions of these theories as introduced in \cite{Cachazo:2016njl,Arkani-Hamed:2023swr}. Indeed for odd-point splitting, all of these theories have poles at infinity. It would be interesting to investigate the connection between extended theories, odd-point splitting and this pole at infinity, similar to the discussion in \cite{Paranjape:2025wjk}.
	
Another limitation of our residue theorems is that they only apply to scalar theories. Understanding the smooth splitting properties of $d$-dimensional YM and gravity from residue theorems is an important future direction. This could either be via the scaffolding residue of YMS \cite{Arkani-Hamed:2023jry} and newly introduced ``scalar-scaffolded gravity'' \cite{Li:2024qfp} or by introducing shifted kinematics for the polarization vectors directly.
	
Also, hidden zeros and near-zero splitting have recently been found to exist in graph contributions to cosmological wavefunction coefficients \cite{De:2025bmf}. Whether our discussion of residue theorems as the origin of such properties (especially in the $g$-vector language) extends to these cosmological graphs is a possible direction for future research.
	
Finally, we also studied how hidden zeros manifest in four dimensions in scalar theories, YM and gravity. We found that certain helicity configurations admit hidden zeros and all helicity amplitudes admit so-called helicity zeros. Indeed both these types of zeros makes themselves known at the level of individual BCFW terms in the BCFW expansion of these 4d gluon and graviton amplitudes. This suggests that it is possible to see the YM hidden zeros from the flattening of positive geometry constructions that rely on gluing together BCFW terms. Examples of such geometric constructions include the non-supersymmetric Hodges polytope \cite{Arkani-Hamed:2010wgm} and the supersymmetric amplituhedron \cite{Arkani-Hamed:2013jha}. Indeed this leads to a larger question of whether flattening limits of the amplituhedron exist. It can be shown that each of the R-invariants vanish when projected onto the correct Grassmann sector, and so it may be possible to construct hidden zeros for superamplitudes that now include conditions not only on the spinors, but also on the on-shell superspace variables.

\vspace{3mm}
\noindent \textbf{Acknowledgment}

\vspace{3mm}
We would like to thank Nima Arkani-Hamed, Henriette Elvang, Carolina Figueiredo, Umut Oktem, Marcos Skowronek and Jaroslav Trnka for helpful discussions. SP is supported by Simons Investigator Award \#376208. The research of CRTJ is funded by the European Union (ERC UNIVERSE+, 101118787). Views and opinions expressed are, however, those of the author(s) only and do not necessarily reflect those of the European Union or the European Research Council Executive Agency. Neither the European Union nor the granting authority can be held responsible for them. CRTJ wishes to convey his gratitude to Brown University for their hospitality during the completion of this work.

\appendix

\bibliographystyle{JHEP}
\bibliography{cite.bib}
\end{document}